\documentclass[fleqn,usenatbib]{mnras}

\usepackage{newtxtext,newtxmath}

\usepackage[T1]{fontenc}
\usepackage{ae,aecompl}

\usepackage{graphicx}	
 
\usepackage{amsmath}	
\usepackage{amssymb}	
\usepackage{pifont}

\usepackage{booktabs, blkarray}
\usepackage{color}
\usepackage{lineno}

\definecolor{color27}{rgb}{0.2, 0.6, 0.8}

\newcommand{\gpch}{\,h^{-1}{\rm Gpc}}
\newcommand{\mpch}{\,h^{-1}{\rm Mpc}}
\newcommand{\hmpc}{\,h{\rm Mpc}^{-1}}

\usepackage[dvipsnames]{xcolor}
\usepackage[normalem]{ulem}

\begin{document}

\title[eBOSS DR16 QSOs in configuration space]
{The Completed SDSS-IV extended Baryon Oscillation Spectroscopic Survey: BAO and RSD measurements from anisotropic clustering analysis of the  Quasar Sample in configuration space between redshift 0.8 and 2.2}

\author[J. Hou et al.]{\parbox{\textwidth}{
Jiamin Hou\thanks{Email: jiamin.hou@mpe.mpg.de}$^{1}$,
Ariel G. S\'anchez$^{1}$,
Ashley J. Ross$^{2}$,
Alex Smith$^{3}$,
Richard Neveux$^{3}$,
Julian Bautista$^{4}$,
Etienne Burtin$^{3}$,
Cheng Zhao$^{5}$,
Rom\'an Scoccimarro$^{6}$,
Kyle S. Dawson$^{7}$,
Arnaud de Mattia$^{3}$,
Axel de la Macorra$^{8}$,
H\'elion du Mas des Bourboux$^{7}$,
Daniel J. Eisenstein$^{9}$,
H\'ector Gil-Mar\'in$^{10,11}$,
Brad W. Lyke$^{12}$,
Faizan G. Mohammad$^{13,14}$,
Eva-Maria Mueller$^{15}$,
Will J. Percival$^{13,14,16}$,
Graziano Rossi$^{17}$,
Mariana Vargas Maga\~na$^{8}$,
Pauline Zarrouk$^{3,18}$,
Gong-Bo Zhao$^{4,19}$,
Jonathan Brinkmann$^{20}$,
Joel R. Brownstein$^{7}$,
Chia-Hsun Chuang$^{21}$,
Adam D. Myers$^{12}$,
Jeffrey A. Newman$^{22}$,
Donald P. Schneider$^{23,24}$,
M. Vivek$^{7,25}$
 } \vspace*{4pt} \\ 
\scriptsize $^{1}$ Max-Planck-Institut f\"ur Extraterrestrische Physik, Postfach 1312, Giessenbachstrasse 1, 85748 Garching bei M\"unchen, Germany\vspace*{-2pt} \\ 
\scriptsize $^{2}$ Center for Cosmology and Astro-Particle Physics, Ohio State University, Columbus, Ohio, USA\vspace*{-2pt} \\ 
\scriptsize $^{3}$ IRFU,CEA, Universit\'e Paris-Saclay, F-91191 Gif-sur-Yvette, France\vspace*{-2pt} \\ 
\scriptsize $^{4}$ Institute of Cosmology \& Gravitation, Dennis Sciama Building, University of Portsmouth, Portsmouth, PO1 3FX, UK\vspace*{-2pt} \\ 
\scriptsize $^{5}$ Institute of Physics, Laboratory of Astrophysics, \'Ecole Polytechnique F\'ed\'erale de Lausanne (EPFL), Observatoire de Sauverny, 1290 Versoix, Switzerland\vspace*{-2pt} \\ 
\scriptsize $^{6}$ Center for Cosmology and Particle Physics, Department of Physics, New York University, New York, NY 10003, USA\vspace*{-2pt} \\ 
\scriptsize $^{7}$ Department Physics and Astronomy, University of Utah, 115 S 1400 E, Salt Lake City, UT 84112, USA\vspace*{-2pt} \\ 
\scriptsize $^{8}$ Instituto de F\'isica, Universidad Nacional Aut\'onoma de M\'exico, Apdo. Postal 20-364, M\'exico\vspace*{-2pt} \\ 
\scriptsize $^{9}$ Harvard-Smithsonian Center for Astrophysics, 60 Garden St., MS-20, Cambridge, MA 02138, USA\vspace*{-2pt} \\ 
\scriptsize $^{10}$ Institut de Ci\`encies del Cosmos,  Universitat  de  Barcelona,  ICCUB,  Mart'i  i  Franqu\`es  1,  E08028  Barcelona,  Spain\vspace*{-2pt} \\ 
\scriptsize $^{11}$ Institut  d'Estudis  Espacials  de  Catalunya  (IEEC),  E08034  Barcelona,  Spain\vspace*{-2pt} \\ 
\scriptsize $^{12}$ University of Wyoming, 1000 E. University Ave., Laramie, WY 82071, USA\vspace*{-2pt} \\ 
\scriptsize $^{13}$ Waterloo Centre for Astrophysics, University of Waterloo, Waterloo, ON N2L 3G1, Canada\vspace*{-2pt} \\ 
\scriptsize $^{14}$ Department of Physics and Astronomy, University of Waterloo, Waterloo, ON N2L 3G1, Canada\vspace*{-2pt} \\ 
\scriptsize $^{15}$ Sub-department of Astrophysics, Department of Physics, University of Oxford, Denys Wilkinson Building, Keble Road, Oxford OX1 3RH\vspace*{-2pt} \\ 
\scriptsize $^{16}$ Perimeter Institute for Theoretical Physics, 31 Caroline St. North, Waterloo, ON N2L 2Y5, Canada\vspace*{-2pt} \\ 
\scriptsize $^{17}$ Department of Physics and Astronomy, Sejong University, Seoul 143-747, Korea\vspace*{-2pt} \\ 
\scriptsize $^{18}$ Institute for Computational Cosmology, Dept. of Physics, Univ. of Durham, South Road, Durham DH1 3LE, UK\vspace*{-2pt} \\ 
\scriptsize $^{19}$ National Astronomy Observatories, Chinese Academy of Science, Beijing, 100012, P.R. China\vspace*{-2pt} \\ 
\scriptsize $^{20}$ Apache Point Observatory and New Mexico State University, P.O. Box 59, Sunspot, NM 88349, USA\vspace*{-2pt} \\ 
\scriptsize $^{21}$ Kavli Institute for Particle Astrophysics and Cosmology, Stanford University, 452 Lomita Mall, Stanford, CA 94305, USA\vspace*{-2pt} \\ 
\scriptsize $^{22}$ PITT PACC, Department of Physics and Astronomy, University of Pittsburgh, Pittsburgh, PA 15260, USA\vspace*{-2pt} \\ 
\scriptsize $^{23}$ Department of Astronomy and Astrophysics, The Pennsylvania State University, University Park, PA 16802, USA\vspace*{-2pt} \\ 
\scriptsize $^{24}$ Institute for Gravitation and the Cosmos, The Pennsylvania State University, University Park, PA 16802, USA\vspace*{-2pt} \\ 
\scriptsize $^{25}$ Indian Institute of Astrophysics, Koramangala, Bangalore 560034, India\vspace*{-2pt} \\ 
}

\date{Accepted XXX. Received YYY; in original form ZZZ}

\pubyear{2020}


\pagerange{\pageref{firstpage}--\pageref{lastpage}}
\maketitle
\label{firstpage}

\begin{abstract}
We measure the anisotropic clustering of the quasar sample from Data Release 16 (DR16) of the Sloan Digital Sky Survey IV extended Baryon Oscillation Spectroscopic Survey (eBOSS). A sample of $343,708$ spectroscopically confirmed quasars between redshift $0.8<z<2.2$ are used as tracers of the underlying dark matter field.  In comparison with DR14 sample, the final sample doubles the number of objects as well as the survey area. In this paper, we present the analysis in configuration space by measuring the two-point correlation function and decomposing it using the Legendre polynomials. 
For the full-shape analysis of the Legendre multipole moments, we measure the BAO distance and the growth rate of the cosmic structure. At an effective redshift of $z_{\rm eff}=1.48$, we measure the comoving angular diameter distance
$D_{\rm M}(z_{\rm eff})/r_{\rm drag} = 30.66\pm0.88$,
the Hubble distance 
$D_{\rm H}(z_{\rm eff})/r_{\rm drag} = 13.11\pm0.52$,
and the product of the linear growth rate  and the rms linear mass fluctuation on scales of $8 \mpch$, $f\sigma_8(z_{\rm eff}) = 0.439\pm0.048$.
The accuracy of these measurements is confirmed using an extensive set of mock simulations developed for the quasar sample. The uncertainties on the distance and growth rate measurements have been reduced substantially ($\sim$45\% and $\sim$30\%) with respect to the DR14 results.
We also perform a BAO-only analysis to cross check the robustness of the methodology of the full-shape analysis. Combining our analysis with the Fourier space analysis, we arrive at 
$D^{\bf{c}}_{\rm M}(z_{\rm eff})/r_{\rm drag} = 30.21 \pm 0.79$,
$D^{\bf{c}}_{\rm H}(z_{\rm eff})/r_{\rm drag} = 13.23 \pm 0.47$, and
$f\sigma_8^{\bf{c}}(z_{\rm eff}) = 0.462 \pm 0.045$.
\end{abstract}




\section{Introduction}

In the current standard cosmological model, a component known as dark energy is 
believed to drive the accelerated expansion of the Universe. While various 
observations indicate that dark energy is consistent with a cosmological 
constant, $\Lambda$ \citep{Riess1998,collaboration2018planck},  there is no 
satisfying explanation to the nature of this component so far. 
Addressing this fundamental question requires
accurate measurements of the expansion history of the Universe and 
the cosmic structure growth rate. Observations of the large-scale structure 
(LSS) of the Universe are a powerful tool to obtain these measurements. 

In the early universe, photons, electrons and baryons were tightly coupled via 
Compton scattering and Coulomb interaction. Around over-density regions, the radiation pressure 
sourced spherical 
waves, which propagated outwards, dragging the matter with it. Later on, as the 
Universe cooled down, neutral atoms formed, and the photons streamed freely 
away while leaving the signature of the waves in the matter distribution frozen at 
a characteristic scale of ${\sim}150\, \rm{Mpc}$. 
This feature, known as the baryon acoustic 
oscillation~\citep[BAO;][]{Peebles1970, Sunyaev1970, Bond1984, Hu1996}, 
is mapped onto the late time galaxy distribution in both Fourier and 
configuration space \citep{Cole2005, Eisenstein2005}. 
BAO measurements at different redshifts can be used as a standard ruler to 
measure the expansion history of the Universe. 

In galaxy redshift surveys, the distances to individual objects are inferred 
from their measured redshifts, which also contain a component due to their 
peculiar velocities. This extra component is responsible for the angular 
dependency of the clustering amplitude (with respect to the line of sight 
direction), and gives rise to a phenomenon known as redshift-space distortions 
\citep[RSD;][]{Jackson1972, Kaiser1987}. 
As the peculiar velocities of the galaxies are sourced by the gravitational 
attraction of the surrounding matter, the strength of the anisotropic clustering 
is tightly related to the matter density fluctuation, which in turn can be  
used as a probe of the growth of structure 
\citep{Guzzo2008}. Besides constraining the properties of dark energy, 
measurements of structure growth can be used to test alternative models of 
gravity on large scales~\citep{Jennings2012,Barreira2016, Hernandez2019}. 

The Sloan Digital Sky Survey \citep[SDSS;][]{York2000} has provided many 
spectroscopic samples of galaxies and quasars for mapping the distribution of 
large scale structure at different redshifts. The eBOSS 
program~\citep{Dawson2016}, which is a successor of BOSS~\citep{Dawson2013}, 
was performed during the fourth phase, SDSS-IV~\citep{Blanton2017}. There are 
four main tracers in the eBOSS program: luminous red galaxies (LRGs), emission 
line galaxies (ELGs), quasars which can be used as direct tracers of the matter 
field (QSOs), and another higher redshift quasar sample for studies of 
the Ly$\alpha$ forest. Together, they cover a wide redshift range. 
The first BAO detection using quasars as tracers at $0.8 < z < 2.2$ was from 
the eBOSS Data Release 14 (DR14) sample \citep{Ata2017}. This quasar sample 
bridges the gap between lower redshift SDSS galaxy measurements 
\citep{Kazin2014, Alam2017} and those from the Ly$\alpha$-forest 
\citep{Bautista2017, du_Mas_des_Bourboux2017}. 
In the DR14 full-shape analysis \citep[]{Hou2018, Zarrouk2018, Gil-Marin2018}, 
it was demonstrated that quasars can be used as robust tracers of the underlying 
matter field, extending growth rate measurements to redshift $z\sim 1.5$. 

The eBOSS program concluded observations on March 1st 2019.
This work is one of a series of papers presenting an analysis of the final eBOSS 
data release 16 (DR16) quasar sample, which approximately doubles the number of quasars
of the previous DR14 release.
The DR16 quasar catalogue is presented in~\citet{Lyke2020}.
The clustering catalogue used for this analysis is described in~\citet{Ross2020}. 
The quasar mock challenge used to assess modelling systematics is described 
in~\citet{Smith2020}. N-body simulations for assessing systematic errors are presented in~\citet{Rossi2020} for LRG, ~\citet{Alam2020} and ~\citet{Avila2020} for ELG.
The approximate mocks used to estimate the covariance 
matrix and assess the observational systematics are presented in~\citet{Zhao2020}. 
A complementary quasar clustering analysis in Fourier space is performed 
by~\citet{Neveux2020}. The BAO and RSD analyses of the QSO sample from this 
work and the one from~\citet{Neveux2020}, LRG sample 
\citep{GilMarin2020, Bautista2020}, ELG sample~\citep{Tamone2020, deMattia2020}, 
together with the BAO analyses of Ly$\alpha$ forest~\citep{2020duMasdesBourbouxH} 
will enter~\citet{Mueller2020} for the cosmological implications from eBOSS\footnote{A summary of all SDSS BAO and RSD measurements with accompanying legacy
figures can be found here: 
https://www.sdss.org/science/final-bao-and-rsd-measurements/.  The full
cosmological interpretation of these measurements can be found here: 
https://www.sdss.org/science/cosmology-results-from-eboss/.}.

This paper is arranged as follows:
Section~\ref{sec:data} provides an introduction to the eBOSS survey and focus on the quasar sample. Section~\ref{sec:methodology} describes the methodology used to infer the cosmological constraints. 
Section~\ref{sec:model} describes the modelling of the full-shape analysis for the two-point statistics. 
Section~\ref{sec:model_bao} describes the BAO-only modelling.
Section~\ref{sec:sys_uncertaintity} discusses the model validation and our estimation on various systematics. 
Section~\ref{sec:constraints_geopar_growthrate} provides the constraints obtained with the final sample from the full-shape analysis, BAO-only analysis, and the combination of the configuration with the Fourier space analysis.
Section~\ref{sec:robustness_tests} discusses the robustness of our analysis. Our conclusions are summarised in Section~\ref{sec:conclusions}.

\section{Data}
\label{sec:data}
\subsection{Overview of the eBOSS survey}
\label{sec:overview_eboss}

\begin{figure*}
	\includegraphics[width=\columnwidth]{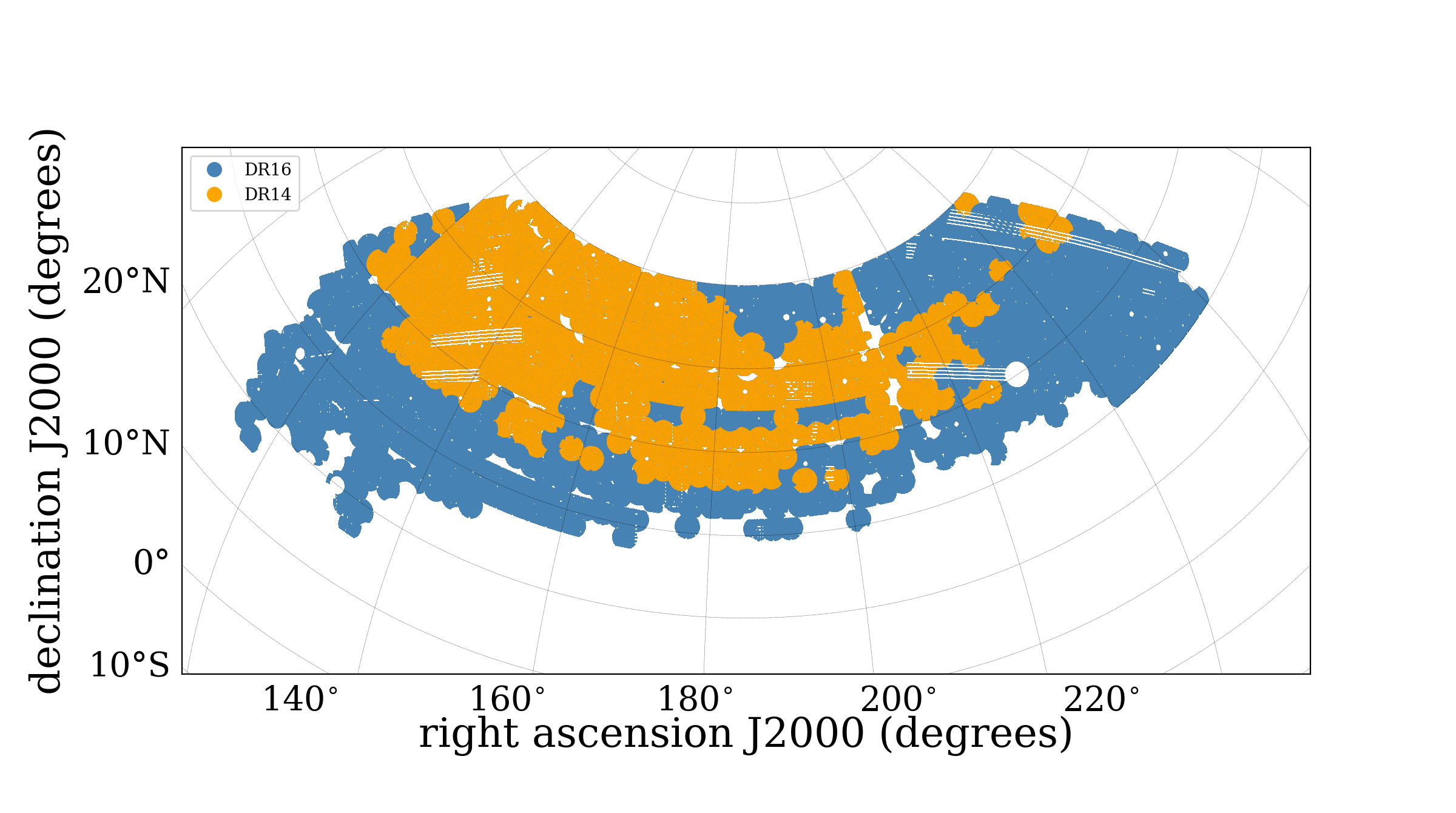}
	\includegraphics[width=\columnwidth]{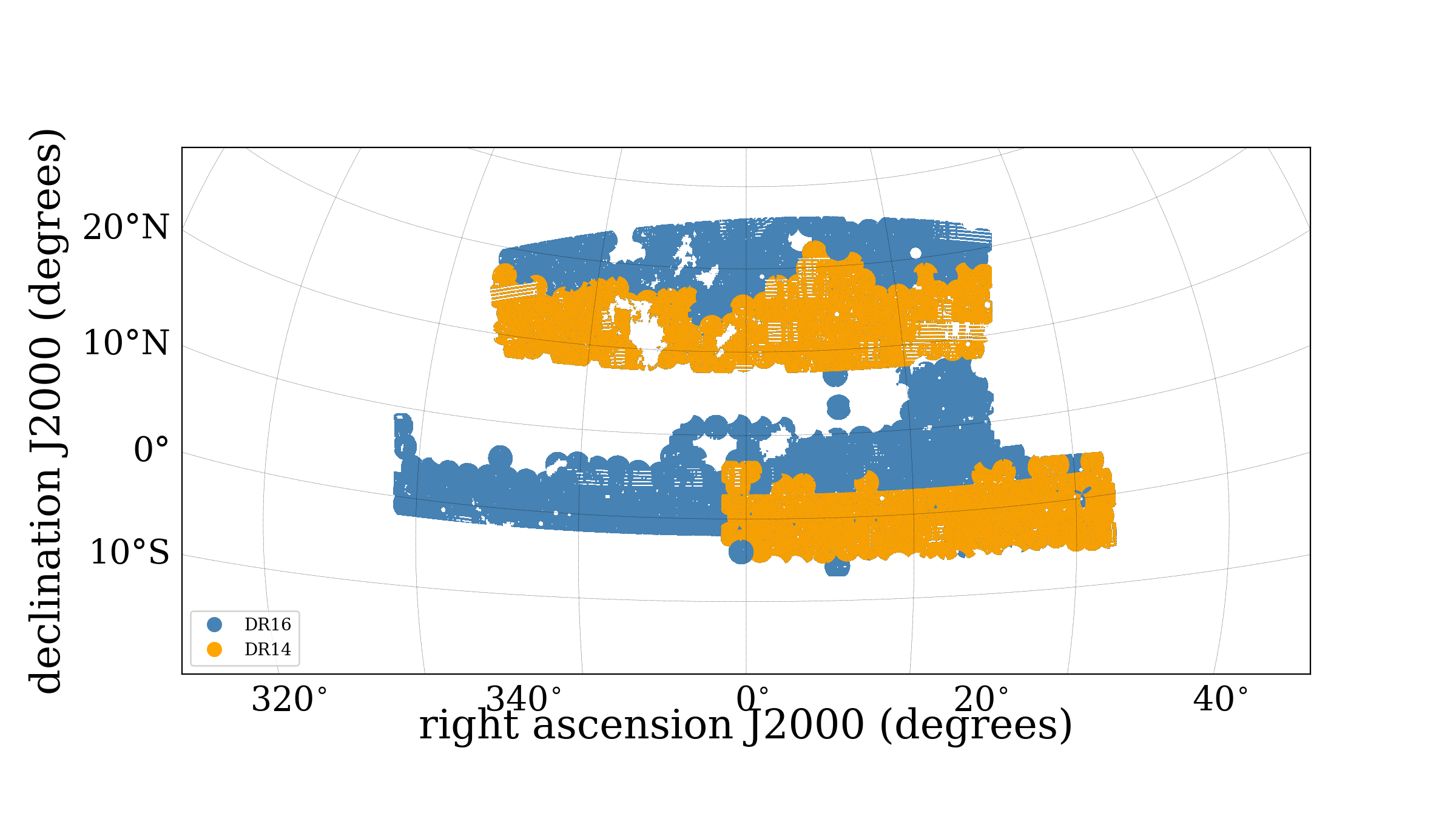}
    \caption{Footprint of the eBOSS QSOs, split into the NGC (left) and SGC (right). The DR14 sample is shown in orange, while the DR16 sample is shown in blue (and also includes the entire orange region).}
    \label{fig:footprint_dr14_dr16}
\end{figure*}

The eBOSS program, which began in July 2014, was performed using the Sloan Foundation Telescope at Apache Point Observatory~\citep{Gunn2006}, and inherited inherited the double-armed spectrographs from BOSS~\citep{Smee2013}.
These spectrographs are fed by a total of 1000 optical fibres (500 each), where the diameter of each fibre subtends an angle of $62^{''}$ on the sky. This paper focuses on the quasar sample that covers the redshift range of $0.8<z<2.2$. Table~\ref{tab:eboss_tracer_summary} summarizes the statistics for the sample, including the number of quasars used for the clustering analysis ($N_{\rm qso}$), number of quasars suffered from the fiber collision ($N_{\rm cp}$),
the effective volume~\citep[equation 5 of][]{Tegmark1997}, and the weighted area of the north galactic cap (NGC) and south galactic cap (SGC).  
Fig.~\ref{fig:footprint_dr14_dr16} shows the footprint of the final DR16 QSO sample for the NGC and SGC. In DR16 a mean completeness of $C_{\rm comp} \sim 0.98$ for both galactic caps is achieved. The final data release doubles the total number of objects, as well as the survey area, compared to DR14 released two years ago.

The details of the catalogue are described in the companion paper~\citep{Ross2020}. Here we briefly summarize the target selection and the spectroscopic observations, which are the two steps needed to construct the quasar catalogue.
The quasar target selection is documented in \citet{Myers2015}. We used the optical imaging data from SDSS-I/II/III, together with a mid-infrared cut from the Wide Field Infrared Survey Explorer \citep[WISE;][]{Wright2010}.

In the DR14 analysis, we corrected for the trends in the $g$-band depth and Galactic extinction. In our final analysis, we also correct for the sky background and seeing. The weight $w_{\rm sys}$ is introduced to mitigate the imaging systematics~\citep[for details see section 5.5 in][]{Ross2020}. The impact of these additional corrections on our final results is discussed in Section~\ref{sec:robustness_tests}.

After the target selection, the quasar candidates are observed spectroscopically. This introduces two new sources of systematics, which need to be corrected.
First, the minimum angular projected distance between two neighbouring quasar targets in each observation is limited by the ferrules (a small bracelet) that supports the fibres, which have a projected size of $62^{''}$. When two objects fall within such angular separation, they are denoted as ``collided objects" and corrected using the close pair (fibre collision) weight, $w_{\rm cp}$, where objects are up-weighted according to the colliding fraction of each group. Second, the redshift efficiency varies between different fibres, showing a dependency on the fibre ID number. Fibres falling near the edge of the spectrograph have lower efficiencies, and this is accounted for with the spectroscopic weight, $w_{\rm noz}$. Section~\ref{sec:robustness_tests} investigates how different definitions of the spectroscopic weight affect our results.  The imaging weights, $w_{\rm sys}$, are iteratively corrected for the spectroscopic weights.
These weights are then combined to correct for the observing and targeting systematics. The final weight that is applied to each object is defined as
\begin{equation}
w_{\rm tot}=w_{\rm FKP}\, w_{\rm sys}\, w_{\rm cp}\, w_{\rm noz},
\label{eqn:weight_tot}
\end{equation}
where the FKP weight ~\citep{Feldman1994} is applied to minimise the variance of the measurement,
\begin{equation}
w_{\rm FKP}=\left({1+P_0 n(z)}\right)^{-1},
\label{eqn:wfkp} 
\end{equation}
with $P_0=6000\, h^{-3}{\rm Mpc}^3$ and $n(z)$ is the volume number density in each redshift bin. Finally, instead of downsampling the random catalogue, the completeness in each sector is used as a weight.

\begin{table}
\centering
\caption{Summary of statistics for the eBOSS DR16 QSOs clustering catalogue. The quasaras used for the clustering analysis are obtained with a cut in redshift $0.8<z<2.2$,  completeness $C_{\rm eBOSS}>0.5$ and sector success rate $C_{\rm z}>0.5$. }
\begin{tabular}{|l|l|l|l|}
\toprule
 & NGC & SGC & Total \\ \midrule
 $N_{\rm qso}$ & 218,209 & 125,499 & 343,708 \\ 
 $N_{\rm cp}$ & 6878 & 4832 & 11,710 \\
 Effective volume ($\text{Gpc}^3$) & 0.39 & 0.21 & 0.60 \\ 
 Area (weighted, $\text{deg}^2$) & 2860 & 1839 & 4699 \\ \bottomrule
\end{tabular}
\label{tab:eboss_tracer_summary}
\end{table}
The redshift estimation is based on the \textsc{REDVSBLUE} algorithm\footnote{https://github.com/londumas/redvsblue} that is detailed in~\citet{Lyke2020}. 
The final clustering catalogue is composed of redshift sources from three classes. i). \textbf{Legacy}: These are quasars with reliable redshifts obtained during SDSS I/II/III. Within this category, the objects that were observed before BOSS were obtained from combining the fifth edition of the SDSS QSO catalog (based on SDSS DR7)~\citep{Schneider2010} with a catalog of known stellar spectra from SDSS-I/II. 
ii). {\textbf{SEQUELS}}: At the end of the BOSS program, Sloan Extended Quasar, ELG, and LRG Survey (SEQUELS) was designed as a pilot survey for eBOSS. 
SEQUELS used a less constrained quasar selection algorithm than that which was adopted for eBOSS, and a subsample of the SEQUELS objects that pass the eBOSS target selection entered the final eBOSS catalogues. These objects are treated the same as eBOSS objects. iii). {\textbf{eBOSS}}: This is the main source of QSOs for the program. During DR14, over 75 percent of the new redshifts were observed during the eBOSS program. In the final data release, this number has increased to ${\sim}80$ percent.

\subsection{Two-point correlation function}
\begin{figure*}
    \centering
    \includegraphics[width=\columnwidth]{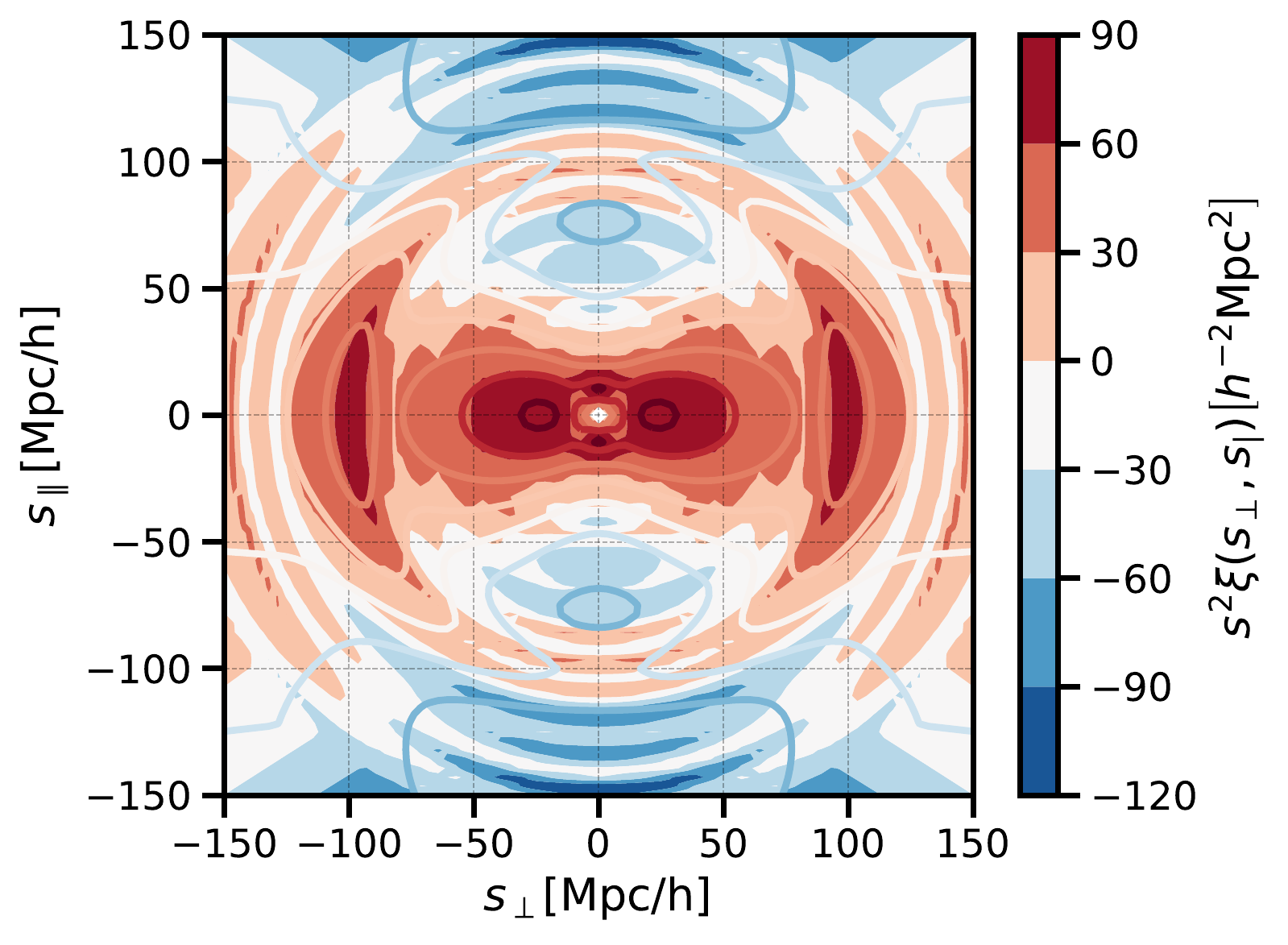}
    \includegraphics[width=0.95\columnwidth]{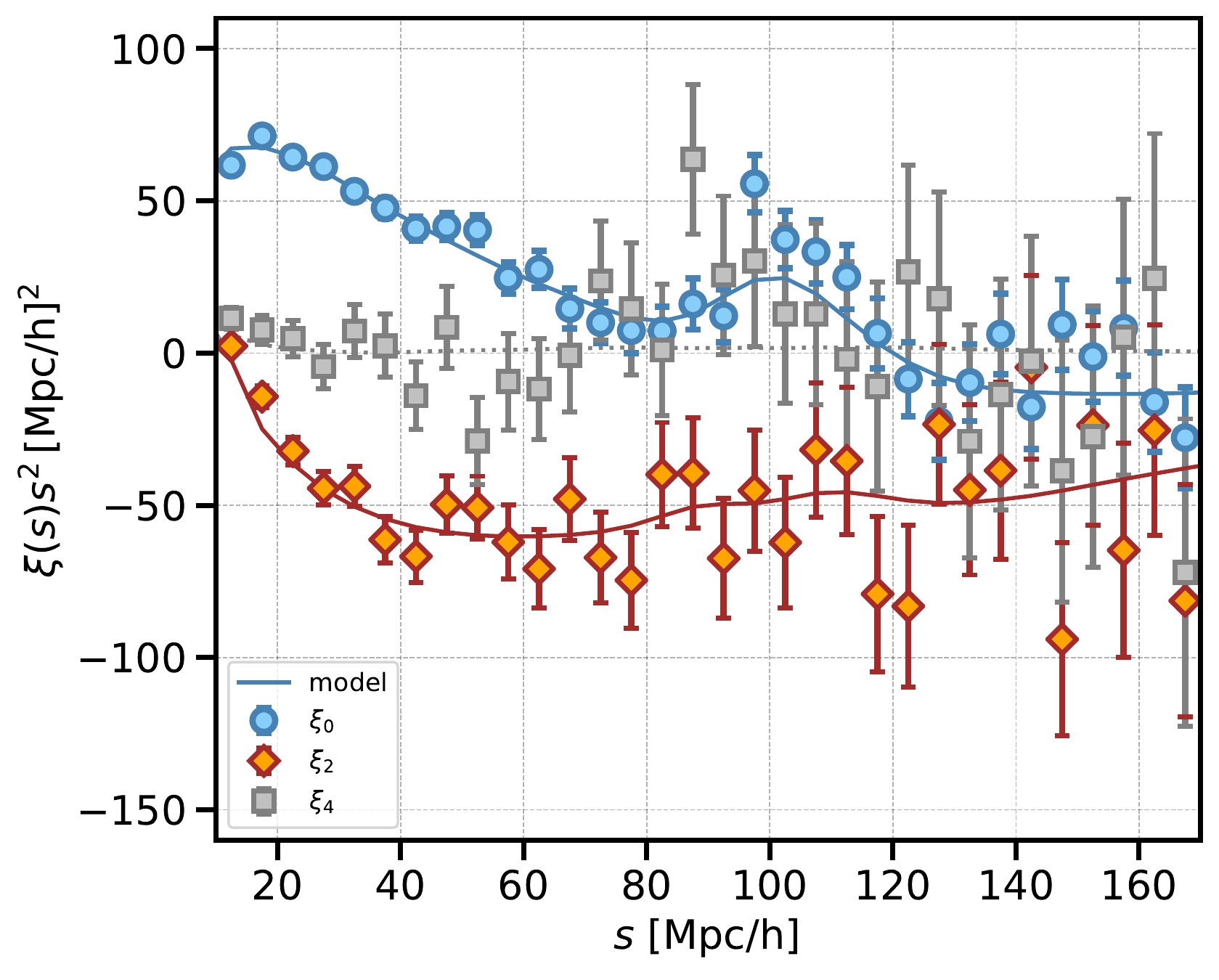}
    \caption[2D correlation function $\xi\left(s_{\perp}, s_{ \|}\right)$ for DR16 quasar]{Left: The 2D correlation function $\xi\left(s_{\perp}, s_{ \|}\right)$ measured from the DR16 quasar sample. The solid contour is from the theory prediction. Right: The measured correlation function for monopole ($\ell=0$, blue), quadrupole ($\ell=2$, red) and hexadecapole ($\ell=4$, gray), with the best fitting full-shape model shown by the solid lines.}
    \label{fig:xi_2dcontour_v7}
\end{figure*}

The two-point correlation function, $\xi(\mathbf{s})=\langle\delta(\mathbf{x}) \delta(\mathbf{x}+\mathbf{s})\rangle$, characterizes the probability excess in observing galaxies pairs as a function of their separation, ${\bf s}$, with respect to a homogeneous distribution. Assuming rotational symmetry along the line of sight direction, the correlation function is reduced to the two-dimensional function $\xi(\mathbf{s})\equiv \xi(\mu,s)$, with $\mu=\cos(\theta)$, where $\theta$ is the angle between the separation vector, $\mathbf{s}$, and the line of sight direction. Fig.~\ref{fig:xi_2dcontour_v7} shows the two dimensional correlation function $\xi(s_{\perp}, s_{ \|})$, which reveals a BAO ring at the scale $s \sim 100 \mpch$. On smaller scales, the correlation function appears to be compressed, due to redshift space distortions.
Analysing the full two-dimensional correlation function $\xi(s, \mu)$ is difficult, due to the low signal-to-noise ratio and a large size of the covariance matrix.

Fortunately, the information in the two-dimensional correlation function can be compressed into a set of one-dimensional projections by choosing different angular-dependent weighting schemes. One of the typical choices is decomposing the correlation function into Legendre polynomials $L_{\ell}(\mu)$

\begin{equation}
\xi(s, \mu) = \sum_{\ell} \xi_{\ell}(s) \mathcal{L}_{\ell}(\mu).
\label{eqn:legendre_decomp_xi}
\end{equation}
Using the orthogonality of the Legendre polynomials $\int_0^1 \mathcal{L}_{\ell}(\mu) \mathcal{L}^{\prime}_{\ell}(\mu) = \delta_{\rm D, \ell \ell^{\prime}}$ one arrives at

\begin{equation}
\xi_{\ell}(s)\equiv \frac{2\ell+1}{2} \int^{1}_{-1} \xi(\mu,s) \mathcal{L}_{\ell}(\mu) \,{\rm d}\mu.
\label{eqn:multi_def}
\end{equation}
Due to the symmetry w.r.t $\mu$, only the even multipoles are non-zero and the $\ell = 0,\,  2, \, 4$ terms are referred to as the monopole, quadrupole and hexadecapole, respectively. During the DR14 analysis, we compared the difference between using multipole moments and the clustering wedges and found that the multipoles yield a better constraint for a shot-noise dominated sample \citep{Hou2018}. We therefore do not repeat the same analysis here. 

\begin{figure}
    \centering
    \includegraphics[width=\columnwidth]{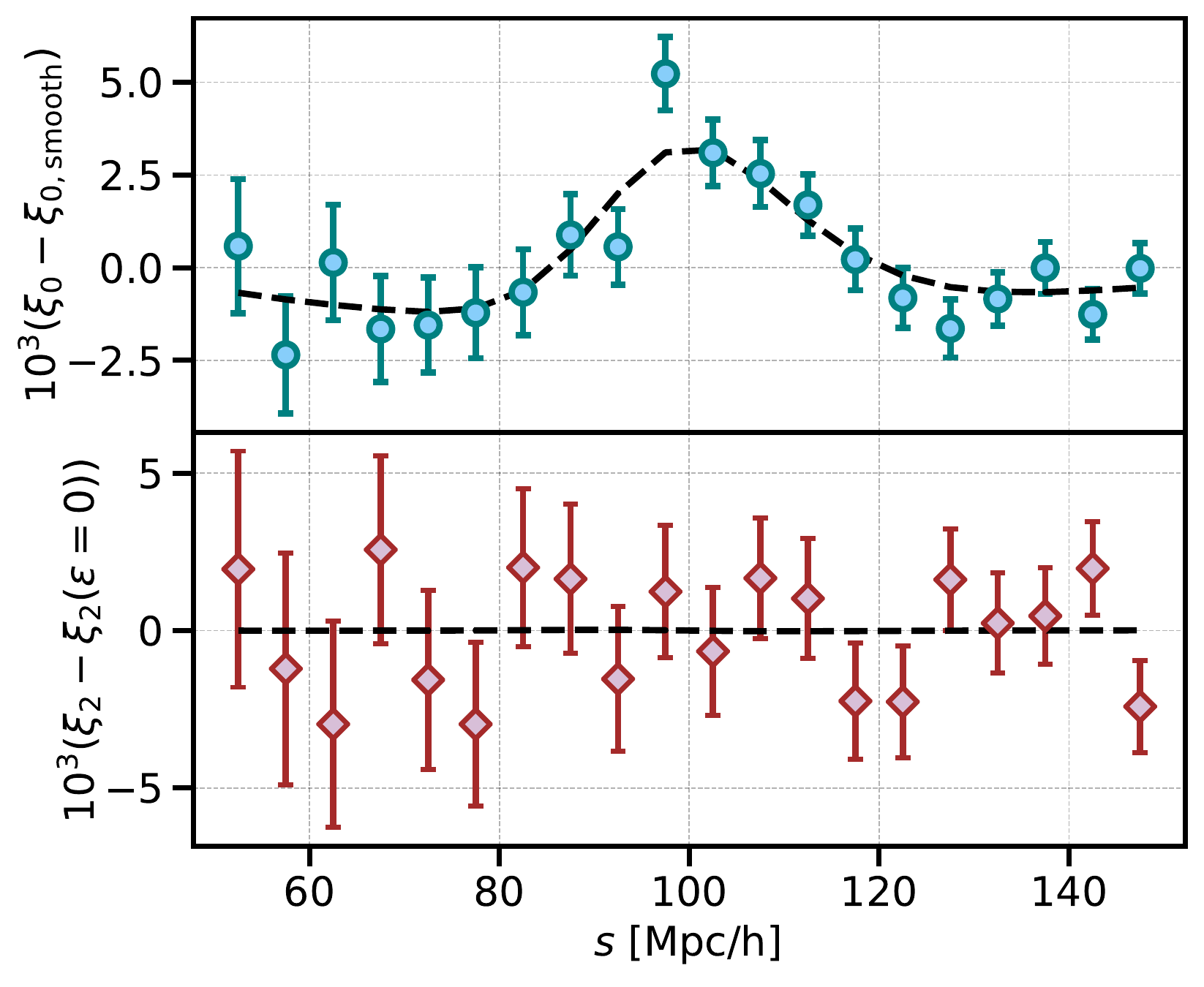}
    \caption{Comparison between our measured correlation function and the best-fit BAO model. In the top panel, we show the monopole, where we have subtracted the smooth component of the model from both the model and the data. In the bottom panel, we display the quadrupole and subtract the quadrupole of a model that has the same parameters as the best-fit, but with $\epsilon = 0$. }
    \label{fig:xibao}
\end{figure}

The right panel in Fig.~\ref{fig:xi_2dcontour_v7} shows the correlation function multipole measurements from the final DR16 data with the best fitting model, which compresses the information from the left-hand panel. In order to highlight the BAO feature, in Fig.~\ref{fig:xibao} the component of the best-fit model with no BAO has been subtracted. The bottom panel displays the result for the quadrupole. In order to highlight the (lack of) difference between $\alpha_{\perp}$ and $\alpha_{\parallel}$, we have subtracted the quadrupole of a model that has the same parameters as the best-fit, but with $\epsilon = \left(\alpha_{\parallel}/\alpha_{\perp} -1\right) = 0$. If $\alpha_{\perp}$ and $\alpha_{\parallel}$ differ,  $\epsilon \neq 0$, a feature is observed in figure 3 from \citet{Alam2017}. Here, we see that the BAO transverse and along the line of sight are consistent with each other with respect to our fiducial model.

{} 

\section{Methodology}
\label{sec:methodology}
\subsection{Inference of cosmological parameters}

In order to infer the best-fit cosmological parameters from a theoretical model, we aim to maximize the likelihood function. Given Bayes's theorem, the posterior distribution of a set of parameters $\{\lambda\}$ is proportional to the product of the likelihood function and the prior $\mathcal{P}(\lambda | \boldsymbol{\xi}) \propto \mathcal{L}(\boldsymbol{\xi} | \lambda) \mathcal{P}(\lambda)$. In our case, the data vector $\boldsymbol{\xi}$ stands for the two-point correlation function.
The likelihood for the Gaussian-distributed data is
\begin{equation}
\mathcal{L}(\boldsymbol{\xi} | \lambda) \propto \exp \left[-\frac{1}{2}\left(\boldsymbol{\xi}-\boldsymbol{\xi}_{\mathrm{model}}(\lambda)\right)^{\mathrm{T}} \Psi\left(\boldsymbol{\xi}-\boldsymbol{\xi}_{\mathrm{model}}(\lambda)\right)\right],
\end{equation}
where $\boldsymbol{\xi}_{\mathrm{model}}$ is the theoretical model for the two-point correlation function  (see Section~\ref{sec:model}), and the precision matrix is the inverse of the true covariance matrix, $\Psi\simeq\mathbf{C}_{\mathrm{true}}^{-1}$, which follows the inverse Wishart distribution. We will discuss the estimation of the covariance matrix in Section~\ref{sec:covariance_matrix}.  
The two-point correlation function for the data vector and the model are expressed in the spatial coordinates. In order to transform the observed redshift into distance, a fiducial cosmology is required. A difference between the true and fiducial cosmological parameters results in a rescaling of cosmological distances~\citep{Padmanabhan2008b, Kazin2012}
\begin{equation}
\begin{aligned} 
s_{\perp} &=\frac{D_{\mathrm{M}}\left(z_{\mathrm{m}}\right)}{D^{\prime}_{\mathrm{M}}\left(z_{\mathrm{m}}\right)} s_{\perp}^{\prime} = q_{\perp} s_{\perp}^{\prime}, \\ 
s_{\parallel} &=\frac{D_{\mathrm{H}}\left(z_{\mathrm{m}}\right)}{D^{\prime}_{\mathrm{H}}\left(z_{\mathrm{m}}\right)} s_{\parallel}^{\prime} = q_{\parallel} s_{\parallel}^{\prime}, \end{aligned}
\label{eqn:ap_dist_par}
\end{equation}
where $D_{\rm M}$ is the comoving angular diameter distance (see Appendix~\ref{appendix:alter_basis}) and $D_{\rm H}=c/H$ is the Hubble distance defined as the ratio of the speed of light in vacuum, $c$, and the Hubble parameter, $H$. $s_{\perp}$ and $s_{\parallel}$ are distances perpendicular and parallel to the line of sight, the prime $\prime$ denotes the distance inferred from the fiducial cosmology, and $q_{\perp, \parallel}$ are the geometric distortion parameters.

The BAO scale is tightly related to the comoving sound horizon at the drag epoch, $r_{\mathrm{drag}}$, which depends on the ratio of the baryon to radiation density. The geometric distortion parameters need further to be rescaled by the ratio of the sound horizon
\begin{equation}
\alpha_{\perp}=q_{\perp} \frac{r^{'}_{\mathrm{drag}}}{r_{\mathrm{drag}}} \quad \text { and } \quad \alpha_{\|}=q_{\|} \frac{r^{'}_{\mathrm{drag}}}{r_{\mathrm{drag}}},
\label{eqn:ap_rd}
\end{equation}
where $\alpha_{\perp}$ and $\alpha_{\parallel}$ are commonly referred to as the Alcock-Paczynski (AP) parameters~\citep{Alcock1979}. This method of compressing the cosmological information is only an approximation, which we test in Section~\ref{subsec:results_bao_only}.

The rescaling of the 2D correlation function $\xi(s, \mu) \rightarrow \xi\left(s^{\prime}, \mu^{\prime}\right)$ can be expressed as
\begin{equation}
\begin{array}{l}
{s=s^{\prime} \sqrt{\alpha_{\|}^{2}\left(\mu^{\prime}\right)^{2}+\alpha_{\perp}^{2}\left(1-\mu^{\prime 2}\right)}} \\ {\mu=\frac{\alpha_{\|} \mu^{\prime}}{\sqrt{\alpha_{\|}^{2}\left(\mu^{\prime}\right)^{2}+\alpha_{\perp}^{2}\left(1-\mu^{\prime 2}\right)}}}.
\end{array}
\end{equation}

\subsection{Estimation of the Covariance matrices}
\label{sec:covariance_matrix}
\subsubsection{Covariance matrices from the \textsc{EZmocks}}
\label{subsec:cov_ezmock}
We use the effective Zel'dovich mock catalogues to estimate the covariance matrices~\citep[EZmocks;][]{Chuang2015}. A detailed description of the methodology for eBOSS QSO mock catalogue is presented in \citet{Zhao2020}. We briefly summarize the steps in the following.
The initial displacement field in the \textsc{EZmocks} is constructed using the Zel'dovich approximation. The probability density function (PDF) of the tracers is linked to the dark matter field using an effective bias model, then further calibrated with respect to the real data. Afterwards, galaxies are assigned to the dark matter particles. The \textsc{EZmocks} cubic boxes for quasars were generated at 7 different redshift snapshots using the same initial condition, each boxes is of side length $5 \gpch$. The boxes at different redshift slices are transformed into sky coordinates, trimmed by the angular geometry, selected by the radial distribution, and then trivially combined in redshifts.
The light-cone mocks constructed out of this way intrinsically captures the redshift uncertainty. The mock catalogues are tuned independently for the NGC and SGC. The \textsc{EZmocks} were constructed using a flat $\Lambda$CDM cosmology, with matter density parameter $\Omega_{\rm{m}}=0.307$, baryon density 
$\Omega_{\rm{b}}h^2=0.022$, a dimensionless Hubble parameter $h=0.678$, and no contribution from massive neutrinos.
The power spectrum of these mocks is characterized by a scalar spectral index $n_{\rm  s}=0.96$, normalized to a value of $\sigma_8(z=0)=0.8225$.

When using the mocks to estimate the covariance matrix, the limited number of mocks will add extra noise to the covariance matrix. This extra noise can lead to a \textit{biased} estimation of the inverse of the covariance matrix. Consequently, the precision matrix needs to be corrected following~\citet{Anderson2003} and \citet{Hartlap2007}
\begin{equation}
C_{\rm debiased}^{-1} = \frac{N_{\rm m}-N_{\rm b}-2}{N_{\rm m}-1} \left\langle\hat{\mathrm{C}}_{*}^{-1}\right\rangle
\quad \text { for }\quad N_{\rm b}<N_{\rm m}-2,
\label{eqn:cov_m_corr}
\end{equation}
where $N_{\rm b}$ represents the number of bins in the data vector and $N_{\rm m}$ is the number of synthetic mocks. Although the bias can be easily corrected by this factor, it does not however correct for the error in the covariance. The uncertainty in the covariance can lead to additional variance in the inferred parameters.
In ~\citet{Dodelson2013} it was shown that if the precision matrix is contaminated by the error $\Psi = \Psi_{\rm true} + \Delta\Psi$, it leads to an additional term in the covariance when expanding the covariance to second order. When the best fitting parameters are estimated from a set of independent mock catalogues, the actual scattering of the best fitting parameters is inflated at the second order given by $B \cdot \left(N_{\rm b}-N_{\rm p}\right)$, with $N_{\rm p}$ being the number of parameters and $B$ is given by
\begin{equation}
B =\frac{N_{\mathrm{m}}-N_{\mathrm{b}}-2}{\left(N_{\mathrm{m}}-N_{\mathrm{b}}-1\right)\left(N_{\mathrm{m}}-N_{\mathrm{b}}-4\right)}.
\end{equation}
When inferring the parameters from the data, the error is derived by integrating the likelihood function, and the noise in the covariance leads to a modified variance estimator that involves additional parameter,
\begin{equation}
A=\frac{2}{\left(N_{\mathrm{m}}-N_{\mathrm{b}}-1\right)\left(N_{\mathrm{m}}-N_{\mathrm{b}}-4\right)}.
\end{equation}
Therefore the final parameter matrix needs to be rescaled following~\citet{Percival2014}
\begin{equation}
M={\frac{1+B\left(N_{\mathrm{b}}-N_{\mathrm{p}}\right)}{1+A+B\left(N_{\mathrm{p}}+1\right)}},
\label{eqn:m_factor}
\end{equation}
where such a correction is suitable under the assumption of a Gaussian likelihood.

\subsubsection{Covariance matrices from the Gaussian analytical approximation}
\label{subsec:gaussian_covmat}
For the mock challenge (see Section~\ref{sec:sys_uncertaintity}), a problem that we face is that the number of simulation doesn't fulfill $N_{\rm s} \gg N_{\rm b}$, where $N_{\rm s}$ is the number of simulations. Consequently, the noise in the covariance matrix will propagate into the parameter estimation and the error bar can be overestimated. A more general problem associated with the brute force method is that for a large survey with high number density, it can be computationally very expensive to run the simulations. Therefore, alternative methods such as an analytical expression of the covariance can be very helpful. We follow the prescription of~\citet{Grieb2016} to estimate the covariance for the \textsc{OuterRim} mocks. A description of the implementation of the analytical method with Gaussian approximation can be found in Appendix~\ref{appendix:analytical_gaussian_covmat}.

\section{Modelling the full-shape of the two-point correlation function}
\label{sec:model}
The modelling of the two-point statistics requires three main ingredients: 1) the nonlinear evolution of the density field, 2) the LSS bias that establishes the relation between the luminous tracers and the underlying matter field, and 3) the modelling of the redshift space distortions. In the following, we will first describe the power spectrum modelling in redshift space, $P^{s}({\bf k})$, and the recipe we use for the LSS bias expansion. To calculate $P^{s}({\bf k})$, we need to input the nonlinear matter power spectrum $P_{\delta\delta}$ (Section~\ref{subsec:respresso}), the matter-velocity divergence cross power spectrum, $P_{\delta\theta}$, and the auto velocity divergence power spectrum, $P_{\theta\theta}$ (Section~\ref{subsec:fitting_function}).

\subsection{Bias and redshift-space distortions}
The model is constructed in the Fourier space and Fourier transformed to obtain the two-point correlation function. The full model in redshift space can be expressed as
\begin{equation}
P^{s}(k, \mu)=F_{\mathrm{FOG}}(k, \mu)  \exp \left[-\left(k \mu \sigma_{\mathrm{zerr}}\right)^{2}\right] P_{\mathrm{novir}}(k, \mu),
\label{eqn:pk_full_zs}
\end{equation}
where the first term $F_{\mathrm{FOG}}(k, \mu)$ denotes the finger-of-god (FoG) factor, which arises from the moment generating function for the line of sight velocity difference and characterizes the random motion of galaxies on small scales. This is given by
\begin{equation}
F_{\mathrm{FOG}}(\mu, k) \equiv \frac{1}{\sqrt{1+ \mu^{2} k^{2} a_{\mathrm{vir}}^{2}}} \exp \left(\frac{- \mu^{2} k^{2} \sigma_{v}^{2}}{1+ \mu^{2} k^{2} a_{\mathrm{vir}}^{2}}\right),
\end{equation}
where $a_{\mathrm{vir}}$ is a free parameter that represents the kurtosis of the small-scale velocity distribution. The one-dimensional linear velocity dispersion is given by
$\sigma_{v}^{2} = \frac{1}{3}\int {P^{\rm lin}_{\theta \theta}(k)}/{k^{2}} d^{3} k$.
In linear theory, we have $P^{\rm lin}_{\delta \delta}=P^{\rm lin}_{\delta \theta}=P^{\rm lin}_{\theta \theta}$~. 
Such a FoG treatment, which takes into account the nonlinear corrections, can also be found in~\citet{Sanchez2016b}, \citet{Grieb2017}, \citet{Hou2018}. We do not explicitly express the $f$ dependence, as done in the previous paper. Instead, since $f$ and $\sigma_8$ are degenerate, we fit the combination of these two parameters.
The second term in Eq.~\eqref{eqn:pk_full_zs}, $\exp \left[-\left(k \mu \sigma_{\mathrm{zerr}}\right)^{2}\right]$, describes the redshift uncertainty of quasars. As tested in ~\citet{Hou2018} this parameter yields a less biased estimation of the cosmological parameters in the presence of the redshift uncertainty.
The final term, $P_{\mathrm{novir}}(k, \mu)$, can be further decomposed into three terms, and this large-scale RSD modelling treatment can be found in~\citet{Scoccimarro2004} and \citet{Taruya2010}
\begin{equation}
\begin{aligned} 
P_{\text {novir }}(k, \mu)=& P_{\text {novir }}^{(1)}(k, \mu)+(k \mu f) P_{\text {novir }}^{(2)}(k, \mu) \\ &+(k \mu f)^{2} P_{\text {novir }}^{(3)}(k, \mu), 
\end{aligned}
\end{equation}
where the first term, $P_{\text {novir }}^{(1)}$, is the non-linear version of the Kaiser formula~\citep{Kaiser1987}, which is given by
\begin{equation}
P_{\text {novir }}^{(1)}(k, \mu)=P_{\mathrm{gg}}+2 f \mu^{2} P_{\mathrm{g} \theta}+f^{2} \mu^{4} P_{\theta \theta},
\label{eqn:p_novir_1}
\end{equation}
where the velocity divergence is defined as $\theta \equiv \nabla \cdot \boldsymbol{v}(\boldsymbol{x}, \tau)$.
The two higher-order terms $P_{\text {novir }}^{(2)}$ and $P_{\text {novir }}^{(3)}$ depend on the cross bispectrum at tree level and the cross-spectrum,
\begin{equation}
P_{\mathrm{novir}}^{(2)}(k, \mu)=\int \mathrm{d}^{3} \mathrm{p} \frac{p_{z}}{p^{2}}\left[B_{\sigma}(\mathbf{p}, \mathbf{k}-\mathbf{p},-\mathbf{k})-B_{\sigma}(\mathbf{p}, \mathbf{k},-\mathbf{k}-\mathbf{p})\right],
\label{eqn:p_novir_2}
\end{equation}
and
\begin{equation}
\begin{aligned}
P_{\mathrm{novir}}^{(3)}(k, \mu)&=\int \mathrm{d}^{3} \mathrm{p} \frac{p_{z}\left(k_{z}-p_{z}\right)}{p^{2}(\boldsymbol{k}-\boldsymbol{p})^{2}}\left(b_{1}+f \mu_{p}^{2}\right)\\
&\left(b_{1}+f \mu_{k-p}^{2}\right)P_{\delta \theta}(\mathbf{p}) P_{\delta \theta}(\mathbf{k}-\mathbf{p}),
\label{eqn:p_novir_3}
\end{aligned}
\end{equation}
where the cross bispectrum is defined as
\begin{equation}
\begin{array}{l}{\left\langle\theta\left(\boldsymbol{k}_{1}\right)\left\{\delta_{g}\left(\boldsymbol{k}_{2}\right)+f \frac{k_{2 z}^{2}}{k_{2}^{2}} \theta\left(\boldsymbol{k}_{2}\right)\right\}\left\{\delta_{g}\left(\boldsymbol{k}_{3}\right)+f \frac{k_{3 z}^{2}}{k_{3}^{2}} \theta\left(\boldsymbol{k}_{3}\right)\right\}\right\rangle} \\ {\quad=(2 \uppi)^{3} \delta_{D}\left(\boldsymbol{k}_{1}+\boldsymbol{k}_{2}+\boldsymbol{k}_{3}\right) B_{\sigma}\left(\boldsymbol{k}_{1}, \boldsymbol{k}_{2}, \boldsymbol{k}_{3}\right)}.
\end{array}
\end{equation}

The LSS bias represents the statistical relation between the distribution of the luminous tracers and the underlying matter field. Down to the quasi-linear scales, this statistical relation can be described as a perturbative bias expansion, which encompasses complicated galaxy formation processes dominated by local gravitational effects. The perturbative expansion of the galaxy density fluctuation, $\delta_{\mathrm{g}}$, in terms of the matter fluctuation, $\delta$, can be generalized as a series of operators with associated coefficients. One efficient way of expressing the operators is in terms of Galileons. If we consider all scalar invariants of the tensor $\nabla_{i j} \Phi(\boldsymbol{x}, \tau)$ for the gravitational potential, and $\nabla_{i j} \Phi_{\rm v}(\boldsymbol{x}, \tau)$ for the velocity potential, only three invariants exists in three dimensions~\citep[see][]{Chan2012b, Eggemeier2019}. The first two terms are
\begin{equation}
\begin{array}{c}{\mathcal{G}_{1}(\Phi) \equiv \nabla^{2} \Phi \equiv \delta}, \\ {\mathcal{G}_{2}(\Phi) \equiv\left(\nabla_{i j} \Phi\right)^{2}-\left(\nabla^{2} \Phi\right)^{2}},
\label{eqn:galilion_operator}
\end{array}
\end{equation}
and similar relations also exist for the velocity potential. The second line in Eq.~\eqref{eqn:galilion_operator} can be associated with the tidal field.
At linear order, the gravitational potential $\Phi$ and velocity potential $\Phi_{\rm v}$ are equal. At higher order, these two potential terms are not equal, and an additional operator emerges from the second Galileon operator $\mathcal{G}_{2}^{(3)}$ at the third order:
\begin{equation}
\Delta_{3} \mathcal{G}_{2}=\mathcal{G}_{2}(\Phi)-\mathcal{G}_{2}\left(\Phi_{\mathrm{v}}\right)=\mathcal{G}_{2}^{(3)}(\Phi)-\mathcal{G}_{2}^{(3)}\left(\Phi_{\mathrm{v}}\right).
\end{equation}
Combining these ingredients, we arrive at the bias expansion following~\citet{Chan2012b}, which is given by
\begin{equation}
\delta_{\mathrm{g}}=b_{1} \delta+\frac{b_{2}}{2} \delta^{2}+\gamma_{2} \mathcal{G}_{2}+\gamma_{3}^{-} \Delta_{3} \mathcal{G}_2+\ldots,
\end{equation}
where $b_{1}$ and $b_{2}$ are the bias parameter at linear and second order. We use the local Lagrangian relation to fix $\gamma_2 = -2/7\left(b_{1}-1\right)$ and we leave $\gamma^{-}_{3}$ as a free parameter. We have ignored the higher-derivative bias in our bias expansion. The effect of this is expected to be suppressed on the scales much larger than the Lagrangian radius of the hosting halos (a few $\rm Mpc$). The shape of the two-point correlation function of QSOs may potentially be affected by the radiation field or large-scale outflows during its formation~\citep{Desjacques2018}. It therefore remains interesting to potentially include the higher-derivative bias in the future.

\subsection{Matter power spectrum}
\label{subsec:respresso}
\begin{figure}
    \centering
    \includegraphics[width=0.9\columnwidth]{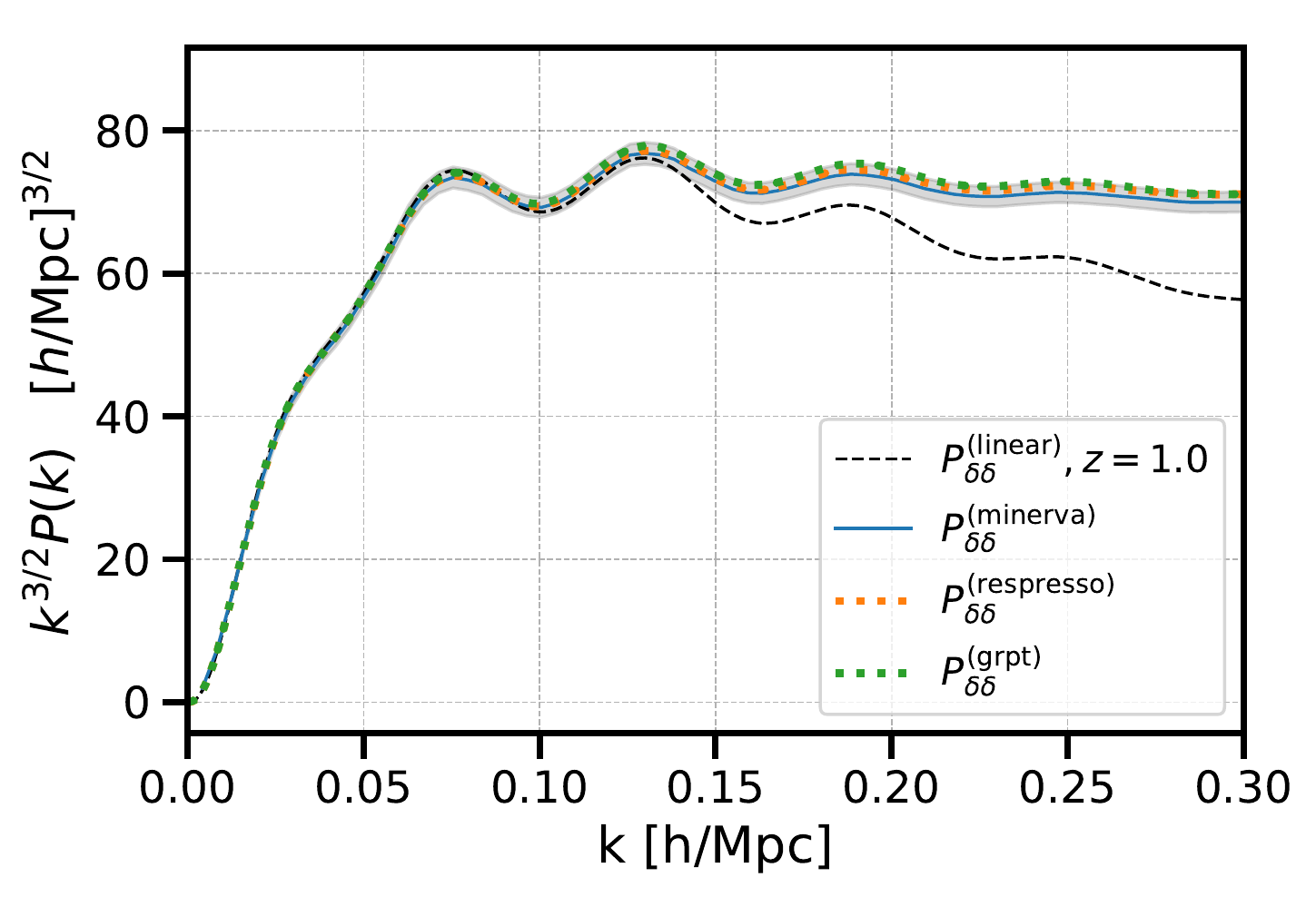}
    \caption{Comparison of matter power spectrum between \textsc{RESPRESSO} (dotted-orange), gRPT (dotted-green) and \textsc{Minerva} N-body simulation (blue, with $2\%$ error indicated by the grey band) at $z=1.0$.}
    \label{fig:matter_power}
\end{figure}

The matter power spectrum is calculated using \textsc{RESPRESSO}~\citep[Rapid and Efficient SPectrum calculation based on RESponSe functiOn; ][]{Nishimichi2017}.
The idea of \textsc{RESPRESSO} is based on the response function at the power spectrum level. The response function characterizes the variation of the nonlinear power spectrum, $\delta P(k, z)$, at redshift $z$ for a given small perturbation of the initial power spectrum, $\delta P_{\rm ini}(q, z)$. The response function is defined as

\begin{equation}
    K(k, q ; z)=q \frac{\delta P(k ; z)}{\delta P_{\text{ini}}(q ; z)}.
    \label{eqn: response_definition}
\end{equation}
Based on the numerical measurements of the response function of the power spectrum, \citet{Nishimichi2017} proposed the following phenomenological model,

\begin{equation}
\begin{split}
     K_{\mathrm{model}}(k, q)=\left[\left(1+\beta_{k, q}+\frac{1}{2} \beta_{k, q}^{2}\right) K_{\mathrm{tree}}^{\mathrm{SPT}}(k, q)\right)\\
    +\left(1+\beta_{k, q}\right) K_{1-\text { loop }}^{\mathrm{SPT}}(k, q)\\
    +K_{2-\text { loop }}^{\mathrm{SPT}}(k, q) ] D\left(\beta_{k, q}\right),
\end{split}
\label{eqn: response_model}
\end{equation}
where the explicit expression for the response function, $K^{\mathrm{SPT}}$, using the standard perturbation theory (SPT) up to 2-loop order, can be found in the original paper.
The damping factor is given by
\begin{equation}
    D(x)=\left\{\begin{array}{ll}{\exp (-x),} & {\text { if } K_{\text {model}}(k, q)>0} \\ {\frac{1}{1+x},} & {\text { if } K_{\text {model}}(k, q)<0}\end{array}\right.,
\end{equation}
with $\beta_{k, q}=\alpha_{k}+\alpha_{q}$, and, \begin{equation}
\alpha_{k}=\frac{1}{2} k^{2} \int \frac{d k}{6 \uppi^{2}} P_{\operatorname{lin}}(k),
\end{equation}
where the one-dimensional integral is the variance of the linear displacement field.
The model is designed to recover the SPT prediction in the low limit of the wavenumber associated with the initial linear power spectrum, and also keep the feature from the regularized perturbation calculation~\citep{Taruya2012}.
Finally, a (multi-step) reconstruction at the power spectrum level is performed. 
\footnote{We modify the \textsc{RESPRESSO} python package into a \textsc{Fortran} version.}
Fig.~\ref{fig:matter_power} compares the matter power spectrum calculated using \textsc{RESPRESSO} and Galilean-invariant RPT~(gRPT, Crocce et al., in prep.) at redshift $z=1.0$.
Both of them agree very well with the measurement from a \textsc{GADGET}-based N-Body simulation~\citep{Springel2005} \textsc{MINERVA}~\citep{Grieb2016}, within $2 \%$. A comparison using \textsc{RESPRESSO} with the empirical fitting function (discussed in the next section) and gRPT in inferring the parameter constraints can be found in Table~\ref{tab:syst_test_dr16}.

{}

\subsection{Auto- and cross-velocity power spectra}
\label{subsec:fitting_function}
\textsc{RESPRESSO} provides the prediction for the auto matter power spectrum. However, the full modelling of the power spectrum for the RSD effects on large scales requires the input of the cross spectrum for the matter and velocity divergence, $P_{\delta\theta}$, as well as the auto-power for the velocity divergence field, $P_{\theta\theta}$. An alternative approach to the perturbative calculation is to model the velocity power spectra using empirical relations measured from N-body simulations. ~\citet{Bel2019} performed a study based on a set of Dark Energy and Massive Neutrinos Universe (DEMNUni) N-body simulations~\citep{Carbone2016}, in the presence of massive neutrinos. The velocity field was reconstructed from the cold dark matter particles using a Delaunay tessellation. Fitting formulae for the velocity power spectra are proposed as

\begin{equation}
P_{\delta \theta}(k)=\left\{P_{\delta \delta}(k) P_{\theta \theta}^{\mathrm{lin}}(k)\right\}^{\frac{1}{2}} e^{-\frac{k}{k_{\delta}}-b k^{6}},
\label{eqn:pdv_fitting_function}
\end{equation}
and
\begin{equation}
P_{\theta \theta}(k)=P_{\theta \theta}^{\operatorname{lin}}(k) e^{-k\left(a_{1}+a_{2} k+a_{3} k^{2}\right)},
\label{eqn:pvv_fitting_function}
\end{equation}
where $P^{\rm lin}_{\theta \theta}$ is the linear auto velocity divergence power spectra, which is equal to the linear matter power.
In our case, the input of the matter power spectra $P_{\delta\delta}$ can be either calculated from \textsc{RESPRESSO} or from \textsc{HaloFit}. The amplitude of $P_{\delta\theta}$ and $P_{\theta\theta}$ is strongly influenced by the amplitude of the matter fluctuation. The free parameters that enter Eq.~\eqref{eqn:pdv_fitting_function} and Eq.~\eqref{eqn:pvv_fitting_function} are given by
\begin{equation}
\begin{aligned} a_{1} &=-0.817+3.198 \sigma_{8, m}, \\ a_{2} &=0.877-4.191 \sigma_{8, m}, \\ a_{3} &=-1.199+4.629 \sigma_{8, m}, \\ 1 / k_{\delta} &=-0.017+1.496 \sigma_{8, m}^{2}, \\ b &=0.091+0.702 \sigma_{8, m}^{2}, \\ 1 / k_{\theta} &=-0.048+1.917 \sigma_{8, m}^{2}, \end{aligned}
\end{equation}
where $\sigma_{8, m}$ is the total matter fluctuation, including cold dark matter as well as massive neutrinos.~\citet{Bel2019} showed that these fitting functions can provide an accuracy of $\sim3\%$ in $P_{\delta\theta}(k<0.7 \hmpc)$ and $P_{\theta\theta}(k<0.65 \hmpc)$ at redshifts down to $z=0$. 

\begin{figure}
    \centering
    \includegraphics[width=0.9\columnwidth]{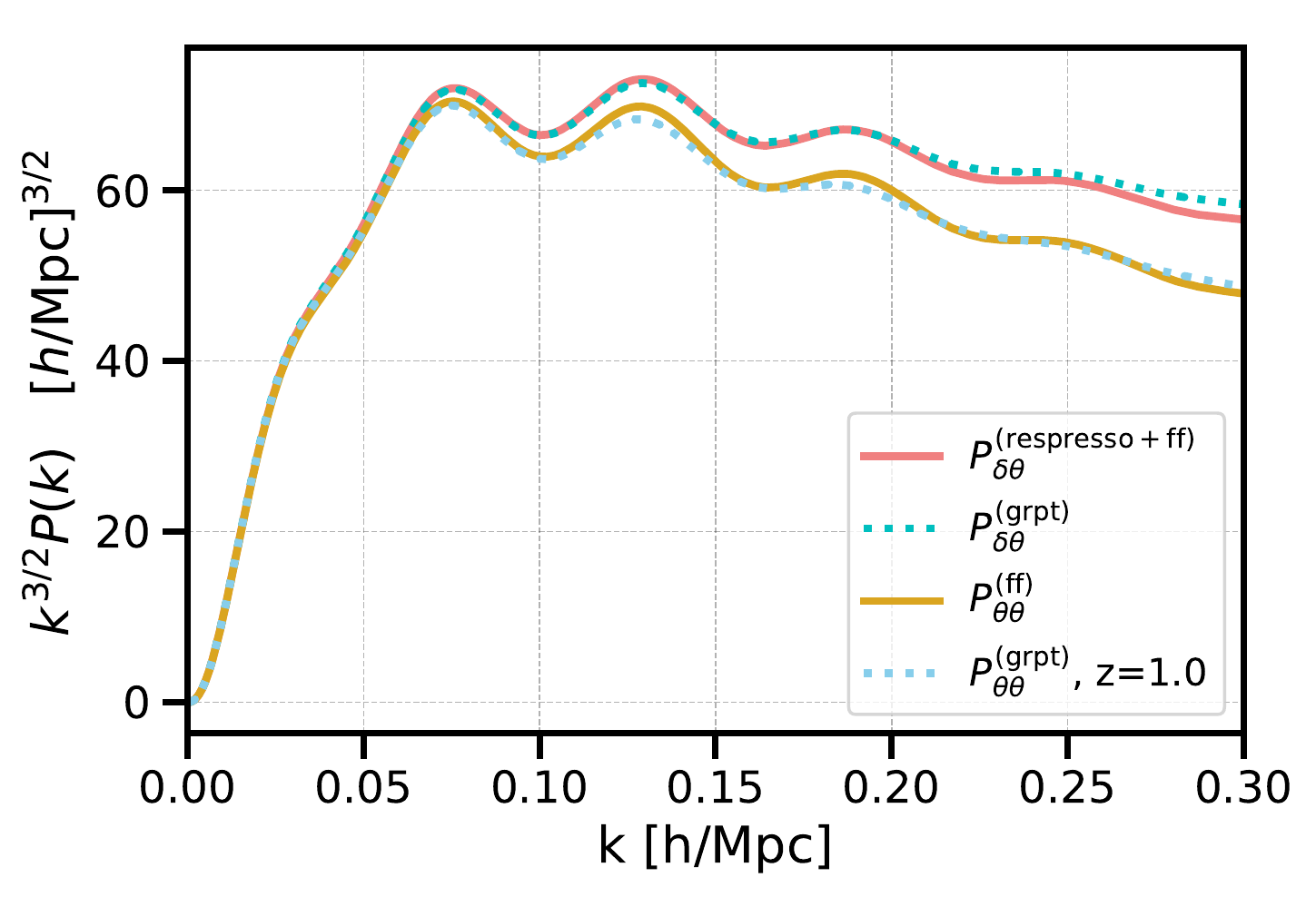}
    \caption{Comparison of the cross matter-velocity divergence power spectrum, $P_{\delta \theta}$, and the auto velocity divergence power spectrum, $P_{\theta \theta}$, at $z=1.0$. Power spectra calculated using the fitting formulae are shown by the solid red and brown curves for $P_{\delta \theta}$ (with input from \textsc{RESPRESSO} for the auto matter power spectrum) and $P_{\theta \theta}$, respectively. Power spectra calculated using gRPT are indicated by the dotted blue curves. }
    \label{fig:velocity_power}
\end{figure}

Fig.~\ref{fig:velocity_power} shows the power spectra that involve the velocity. The velocity power spectra are suppressed in comparison to the amplitude of the matter power in Fig.~\ref{fig:matter_power} due to the nonlinear correction.
At redshift $z=1.0$, we observe a good agreement between the empirical fitting formula and the perturbative calculation by gRPT, for both $P_{\delta \theta}$ and $P_{\theta \theta}$. For the cross matter-velocity power spectrum $P_{\delta \theta}$, we have input \textsc{RESPRESSO} as the nonlinear matter power spectrum $P_{\delta \delta}$ (red). The auto velocity divergence power spectrum $P_{\theta \theta}$ depends only on the linear matter power spectrum and uses the direct input from \textsc{CAMB}~\citep{Lewis1999}.

{}

\section{BAO-only modelling}
\label{sec:model_bao}
In addition to the full-shape analysis, we also present BAO-only measurements of the geometric parameters $\alpha_{\perp}$ and $\alpha_{\parallel}$ as an additional consistency check. These measurements attempt to isolate the BAO information such that none of the constraining power comes from information in the broad-band amplitude of the correlation function. 
We follow the same methodology as in~\citet{Ross2016}, which was itself based on \citet{Xu2013} and \citet{Anderson2014a}. The BAO feature is isolated in Fourier-space and damped as a function of $\mu$ in order to approximate the effects of non-linear structure formation and redshift-space distortions
\begin{equation}
P_{\rm BAO}(k,\mu) =  \left(P_{\rm {lin}}-P_{\mathrm{nw}}\right) e^{-k^{2}. \sigma_{s}(\mu)^{2}}+P_{\mathrm{nw}}.
\label{eqn:pk_bao}
\end{equation}
The linear power spectrum, $P_{\rm {lin}}$, is calculated using \textsc{CAMB}~\citep{Lewis1999}, while the ``no-wiggle" power spectrum is obtained from the fitting formulae of~\citet{Eisenstein1998}. In the exponential term of Eq.~\eqref{eqn:pk_bao}, $\sigma_{s}$ captures the nonlinear damping of the BAO feature, which is anisotropic, and given by
\begin{equation}
\sigma_{s}^{2}=\left(1-\mu^{2}\right) \Sigma_{\perp}^{2} / 2+\mu^{2} \Sigma_{\|}^{2} / 2.
\end{equation}
The damping parameters are set fixed to $\Sigma_{\perp} = 3\, \mpch$ and $\Sigma_{\parallel} = 8 \mpch$ in order to match those adopted by the Fourier space analysis of \cite{Neveux2020} for the Fourier-space analysis.

The effect of redshift space distortions on the power spectrum are modelled using
\begin{equation}
P(k, \mu)=\left(\frac{1+\mu^{2} \beta}{1+k^{2} \mu^{2} \Sigma_{s}^{2} / 2}\right)^2P_{\rm BAO}(k).
\label{eqn:pk_bao_tot}
\end{equation}
Broad-band polynomial terms are included in the model as they allow considerable freedom in fitting the broadband, but the inclusion of the factor above allows the fiducial model to be in reasonable agreement before their inclusion. The factor $\beta$ is fixed to be 0.4. For a physical redshift-space distortion model this is the ratio between the growth rate and the linear bias, $\beta=f/b_1$. Here, it controls the overall amplitude of the quadrupole, which is allowed to vary in the BAO fits through the $B_0$ and $B_2$ terms defined below. The term $\Sigma_{s}$ is included to model the effect of redshift smearing and redshift errors, and we set it fixed to a value $\Sigma_s=4 \mpch$, again matching the choice adopted by \cite{Neveux2020}.

The correlation function BAO template, $\xi_{\rm temp}$, is then the Fourier transform of Eq.~\eqref{eqn:pk_bao_tot}. As a generalization of Eq.~\eqref{eqn:multi_def} we have
\begin{equation}
     \xi_{\rm temp}(s)_F = \int_0^1{\rm d}\mu\,  F(\mu)\, \xi_{\rm temp}(s, \mu ),
 \end{equation}
where $F$ is a weighting function over $\mu$ defined for particular case. For example, $F(\mu)$ can be Legendre polynomials for $\xi_{\ell} = 0, 2$, or $F(\mu) = 3\mu^2$ for $\xi_{\mu 2}$ (see below).

We fit for the monopole $\ell=0$ and quadrupole $\ell=2$, which are given by
\begin{equation}
\xi^{\bmod}_{0}(s)=B_{0} \xi^{\rm temp}_{0}\left(s \right)+A_{0}(s),
\end{equation}
and
\begin{equation}
\xi^{\bmod }_{2}(s)=\frac{5}{2}\left(B_{2} \xi^{\rm temp}_{\mu 2}\left(s\right)-B_{0} \xi^{\rm temp}_{0}\left(s\right)\right)+A_{2}(s),
\end{equation}
where the polynomial $A_{x}(s)=a_{x, 1} / s^{2}+a_{x, 2} / s+a_{x, 3}$ removes information from the broad-band shapes of the $\xi_{\ell}$, $B_{x}$ adjusts the amplitude of the BAO feature.
In order to obtain the likelihood for $\alpha_{\parallel}$ and $\alpha_{\perp}$, we find the minimum $\chi^2$ over a grid of values in the range $0.8 < \alpha_{\parallel} < 1.2$ and $0.8 < \alpha_{\perp} < 1.2$.

We also obtain BAO results only fitting to $\xi_0$. In this case we use the same model and nuisance parameters for $\xi_0$, but we assume spherical symmetry, so we simply have
\begin{equation}
    \xi_{0}^{\rm temp}(s,\alpha_{\rm iso}) = \xi_{0}^{\rm temp}(s\alpha_{\rm iso}).
\end{equation}
The parameter $\alpha_{\rm iso}$ is defined as $\alpha_{\rm iso} = \alpha^{2/3}_{\perp}\alpha^{1/3}_{\parallel}$, which is the best constrained combination of the BAO information. We obtain the likelihood for $\alpha_{\rm iso}$ by finding the $\chi^2_{\rm min}(\alpha_{\rm iso})$ on a grid in the range $0.8 < \alpha_{\rm iso} < 1.2$.
We test the BAO template on both the \textsc{OuterRim} (blind and non-blind) mocks and the \textsc{EZmocks}. 

A comparison between our BAO analysis in configuration space on the \textsc{EZmocks} and a Fourier space BAO analysis can be found in the companion paper~\citep{Neveux2020}.

\section{Assessing the systematic uncertainty}
\label{sec:sys_uncertaintity}
In this section, we describe how we assess the systematic uncertainties in our measurements. We split the systematic uncertainties into modelling and observational systematics.

To assess the modelling systematics, we perform a N-body mock challenge~\citep{Smith2020}, using the \textsc{OuterRim} simulation~\citep{Heitmann2019}.
The \textsc{OuterRim} simulation was run in a cubic box of side length $L_{\rm b} = 3\gpch$, with $10,240^3$ dark matter particles and a force resolution of $6\, h^{-1}{\rm kpc}$, corresponding to a mass resolution of $m_{\rm p} = 1.85\, 10^9 \times M_{\odot}$. The cosmology for the \textsc{OuterRim} simulation is consistent with WMAP7 cosmology~\citep{Komatsu2011}, with $h_{0} = 0.71$, $\Omega_{\rm b} h^2 = 0.02258$, $\Omega_{\rm cdm} h^2 = 0.1109$, $\sigma_{8} = 0.8$, $n_s = 0.963$, and zero neutrino mass.
The mocks are constructed from a cubic box using a single snapshot at $z=1.433$, and are populated with quasars using halo occupation distribution (HOD) models. 
The goal of the mock challenge is two-fold: first, it serves to provide an estimate of the systematics in the modelling of the two-point statistics. Second, it is used to assess the impact of the assumption of the fiducial cosmology\footnote{Fiducial cosmology here refer to both the set of cosmological parameters for the coordinates transformation and the ones for the generation of the template for the two-point correlation function. We do not distinguish the terminology because we always keep the same set of cosmological parameters for both.}. In the first stage of the mock challenge (Section~\ref{subsec:nonblind_mock_challenge}), we test our model on a `non-blind' set of mocks, where we know precisely the underlying cosmology. In order to test the full analysis pipeline, in the second stage, we test our methodology on a set of `blind' mocks which have been rescaled to different cosmologies. The true cosmological parameters of these mocks are unknown during the analysis (Section~\ref{subsec:blind_mock_challenge}).
The mock challenge is described in detail in the companion paper~\citep{Smith2020}.

The observational systematics are quantified using a set of approximate~\textsc{EZmocks}  (Section~\ref{subsec:obs_sys_ezmocks}).
In the following sections, we summarize the tests we performed and our main conclusions.

\subsection{Modelling systematics: Non-blind mock challenge}
\label{subsec:nonblind_mock_challenge}
The mock catalogues for the non-blind part of the mock challenge are created using 20 different HOD models, and we generate 100 random realizations of each. To test the flexibility of our model, we use a wide range of HOD models, including some more extreme models that are not motivated by quasar physics. We do not explicitly include effects such as assembly bias or star formation rate, but their impacts are partially degenerate with the wide range of HOD models.
We validate our model using mocks with and without different observational effects. For each mock, we create a version with no redshift smearing, with Gaussian redshift smearing, and with a double-Gaussian smearing (see equation (4) in \citet{Smith2020}) that matches the redshift distribution seen in the data. We also create an additional catalogue with catastrophic redshift failure objects, using the estimated catastrophic redshift failure rate from the data, of $1.5\%$.

Covariance matrices are calculated analytically using the method described in Section~\ref{subsec:gaussian_covmat}. This requires the power spectrum, which we directly calculate from each mock, and the effective volume, which is estimated using Eq.~\eqref{eqn:veff}. We fit our model to the correlation function multipoles calculated from each mock, on comoving scales in the range $s = [20, 160]\mpch$, with bin separation $\Delta s = 8 \mpch$. The fitting parameters of our model can be found in Table~\ref{tab:param_space_nuisance}. 
The takeaway message from these non-blind mock analyses are:
1) we are able to recover $\alpha_{\parallel}$ and $\alpha_{\perp}$ to within an accuracy of $1\%$, and $3\%$ for $f\sigma_8$. 2) When adding the effect of $1.5\%$ catastrophic redshift failures to the mocks, we observe a $-3\%$ shift in $f\sigma_8$. The redshift of an object is completely randomized by a catastrophic redshift failure, removing some of the structure growth information, which results in the shift in $f\sigma_8$. 3) The exact choice of the HOD formalism does not have a strong impact on the geometrical parameters or the growth rate. The impact of the extreme HODs is mostly absorbed by the nuisance parameters which model the effect of the effect RSD through the redshift randomization and the satellite fraction.
The systematic error is quantified from the mocks by taking the root-mean-square (rms) of the difference to the true cosmology. Using the mocks with realistic redshift smearing and catastrophic redshifts failures, we arrive at modelling systematics of $\delta {\alpha_{\perp}}=0.003$, $\delta {\alpha_{\parallel}}=0.004$ and $\delta f\sigma_{8}=0.008$.

\subsection{Fiducial cosmology systematics: Blind mock challenge}
\label{subsec:blind_mock_challenge}
To test the full analysis pipeline, we go one step further by testing our model ``blindly". Since the \textsc{OuterRim} simulation is in a known cosmology, we use the method of~\citet{Mead2014a} to rescale the halo positions and velocities, in order to mimic a simulation of a different cosmology. The method has two aims: 
1) rescale the units of the simulation to match the halo mass function of the new cosmology,
2) use the displacement field to adjust positions and velocities to match the linear clustering.
We produced in total 8 different cosmologies, with 3 HOD configurations for each. 
The choice of the cosmological parameters, as well as the validation of the rescaling method, can be found in the companion paper~\citep{Smith2020}, which justifies the parameter range being tested. 
We find that the inferred parameters are sensitive to the choice of the fiducial cosmology.
For our final results, we decide to add the effect of an incorrect fiducial cosmology as additional source of systematic error. To calculate a systematic error due to the fiducial cosmology, we calculate the rms of the set of 24 blind mocks, which are then added in quadrature to the modelling systematic error calculated from the non-blind mocks.
Although the technique of \citet{Mead2014a} is also applicable to more general cases, e.g. dynamical dark energy models, we restrict our blind analysis to standard $\Lambda$CDM cosmologies. Our estimate of the systematic error budget is not affected by this choice. The range of cosmological parameter values explored, e.g. varying $\Omega_{\rm m} h^2$ by ${\sim}20\%$, the spectral index $n_{\rm s}$ by ${\sim}10\%$, and the baryon density parameter $\Omega_{\rm b}$ by ${\sim}35\%$, represent models with a wide range of power spectrum shapes, expansion and growth of structure histories, corresponding to different values of $D_{\rm M}$, $D_{\rm H}$ and $f\sigma_8$, which are the quantities that are most relevant for our analysis. 
The rms we find with the blind mocks challenge is $\delta {\alpha_{\perp}}=0.007$, $\delta {\alpha_{\parallel}}=0.011$, and $\delta f\sigma_{8}=0.010$.

\subsection{Observational systematics: \textsc{EZmocks}}
\label{subsec:obs_sys_ezmocks}
\begin{figure*}
    \centering
    \includegraphics[width=0.65\columnwidth]{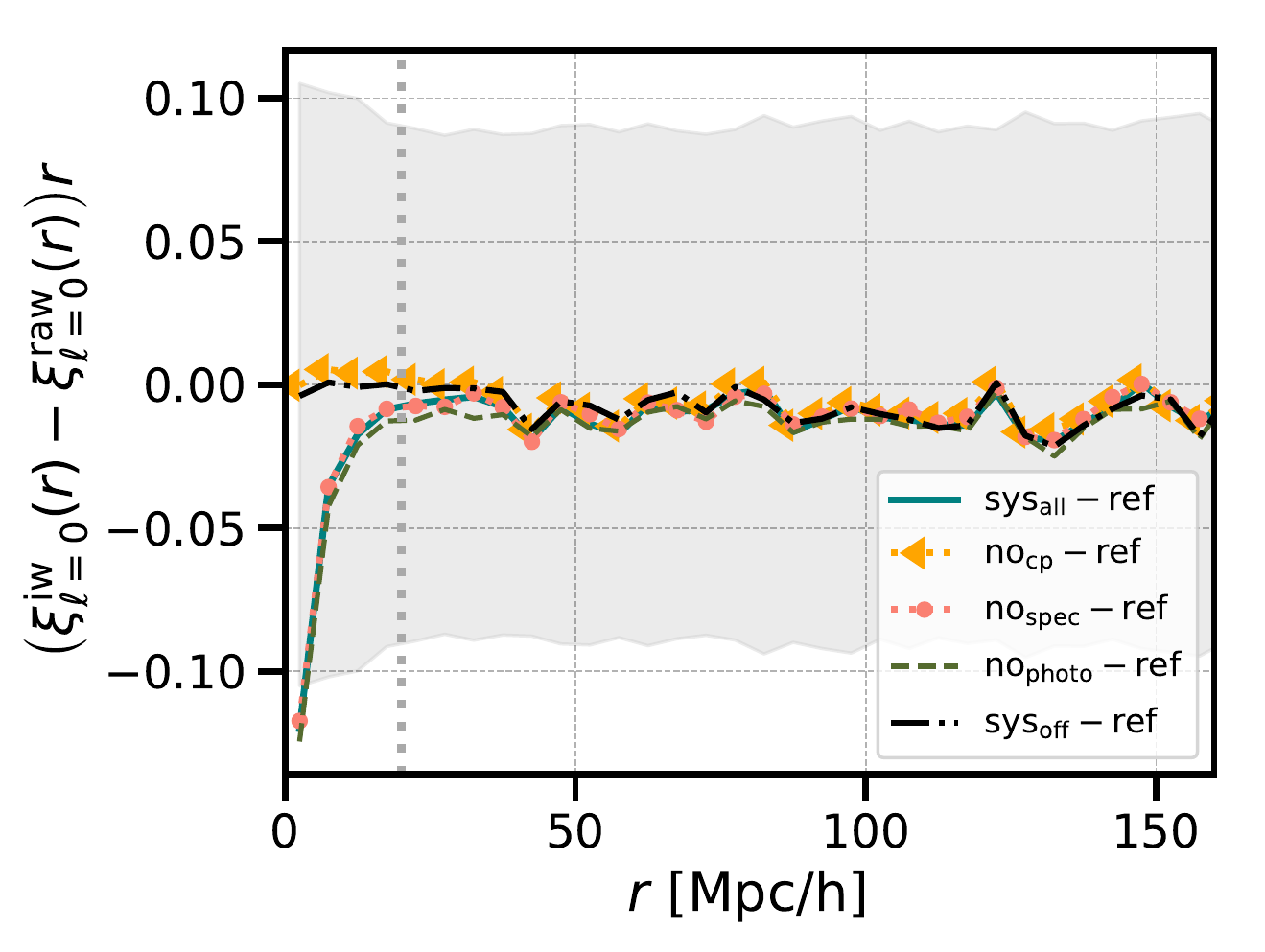}
    \includegraphics[width=0.65\columnwidth]{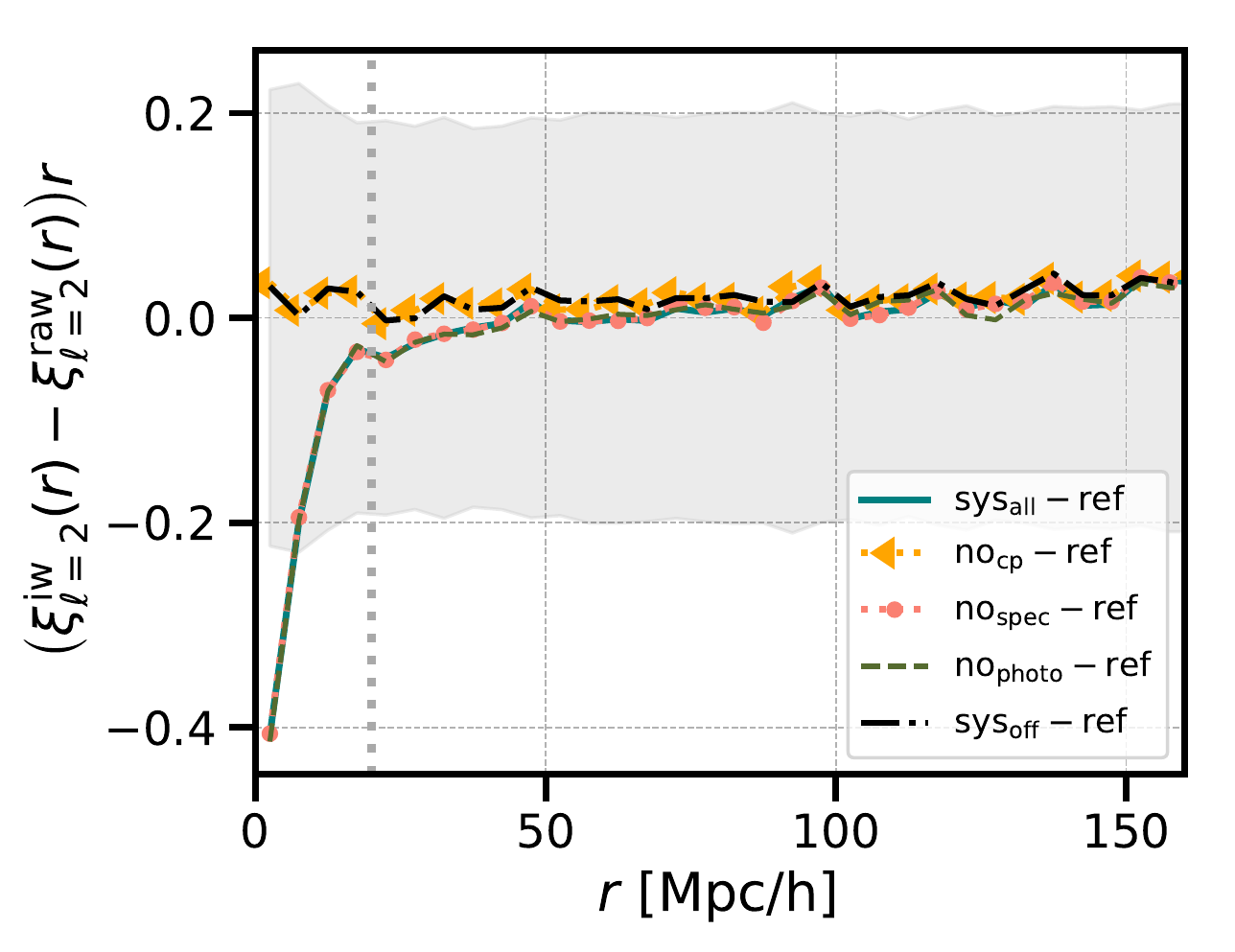}
    \includegraphics[width=0.65\columnwidth]{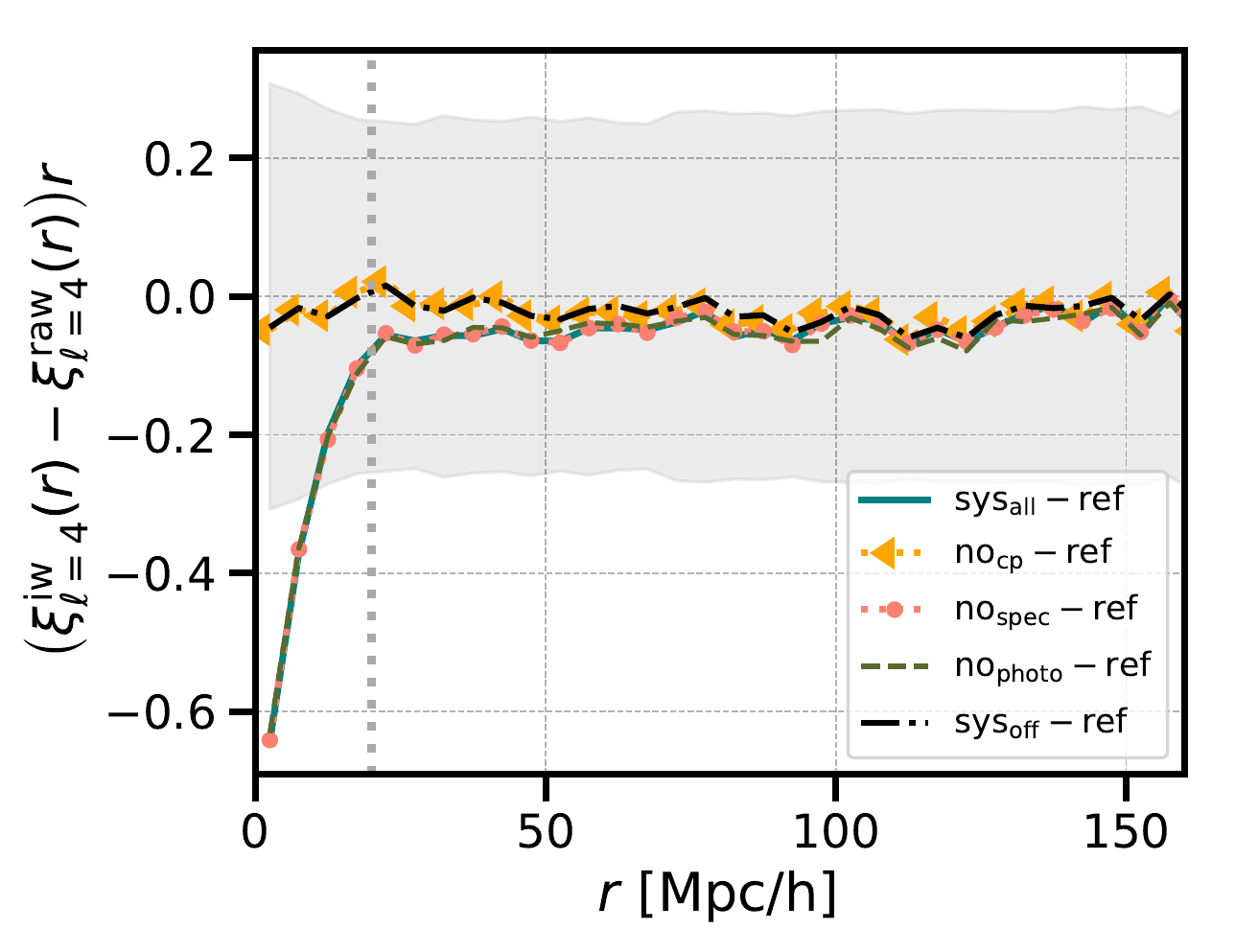}
    \caption{Difference between the correlation function monopole (left), quadrupole (centre) and hexadecapole (right) from the \textsc{EZmocks} with different systematics, with respect to a reference correlation function with no systematics. The black dash-dotted curves denote the difference in the case without the systematic effects compared to the reference case (also show the impact on the radial integral constraint effects). The difference in the mocks including all the systematics is given in the green solid curve. The cases without including the spectroscopic effect and photometric effect are shown in the red dotted curve, the grey band indicates the standard deviation of the 1000 mocks, while the vertical dotted line denotes the scale at $s=20 \mpch$.}
    \label{fig:ximulti_syst_compare}
\end{figure*}

We utilize the \textsc{EZmocks} introduced in Section~\ref{subsec:cov_ezmock} to quantify the observational systematics. We consider the impact of the following observational effects: spectroscopic redshift failures, close pairs, and the photometric calibration. The construction of the EZmocks, which include observational effects, are summarized in Appendix~\ref{appendix:postprocess_ezmocks}. The code for post-processing the systematic effects on the mocks is integrated into the clustering analysis toolkit.\footnote{https://github.com/julianbautista/eboss\_clustering}
Fig.~\ref{fig:ximulti_syst_compare} compares the impact of the different systematics on the correlation function multipoles. It can be seen that the largest effect on small scales is due to fibre collisions (orange curve).
For the monopole, the impact is visible from scales $s \lesssim 25 \mpch$. For higher order multipoles, this effect is already visible at scales starting from $s \lesssim 50 \mpch$. 

\begin{figure*}
    \centering
    \includegraphics[width=1.5\columnwidth]{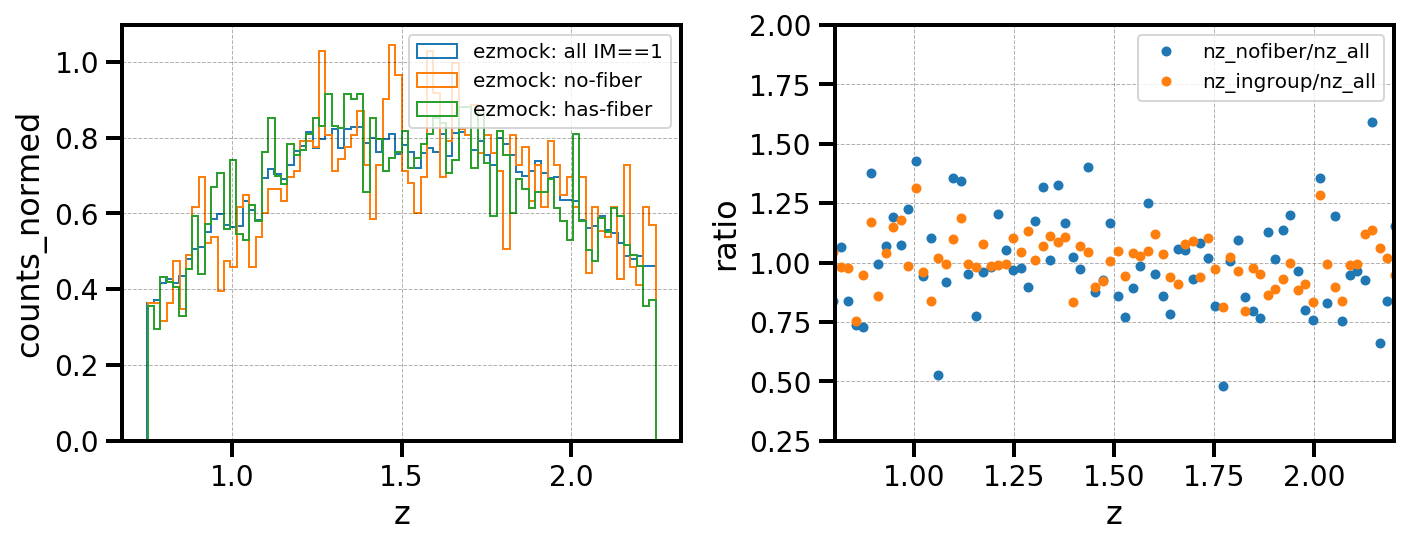}
    \caption{Left panel: distribution of the $n(z)$ in one realization of the \textsc{EZmocks}. The blue histogram is the distribution of the total objects. The green histogram is the distribution for those objects that are assigned with a fibre, and the orange one corresponds to those do not receive a fibre assignment.
    Right panel: blue dots denote the ratio of $n(z)$ between the objects assigned (orange) or not assigned (blue) with a fiber and the total objects. The average difference between the total and ``unobserved" objects in the radial distribution is less than $1\%$}
    \label{fig:nz_ezmock_ngc}
\end{figure*}

There are several methods that can be utilized to correct the small-scale clustering measurements for the effect of fibre collisions. This includes an angular up-weighting~\citep[e.g.][]{Hawkins2003}, modelling the effect of fibre collisions on the correlation function~\citep[][]{Hahn2017}, or an inverse pair weighting scheme~\citep[e.g.][]{Bianchi2017}. To assess the systematics due to the fibre collision, it is required that the radial distribution of the `unobserved' objects is similar to the one of the total objects. It is not necessary that the collided objects which are identified within the same group are physically associated. Therefore, it is not critical whether \textsc{EZmocks} predicts as accurate small scale clustering as the N-body simulations. Fig.~\ref{fig:nz_ezmock_ngc} shows the radial distribution of the unobserved objects and the total objects in one of the \textsc{EZmocks} realizations (left panel), as well as their ratio as a function of redshift (right panel). The similarity of the radial distribution between the unobserved and the total objects in the post-processed mocks making it viable to use these mocks for assessing the systematics.

To correct for the effect of fibre collisions, the method we use is based on~\citet{Hahn2017}, which models the effect of fibre collisions on the correlation function. This method produces similar results to recent pair weighting scheme schemes (see Section~\ref{sec:robustness_tests}).
The effect of fibre collisions is treated as a top-hat function in configuration space. Since our model is built in Fourier space, it is more convenient to modify the power spectrum directly by convolving it with the Fourier transform of the top-hat function. 
We have implemented this method both in configuration and Fourier space and have verified that the difference between the two is very small. 

\begin{figure*}
    \centering
    \includegraphics[width=1.6\columnwidth]{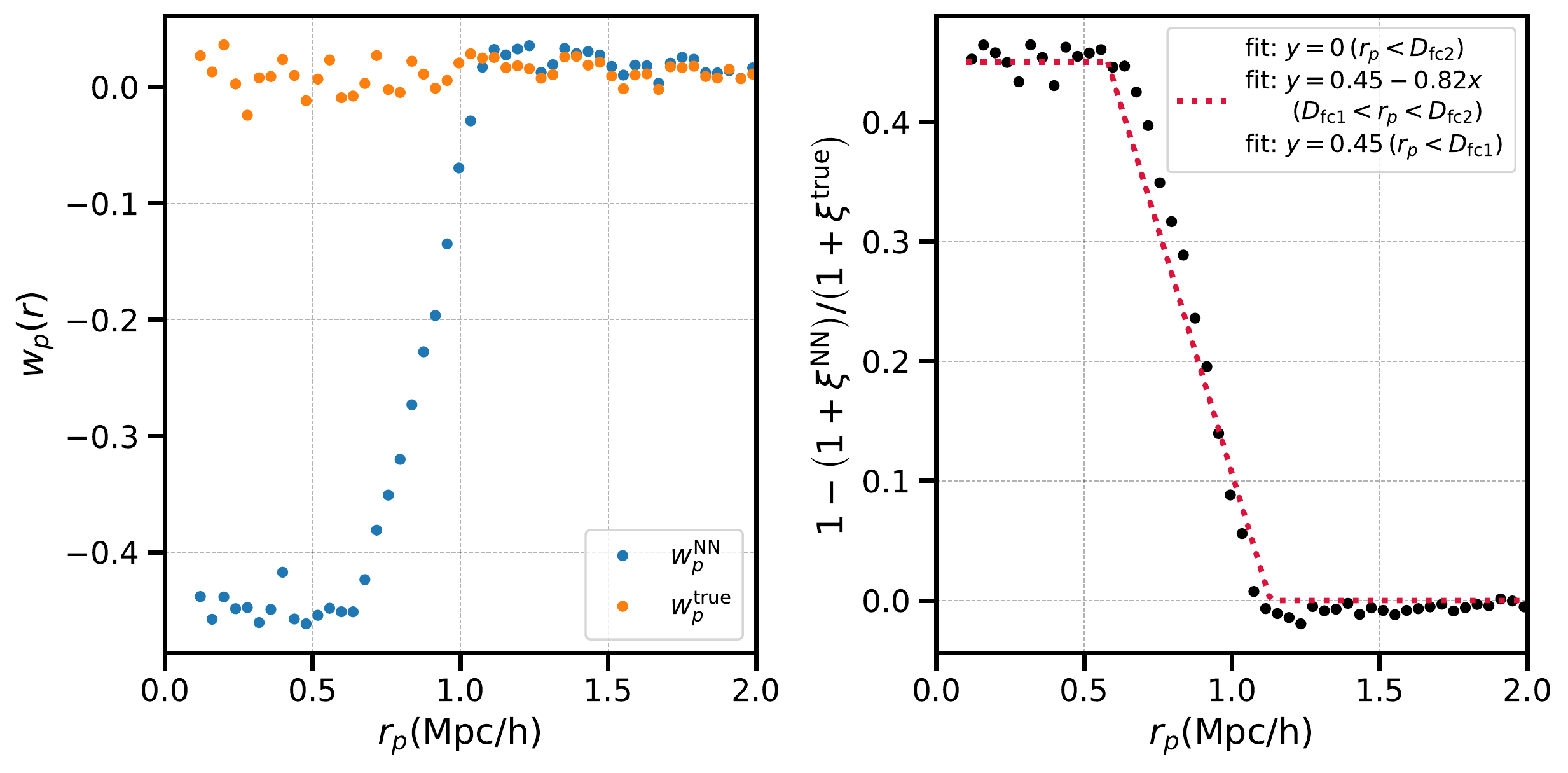}
    \caption[Projected correlation function in \textsc{EZmocks}]{Left panel: projected correlation function measured from the \textsc{EZmocks} with no fibre collisions ($w_p^{\rm true}$, orange), and with fibre collisions and a nearest neighbour weighting ($w_p^{\rm NN}$, blue). Right panel: the ratio $1-(1+\xi^{\rm NN})/(1+\xi^{\rm true})$ (black dots), with a best fit model (dotted red curve). The turnover scales are $D_{\rm fc1}= 0.58 \mpch$ and $D_{\rm fc2} = 1.13 \mpch$.}
    \label{fig:wp_ezmock}
\end{figure*}

The projected correlation function measured from the \textsc{EZMocks} on small scales is shown in the left panel of Fig.~\ref{fig:wp_ezmock}. For the full mock with no fibre collisions ($w_p^{\rm true}$), the clustering amplitude is approximately zero. This is because pairs of physically associated quasars at these separations are very rare, and most of the pairs are due to random alignments on the sky. $w_p^{\rm NN}$ indicates the clustering measured from the mocks with fibre collisions that has been corrected with a nearest neighbour (NN) weight. The negative clustering amplitude indicates an `anti-correlation' due to the fibre collision, but $w_p^{\rm NN}$ does not reach $-1$, since a fraction of closely separated pairs can still be observed, due to the overlapping regimes and the Legacy objects. 

The right panel of Fig.~\ref{fig:wp_ezmock} shows the ratio of the two projected correlation functions. This function is sloped between $0.5 \mpch \lesssim r_{p} \lesssim 1.0 \mpch$, making a top-hat function a poor fit. This is because the fibre collision scale corresponds to a physical scale that depends on redshift, varying from $D_{\rm fc}(z_{\rm min}) = 0.58 \mpch$ to $D_{\rm fc}(z_{\rm max}) = 1.13 \mpch$.

Starting from equation (23) in \citet{Hahn2017}, the correction can be written in terms of the configuration space multipoles,
\begin{equation}
\begin{aligned}
\Delta \xi_{\ell} &=-f_{s} (2\ell + 1)
\int_{0}^{1} W_{\mathrm{fc}}\left(s \sqrt{1-\mu^{2}}\right) \left(\xi(s, \mu)+1\right) \mathcal{L}_{\ell}(\mu) d \mu.
\label{eqn:correction_cp_xi}
\end{aligned}
\end{equation}
We use two different functional forms for $W_{\rm fc}(x)$. The first function we use is the original top-hat function, where the step is at the scale $r_p = D_{\mathrm{fc}}(z_{\rm eff}) = 0.91\mpch$. It is natural to introduce a cut in the line of sight direction, with $\mu_{c}=\sqrt{1-r_{p}^{2}/s^{2}}$, and therefore Eq.~\eqref{eqn:correction_cp_xi} can be simplified to
\begin{equation}
\begin{aligned}
\Delta \xi_{\ell}
= -f_{s} (2\ell + 1) \left[\int_{\mu_{c}}^{1}\left(1+\xi(s, \mu)\right)
\mathcal{L}_{l}(\mu) d \mu\right].\\
\label{eqn:correction_cp_xi_simplified}
\end{aligned}
\end{equation}
We also use a functional form for $W_{\rm fc}(x)$ that is motivated by Fig.~\ref{fig:wp_ezmock}, which we define as
\begin{equation}
W_{\mathrm{fc}}\left(r_{p}\right)=
\left\{
\begin{array}{ll}
{1} & {\text { for } r_{p}\leq D_{\mathrm{fc1}}} \\ 
{t_p-k_p r_{p}} & {\text { for } D_{\mathrm{fc2}}\geq r_{p}>D_{\mathrm{fc1}}}\\
{0} & {\text { for } r_{p}>D_{\mathrm{fc2}}}.\\
\end{array}
\right.
\end{equation}
The slope $k_p$ and intercept $t_p$ are determined by the two characteristic scales, $D_{\mathrm{fc1}}$ and $D_{\mathrm{fc2}}$ as well as the fraction of non-overlapping area, $f_s$, which we leave as a free fitting parameter. 

The systematics obtained from fitting the 1000 \textsc{EZmocks} are summarized in Table~\ref{tab:summary_ezmocks_v7_syst}. We show the systematic shifts in the measurements of $\alpha_{\perp}$, $\alpha_{\parallel}$ and $f\sigma_8$, with respect to the expected values in the cosmology of the mocks.
We divide them into two groups: in the first group, we examine the effect associated to the radial integral constraint~\citep[RIC;][]{deMattia2019}. We used a set of mocks $w_{\rm baseline}$, which are only downsampled by completeness and the redshifts for the random catalogue are drawn from a single global file. Then we added the RIC effect by drawing the redshifts for the random catalogues from each individual data mock $w_{\rm baseline}^{\rm ric}$. In the next line, we correct this effect follow~\citet{deMattia2019} and denote it as $w_{\rm baseline}^{\rm ric{\text -}corr}$. 
In the second group, we examine the effects associated with the observational effects. $w_{\rm no{\text -}sys}$ are mocks without applying any systematics.
Since fibre collisions have the largest impact on the correlation function among the observational systematics (Fig.~\ref{fig:ximulti_syst_compare}), we show in Table~\ref{tab:summary_ezmocks_v7_syst} the results where all systematics
are applied, including and excluding fibre collisions ($w_{\rm all}$ and $w_{\rm nocp}$, respectively). We show results using the top-hat, $w_{\rm all}^{\rm fc(top-hat)}$, or trapezoidal function, $w_{\rm all}^{\rm fc(trapezoid)}$, to apply a correction. When applying the trapezoidal correction, we initially left $f_{\rm s}$ as a free parameter, but found a best fit value of $f_{\rm s}=0.45$ which coincides very well with the predicted value from Fig~\ref{fig:wp_ezmock}. Hereafter, we keep this parameter fixed.
Uncertainties are the standard error of the mean from the 1000 \textsc{EZmocks}. 

To estimate the final observational systematics, we add the RIC effect (${\Delta}^{\rm ric}$) and the observational effects (${\Delta}^{\rm obs}$) in quadrature. We quote the final systematics as the larger value between the systematic bias and two times the standard error of the mean for the mocks, $\delta{\rm sys} = {\rm max} \left\{\Delta_{\rm sys}, 2\sigma_{\rm stat} \right\}$, and we arrive at $\delta {\alpha_{\perp}}=0.003$, $\delta {\alpha_{\parallel}}=0.005$, and $\delta f\sigma_{8}=0.004$.

\begin{table*}
\centering
\caption{Observational systematics measured from the \textsc{EZmocks}, showing the offsets in the measured values of $\alpha_{\perp}$, $\alpha_{\parallel}$ and $f\sigma_8$ to the fiducial values, with different systematics applied. The offset is inferred from the fits to the mean of the 1000 correlation function multipoles. The first group shows the effect associated to the radial integral constraint~\citep[RIC;][]{deMattia2019}. Mocks $w_{\rm baseline}$ are only downsampled by completeness and the redshifts for the random catalogue are drawn from a single global file. $w_{\rm baseline}^{\rm ric}$ are the same, but redshifts in the random catalogues are drawn from the data mocks. The RIC effects are corrected in the model for $w_{\rm baseline}^{\rm ric{\text -}corr}$. 
The second group shows the effects related to the observational effects. 
$w_{\rm no{\text -}sys}$ are mocks without including observational systematics. $w_{\rm nocp}$ includes all systematics except for fibre collisions, while $w_{\rm all}$ includes all systematics. The next rows show the result after applying the correction of~\citet{Hahn2017}, using a top-hat function and a trapezoidal function. 
Uncertainties are taken from the standard error of the mean of the 1000 \textsc{EZmocks}.}
\begin{tabular}{lccc}
\toprule
\textbf{systematics} & \textbf{$\Delta\alpha_{\perp}$} & \textbf{$\Delta\alpha_{\parallel}$} & \textbf{$\Delta f\sigma_{8}$} \\ \midrule
$w_{\rm baseline} $ & 0.002 $\pm$ 0.001 &  -0.003 $\pm$ 0.001 &  -0.009 $\pm$ 0.001 \\ \vspace{1mm}
$w_{\rm baseline}^{\rm ric}$ & 0.006 $\pm$ 0.001 &  -0.005 $\pm$ 0.001 &  -0.013 $\pm$ 0.001 \\ \vspace{1mm}
$w_{\rm baseline}^{\rm ric{\text -}corr}$ & 0.003 $\pm$ 0.001 & -0.004 $\pm$ 0.001 & -0.012 $\pm$ 0.001\\ \cmidrule(r){1-1} \vspace{1mm} 
${\Delta}^{\rm ric}$ & 0.001 $\pm$ 0.001 & -0.001 $\pm$ 0.001 & -0.003 $\pm$ 0.001
\\ \midrule \vspace{1mm} 
$w_{\rm no{\text -}sys}$ & 0.009 $\pm$ 0.001 & 0.002 $\pm$ 0.002 & -0.006 $\pm$ 0.001 \\ \vspace{1mm} 
$w_{\rm nocp}$           & 0.008 $\pm$ 0.001 &  0.002 $\pm$ 0.002 &  -0.006 $\pm$ 0.001  \\ \vspace{1mm}
$w_{\rm all}$            & 0.017 $\pm$ 0.001 &  -0.008 $\pm$ 0.002 &  0.008 $\pm$ 0.002 \\ \vspace{1mm}
$w_{\rm all}^{\rm fc(top-hat)}$ &  0.011 $\pm$ 0.001 &  0.002 $\pm$ 0.002&  -0.003 $\pm$ 0.002  \\ \vspace{1mm}
$w_{\rm all}^{\rm fc(trapezoid)}$  
                         & 0.010 $\pm$ 0.001&   -0.001 $\pm$ 0.002 &  -0.004 $\pm$ 0.002 \\ \cmidrule(r){1-1}
${\Delta}^{\rm obs}$ & 0.001 $\pm$ 0.001 & -0.003 $\pm$ 0.002 & 0.002 $\pm$ 0.002 \\ \midrule
Total & 0.003 & 0.005 & 0.004 \\
\bottomrule
\end{tabular}
\label{tab:summary_ezmocks_v7_syst}
\end{table*}

\section{Constraints on the geometrical parameters and growth rate}
\label{sec:constraints_geopar_growthrate}
In this section we explore the BAO and RSD constraints in terms of comoving angular diameter distance, Hubble distance, and the growth rate of cosmic structure. We estimate the effective redshift using the definition:
\begin{equation}
    z_{\rm eff} = \frac{\sum_{i, j} w_{i} w_{j} (z_{i} + z_{j})}{\sum_{i,j}  2 w_{i} w_{j}},
\label{eqn:zeff_eboss}
\end{equation}
where we sum over pairs with a separation distance between $20\mpch \le ds \le 160\mpch$, the weights $w_i$ are defined as in Eq.~\eqref{eqn:weight_tot}. The exact definition of the pair separation distance has marginal impact on the effective redshift. A comparison using different definition of effective redshift can be found in Section~\ref{appendix:zeff_definition}. 

\subsection{Results in the configuration space: full-shape analysis}
\label{subsec:results_config_full_shape}
\begin{figure}
    \centering
    \includegraphics[width=\columnwidth]{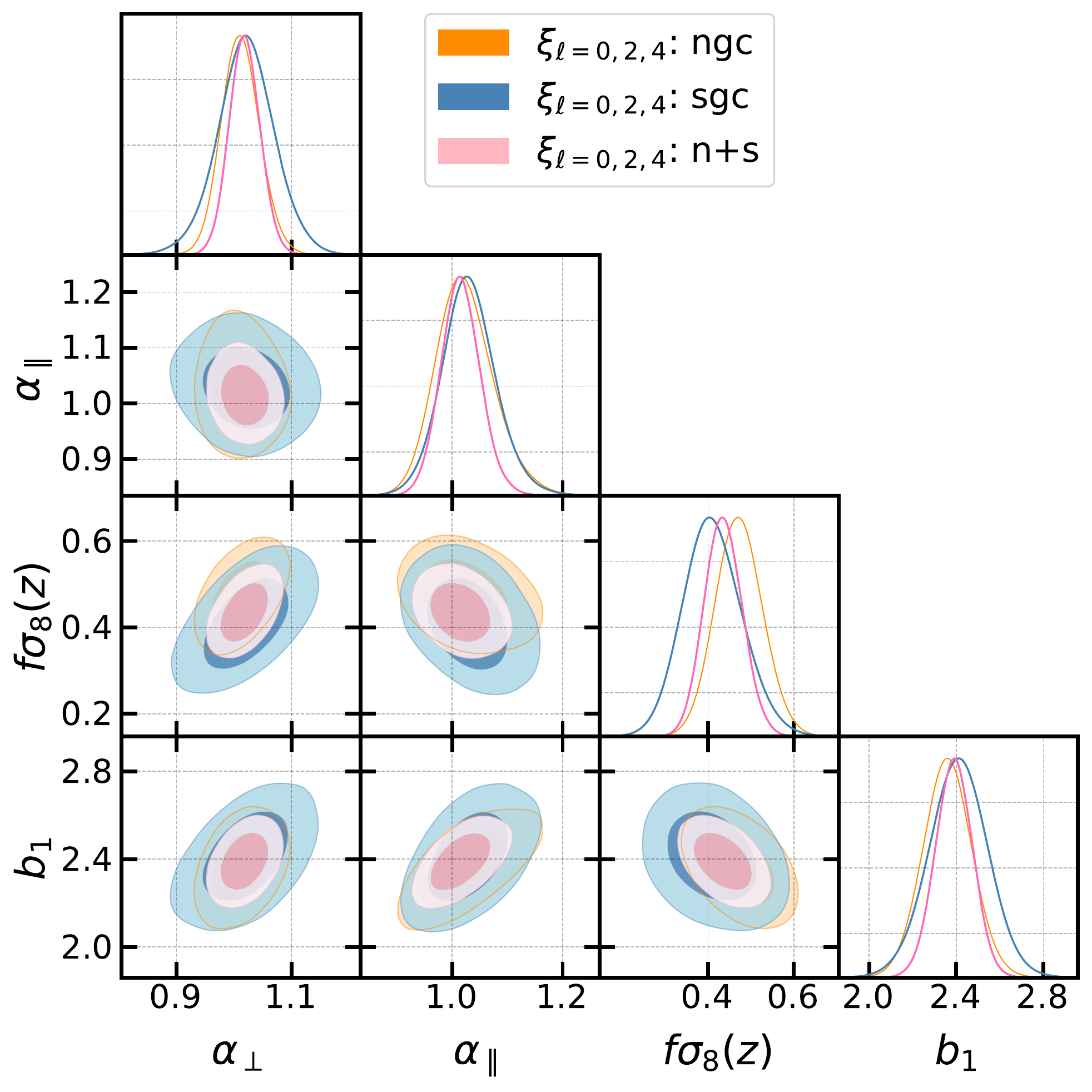}
    \caption{Posterior distribution for AP parameters, $f\sigma_8$ and linear bias $b_1$ for NGC (orange), SGC (blue), and combined (pink).}
    \label{fig:posterior_dr16}
\end{figure}

\begin{figure*}
    \centering
    \includegraphics[width=1.8\columnwidth]{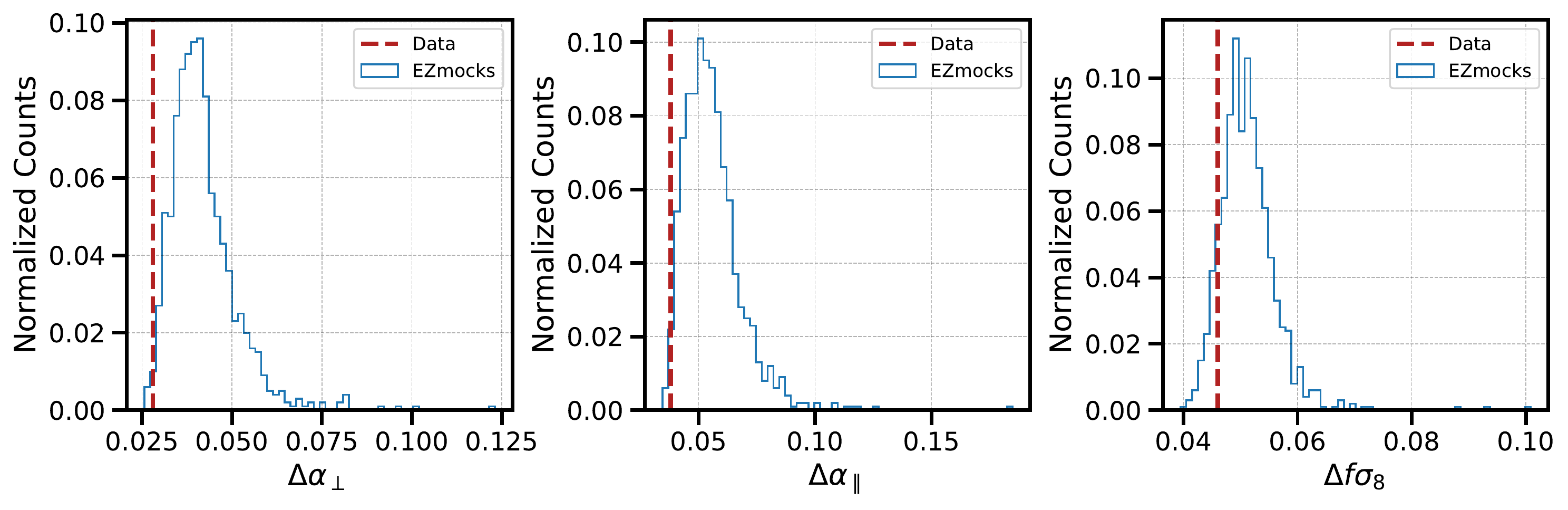}
    \caption{Distribution of the statistical error for $\Delta\alpha_{\perp}$, $\Delta\alpha_{\parallel}$ and $\Delta f\sigma_{8}$ on the \textsc{EZmocks} with all observational effects included (blue histogram) and the statistical error of the DR16 QSO data (red dashed line).}
    \label{fig:err_mock_data}
\end{figure*}

For the full-shape analysis, the final parameter inference is performed using \textsc{RESPRESSO} $+$ fitting function, combined with a RSD model, which is described in Section~\ref{sec:model}. Fibre collisions are corrected using the method described in Section~\ref{subsec:obs_sys_ezmocks}, which is based on~\citet{Hahn2017} but the effect of fibre collisions is modelled by a trapezoidal function. We iteratively find that the parameter $f_{\rm s}=0.4$, which is in good agreement with the measurements of the projected correlation function (see Section~\ref{sec:robustness_tests}).
We perform the analysis on the multipoles $\xi_{\ell}(s)=0,2,4$ within the range $20 \mpch \leq s \leq 160 \mpch$, and with bin separation $\Delta s=5 \mpch$. The 1000 \textsc{EZmocks}, including photometric and spectroscopic systematics, are used to estimate the covariance matrix.
Fig.~\ref{fig:posterior_dr16} shows the posterior distribution of the AP parameters as well as $f\sigma_8$ for the NGC (orange), SGC (blue), and the combination of both (pink). 
Fig.~\ref{fig:err_mock_data} compares the statistical error on the data to the distribution from the \textsc{EZmocks} (combined NGS$+$SGC) for the AP parameters and $f\sigma_8$, where the error inferred from the data sits at the lower tail of the mocks. One reason is that the BAO signal in the data is higher than that in the average of the mocks, which effectively leads to a strong SNR and reduces the statistical error. 
A similar distribution is also observed in the eBOSS LRG sample~\citep{Bautista2020}.
Table~\ref{tab:combined_xipk_constraints} lists our measured values in terms of the AP parameters and $f\sigma_8$. The error bars are derived statistically from the Monte Carlo Markov Chain (MCMC) chain with the correction factor $\sqrt{M}=1.036$ (see Eq.~\eqref{eqn:m_factor}).

{}

We adopt the same fiducial cosmology as for DR14 analysis, $\left\{\Omega_{\rm m},\, \Omega_{\rm b}h^2,\, h, \,n_{\rm s},\, \sigma_8 \right\} = \left\{0.31,\, 0.022, \,0.676,\, 0.97, \, 0.8 \right\}$, where the total matter density parameter also includes a contribution from massive neutrinos $\sum_i m_{\nu}^i = 0.06 {\rm eV}$, corresponding to $\Omega_{\nu} h^2 = 0.0064$. We obtain the fiducial distances, 
$H(z=1.48)=157.40 ~{\rm kms^{-1}Mpc^{-1}}$, 
$D_{\rm M}(z=1.48)=4446.82$ Mpc, 
and $r_{\rm drag}=147.8$ Mpc. 
This corresponds the rms of the mass contained in a sphere of radius $12\, \rm{Mpc}$, $\sigma_{12} = 0.79$, as suggested in~\citet{Sanchez2020}.
Finally, using Eq.~\eqref{eqn:ap_dist_par} and ~\eqref{eqn:ap_rd}, we arrive at the comoving angular diameter distance, Hubble distance, and $f\sigma_8$:
\begin{align}
&D_{\rm M}(z_{\rm eff}=1.48)/r_{\rm drag} = 30.66 \pm 0.84 \pm 0.25, \\
&D_{\rm H}(z_{\rm eff}=1.48)/r_{\rm drag} = 13.11 \pm 0.49 \pm 0.17, \\
&f\sigma_8(z_{\rm eff}=1.48) = 0.439 \pm 0.046 \pm 0.014,
\end{align} 
where the first error denotes the statistical uncertainty including correction factor $\sqrt{M}=1.036$. The second error denotes the systematic uncertainty inferred from the \textsc{OuterRim} mock challenge (including both the blind and non-blind tests) as well as the observational systematics from the \textsc{EZmocks}. The three of them are summed in quadrature. The individual systematic uncertainties are listed in Table~\ref{tab:config_dmh_syst}. The systematic errors are quoted as the larger value between the systematic bias and the $2\sigma$ of the standard deviation of the mean of the mocks.

\begin{table}
\caption{Systematics for full-shape and BAO-only fit to the mocks. On the \textsc{OuterRim} HOD mocks, we took the set with realistic redshift smearing and catastrophic redshift failure for the non-blind mocks. For the blind mocks, we use 24 boxes.
The rms is calculated from the \textsc{EZmocks} sets with all observational systematics included for the full-shape analysis.}
\label{tab:config_dmh_syst}
\centering
\begin{tabular}{@{}cccc@{}}
\toprule
\begin{tabular}[c]{@{}c@{}}Full-shape\end{tabular} & \multicolumn{2}{c}{\textsc{OuterRim}} & \textsc{EZmocks} \\ \midrule
rms & non-blind & blind & all-syst\\ \midrule
$\delta D_{\rm M}/r_{\rm drag}$ & 0.070 & 0.210 & 0.104\\
$\delta D_{\rm H}/r_{\rm drag}$ & 0.057 & 0.145 & 0.057 \\
$\delta f\sigma_{8}$ & 0.008 & 0.011 & 0.004 \\ 
\midrule
\begin{tabular}[c]{@{}c@{}}BAO-only\end{tabular} & \multicolumn{2}{c}{\textsc{OuterRim}} &  \\ \midrule
rms & non-blind & blind & \\ \midrule
$\delta D_{\rm M}/r_{\rm drag}$ & 0.133 & 0.161 &\\ 
$\delta D_{\rm H}/r_{\rm drag}$ & 0.091 & 0.113 & \\ 
\bottomrule
\end{tabular}
\end{table}

\begin{figure*}
    \centering
    \includegraphics[width=0.8\columnwidth]{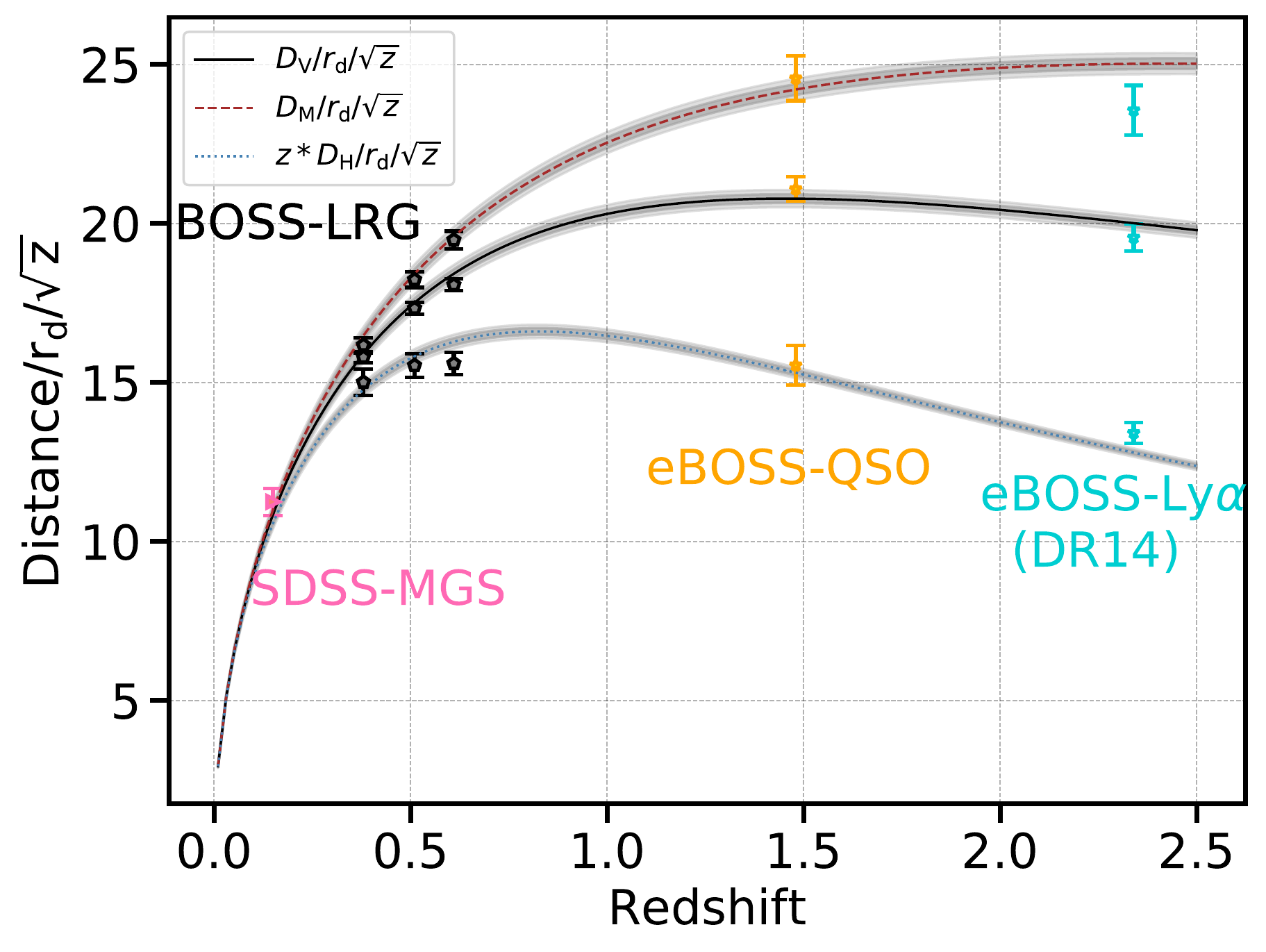}
    \includegraphics[width=0.8\columnwidth]{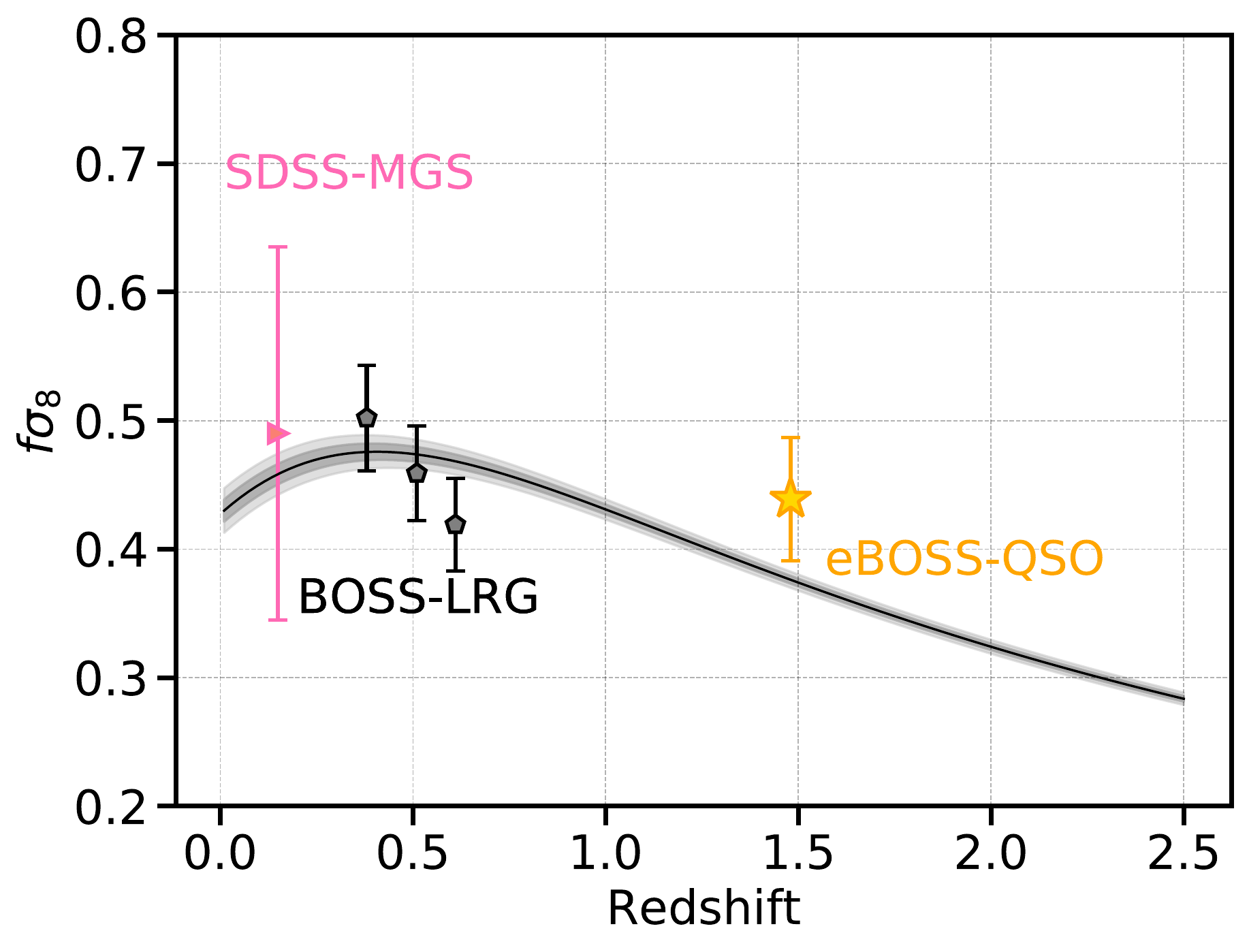}
    \caption{Redshift evolution of the distance parameter and the cosmic growth rate. The $\Lambda$CDM model, with input from the Planck 2018 MCMC chains, is shown by the curves (as indicated in the legend). The grey points are from the BOSS DR12 LRG sample~\citep{Alam2017}, the blue points are from the combined eBOSS DR14 Ly$\alpha$ auto- and cross-correlation function~\citep{deSainteAgathe2019, Blomqvist2019}, the pink points are from an early SDSS MGS sample (\citet{Ross2015} for the distance measurement and~\citet{Howlett2015} for the growth rate measurement). The orange points show the final results from the eBOSS DR16 quasar analysis, in configuration space.}
    \label{fig:param_of_z}
\end{figure*}

Fig.~\ref{fig:param_of_z} shows the redshift evolution of the distance measurements (left panel) and growth rate measurement (right panel).
Our final results from the DR16 quasar sample in the configuration space are shown by the yellow points with error bars. We compare this to the $\Lambda$CDM model inferred from the Planck CMB temperature and polarization measurements. We also show previous results from the SDSS main galaxy sample (MGS) for the distance measurement~\citep{Ross2015} and growth rate measurement~\citep{Howlett2015}, the constraints from BOSS DR12 LRG sample~\citep{Alam2017}, and the combined constraints from eBOSS DR14 Ly$\alpha$ measurements~\citep{deSainteAgathe2019, Blomqvist2019}.
With the final QSO sample, statistically we gain ${\sim}45 \%$ in the distance measurement, and ${\sim}30\%$ in the growth rate measurement compared to our DR14 QSO analysis.

We present the parameter covariance matrix including the statistical error, theoretical modelling systematics and observational systematics in the $D_{\rm M}/r_{\rm drag}$, $D_{\rm H}/r_{\rm drag}$, and $f\sigma_{8}$ basis as
\begin{equation}
\mathbf{C}_{\lambda} = \begin{blockarray}{ccc}
D_{\rm M}/r_{\rm drag} & D_{\rm H}/r_{\rm drag} & f\sigma_{8} \vspace{1mm} \\
\begin{block}{( ccc )}
  7.709\times 10^{-1} & -5.656\times 10^{-2} & 1.750\times 10^{-2} \\ 
  - & 2.640\times 10^{-1} & -6.204\times 10^{-3} \\
  - & -      & 2.308\times 10^{-3}\\
\end{block}
\end{blockarray}
\end{equation}
The results expressed in various alternative basis can be found in the Appendix~\ref{appendix:alter_basis}.

\begin{table}
\centering
\caption{A summary of the parameter space $\lambda$. A flat prior is applied to all parameters with uniform distribution inside the limits and zero otherwise. In all cases, the distortion parameters $q_{\perp}$, $q_{\parallel}$ and $f\sigma_8$ are all free. We vary also the three bias parameters $b_1$, $b_2$, $\gamma_3^{-}$ and $a_{\rm vir}$ for the RSD effect at small scale. $\sigma_{\rm zerr}$ and $f_{\rm s}$ are left optional depending on the feature of the problem we study. }
\label{tab:param_space_nuisance}
\begin{tabular}{ | l | l | c | }
\toprule
Parameter & Description  &  Prior limits \\ 
\midrule
$b_1$ & Linear bias & $[0.25, 6]$ \\
$b_2$ & Second order bias & $[-2, 3]$\\
$\gamma_3^{-}$ & non local bias & $[-2, 2]$ \\
$a_{\rm vir}$ & FoG kurtosis &  $[0.2, 10]$\\
$\sigma_{\rm zerr}$ & Redshift error & $[0, 6]$\\
$f_{\rm s}$ & fibre collision  & $[0.2, 0.7]$\\
\midrule
$q_{\perp}$ &  Distortion ${\perp}$ L.O.S& $[0.5, 1.5]$ \\
$q_{\parallel}$ & Distortion ${\parallel}$ L.O.S &   $[0.5, 1.5]$ \\
$f\sigma_8$ & growth parameter & $[0, 1]$ \\
\bottomrule
\end{tabular}
\end{table}

\subsection{Results in the configuration space: BAO-only analysis}
\label{subsec:results_bao_only}
We apply the model described in Section \ref{sec:model_bao} to the measured eBOSS quasar monopole and quadrupole, $\xi_{\ell = 0, 2}(s)$, in the range $50 \mpch < s < 150 \mpch$, with bin size $\Delta s=5\mpch$. 
The constraints in the basis of AP parameters can be found in Table~\ref{tab:combined_xipk_constraints}. The error bars given here are only derived statistically from the MCMC chain with the correction factor $\sqrt{M}=1.010$ (see Eq.~\eqref{eqn:m_factor}).

The $\chi^2$ value of our fit to $\alpha_{\perp}$ and $\alpha_{\parallel}$ is $\chi^2$/dof = 34.1/30, while for the fit to $\alpha_{\rm iso}$ we have $\chi^2$/dof = 16.5/15. 
The best fitting models are presented in Fig.~\ref{fig:xibao}. The top panel shows the monopole, where we fit $\alpha_{\perp}$ and $\alpha_{\parallel}$. However, the model where we fit $\alpha_{\rm iso}$ looks almost identical.

\begin{figure}
    \centering
    \includegraphics[width=0.8\columnwidth]{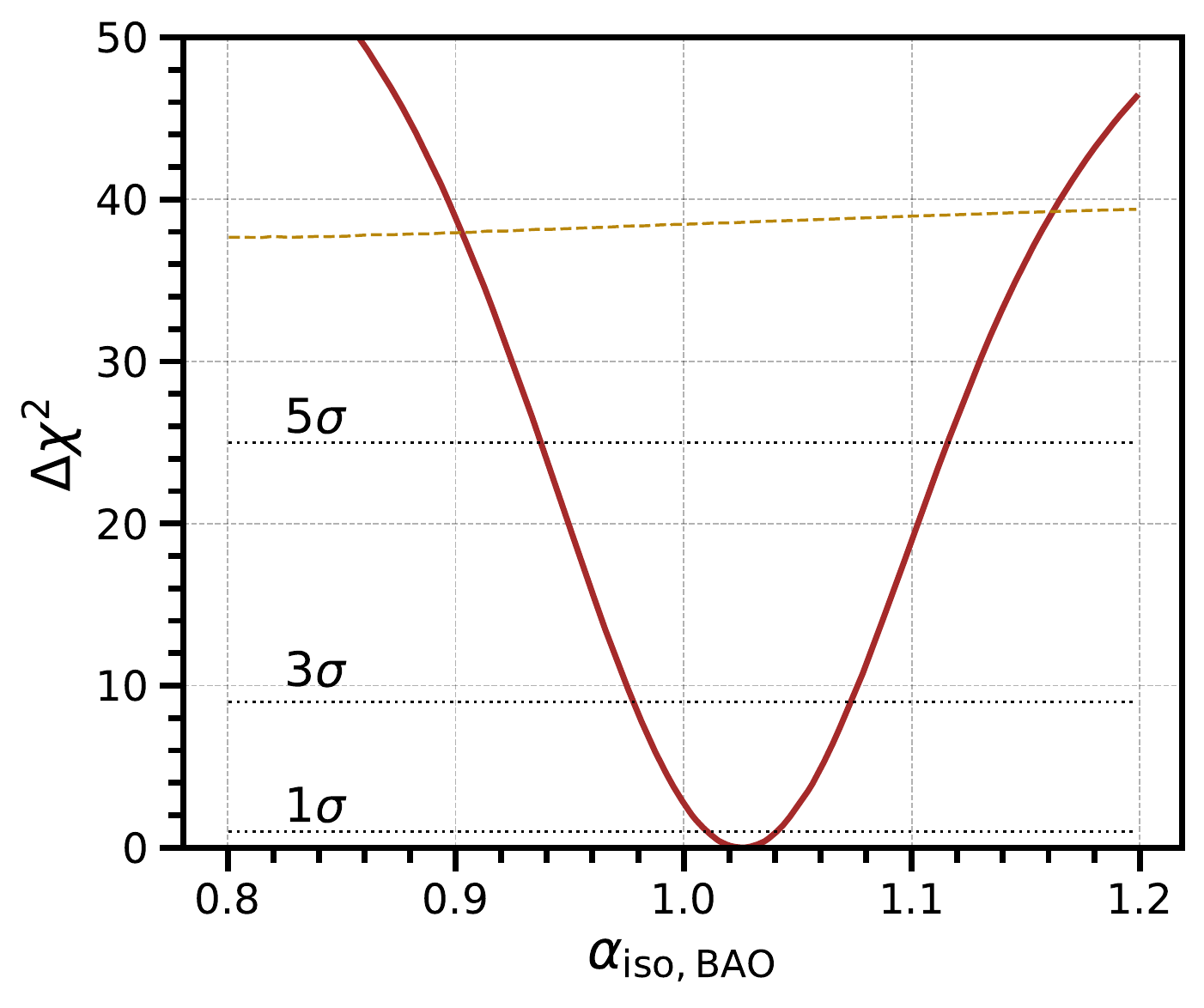}
 \caption{The likelihood of the BAO parameter $\alpha_{\rm iso}$ from the fit to the correlation function monopole, in terms of $\Delta \chi^2$ (solid curve). The dashed curve indicates the likelihood for a model with no BAO feature. The no BAO model has a $\chi^2$ greater than 37 over the full range of $\alpha_{\rm iso}$ values. This implies that the clustering of the eBOSS DR16 quasar sample has a BAO feature at greater than $6\sigma$ significance.}
    \label{fig:baosig}
\end{figure}

We show the likelihood in Fig.~\ref{fig:baosig}, in terms of $\Delta \chi^2$, for our fit to $\alpha_{\rm iso}$. Our BAO measurement is shown by the solid curve, while the dashed curve is the result for a fit to a template that does not include the BAO. This highlights the significance of the BAO feature in the eBOSS DR16 quasar data, as we find that the BAO model is preferred by a significance greater than $6\sigma$.

{}   

We convert the BAO $\alpha_{\perp}$ and $\alpha_{\parallel}$ results to constraints on the comoving angular diameter and Hubble distance:
\begin{align}
&D_{\rm M}(z_{\rm eff}=1.48)/r_{\rm drag} = 30.82 \pm 0.82 \pm 0.21, \\
&D_{\rm H}(z_{\rm eff}=1.48)/r_{\rm drag} = 13.22 \pm 0.56 \pm 0.14.
\end{align} 
The first error denotes the statistical uncertainty including the correction factor $\sqrt{M}=1.010$, while the second error denotes the systematic uncertainty, which is inferred from the mock challenge based on Table~\ref{tab:config_dmh_syst}. We have not explicitly performed the tests on the observational systematics for the BAO-only fit, but the results are expected to be very similar to the ones reported in the Fourier-space analysis at sub-percent level~\citep{Neveux2020}.

\begin{figure}
    \centering
    \includegraphics[width=0.8\columnwidth]{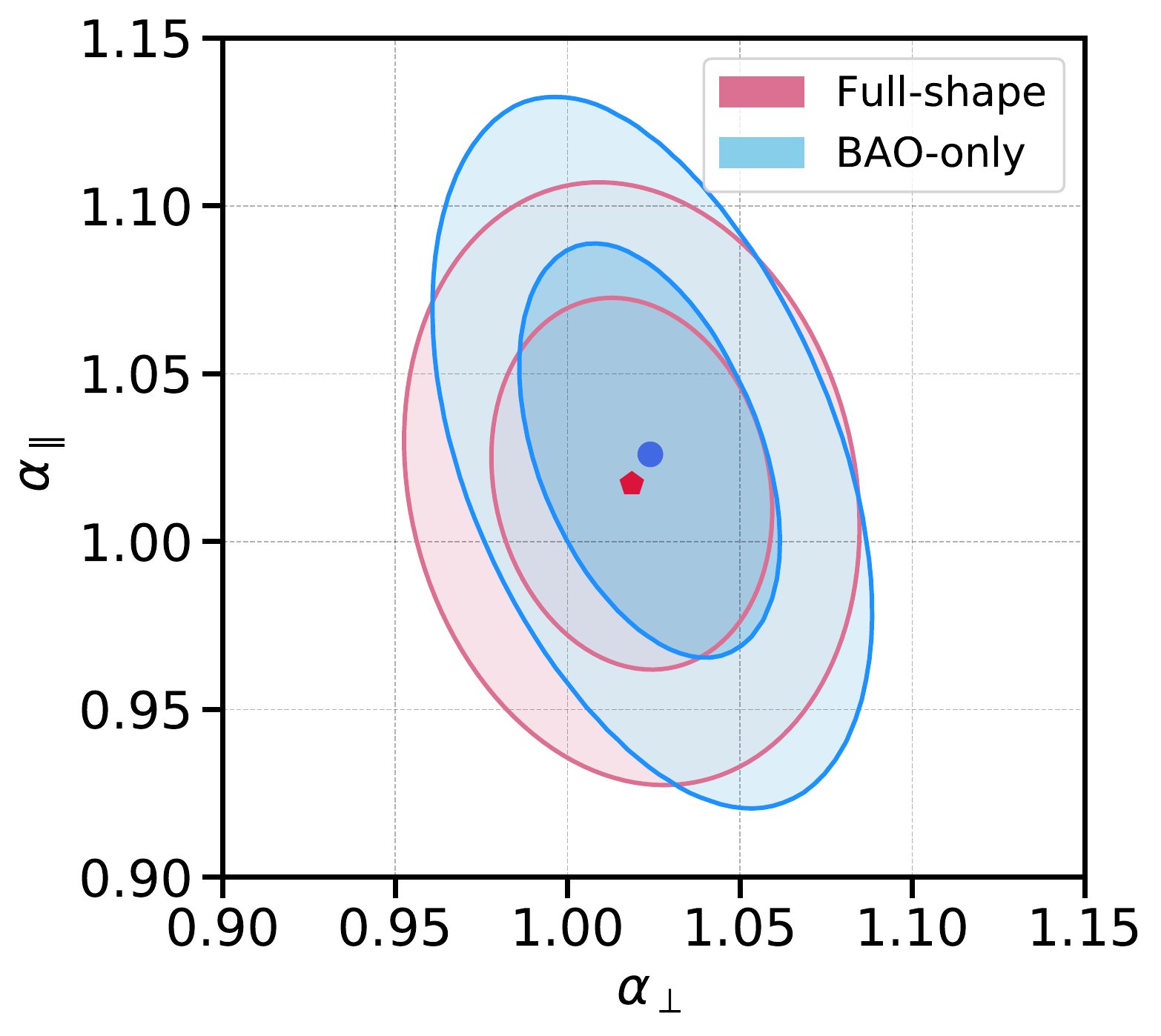}
    \caption{Comparison of full-shape (red contour) and BAO only (blue contour) fits in configuration space.}
    \label{fig:2dcontour_fs_bao}
\end{figure}

Fig.~\ref{fig:2dcontour_fs_bao} compares the posterior distribution of the AP parameters for full-shape and BAO-only analysis. The BAO-only measurements are in good agreement (within $0.5\sigma$) with the full-shape measurements presented in the previous subsection. The degeneracy direction of $\alpha_{\perp}$-$\alpha_{\parallel}$ for the BAO-only fit (blue contour) is precisely predicted in~\citet{Ross2015b}.
The full-shape measurement is expected to obtain improved results on $D_{\rm M}(z)$ and $D_{\rm H}(z)$ through the broad-band modeling of the AP effect. For our results, this manifests as a $14$ percent improvement in the statistical uncertainty on $D_{\rm H}(z)$. 

It has been shown that the BAO-only analysis is robust to the assumption of fiducial cosmology~\citep{Carter2019}. Since the full-shape analysis is potentially sensitive to the shape of the model template, we have performed a detailed analysis using the set of \textsc{OuterRim} mocks in blind cosmologies~\citep{Smith2020} and thus believe the full-shape results, with the inclusion of our systematic uncertainties, are robust to these concerns. The good agreement between the full-shape and BAO-only results further strengthen our confidence. Our BAO results are used, after being combined with those of \citealt{Neveux2020}, for the cosmological tests in \citet{Mueller2020} that only use BAO information.

{}

\subsection{Combination of the configuration space and Fourier space results}
\label{subsec:comb_xi_pk}
We use the method described in ~\citet{Sanchez2016a} to combine the results. The aim is to compress the information obtained from $m$ different of methods into a single set of measurement. Under the Gaussian assumption, such a measurement should always be possible and we should be able to write down the equation
\begin{equation}
\mathbf{D}_{\mathrm{c}}=\mathbf{\Psi}_{\mathrm{c}}^{-1} \sum_{i=1}^{m}\left(\sum_{j=1}^{m} \Psi_{j i}\right) \mathbf{D}_{i},
\end{equation}
where the compressed precision matrix is,
\begin{equation}
{\bf \Psi_{\mathrm{c}}}^{-1} \equiv\left(\sum_{i=1}^{m} \sum_{j=1}^{m} \Psi_{i j}\right)^{-1}.
\end{equation}
In the case the two methods are completely independent from each other, the big precision matrix, ${\bf \Psi_{\rm c}}$, reduces to be block diagonal. The statistical error of the data is directly calculated from the MCMC chain. We use the 1000 \textsc{EZmocks} including the systematic effects to estimate the correlation $C_{m_{i}(\lambda_{n})\otimes m_{j}(\lambda_{l})}$ between the cosmological parameters $\lambda = \{\lambda_{1}, \lambda_{2} \ldots \lambda_{n}\}$ among different methods $m = \{m_1, m_2 \ldots m_{i}\}$ as well as the correlation coefficients between cosmological parameters of the same method $C_{m_{i}(\lambda_{n} \otimes \lambda_{l})}$. The estimation of the correlation between the parameters of the same method is different from the original proposal, and we discuss the difference in Section~\ref{sec:robustness_tests}.

\begin{figure}
    \centering
    \includegraphics[width=0.8\columnwidth]{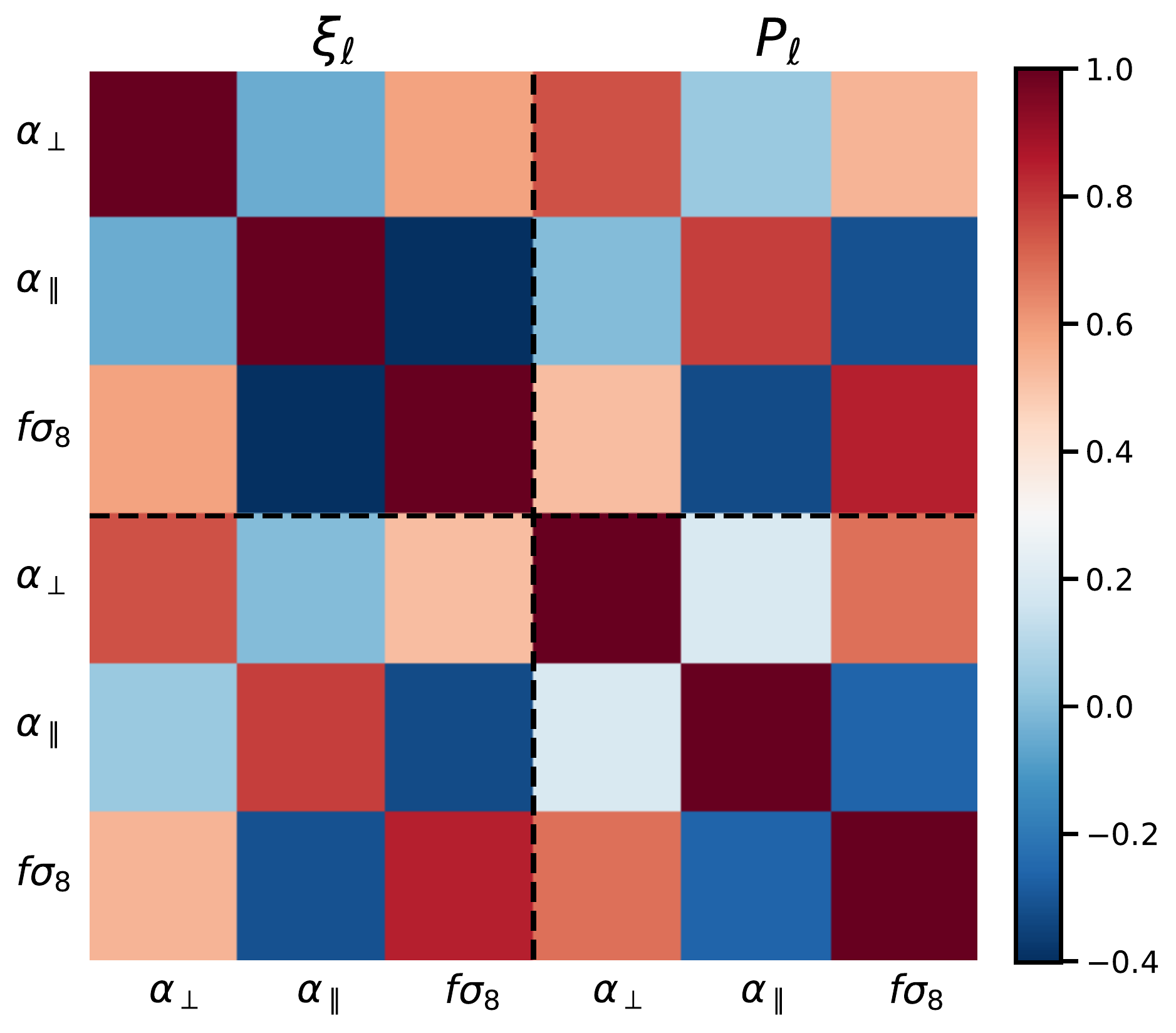}
    \caption{Correlation coefficients for the configuration multipoles and the power spectrum.} 
    \label{fig:corr_coeff}
\end{figure}

The diagonal elements from the real data are rescaled using Eq.~\eqref{eqn:cov_m_corr} for both configuration and Fourier space. The covariance matrix of the \textsc{EZmocks} is estimated from the scattering of the best-fit parameters, which is then normalised using the error inferred from the data.
Fig.~\ref{fig:corr_coeff} shows the correlation coefficients between two methods, with the diagonal terms of the off-diagonal blocks being $0.743$, $0.783$, $0.844$. 

{}

\begin{figure*}
    \centering
    \includegraphics[width=0.6\columnwidth]{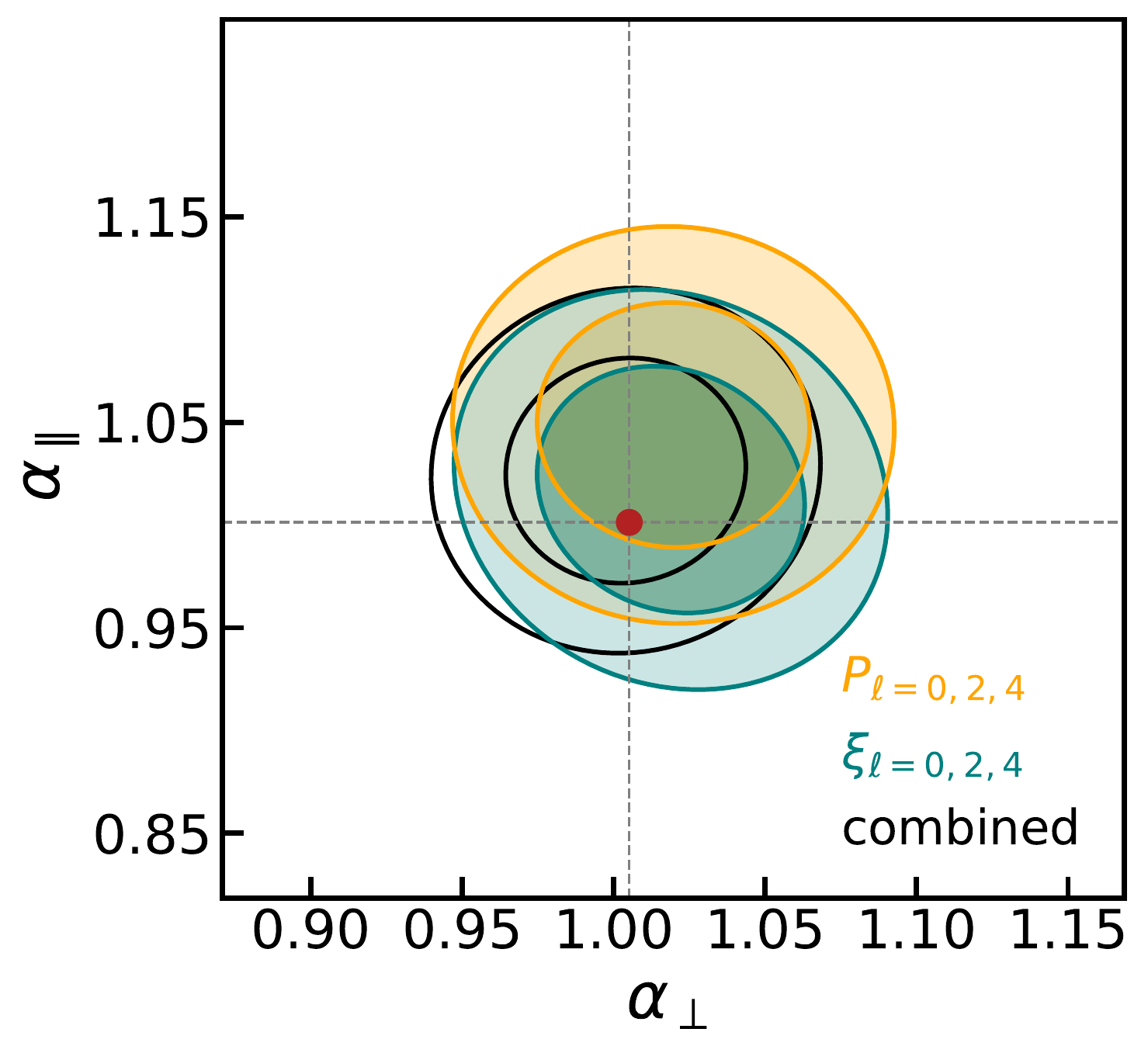}
    \includegraphics[width=0.6\columnwidth]{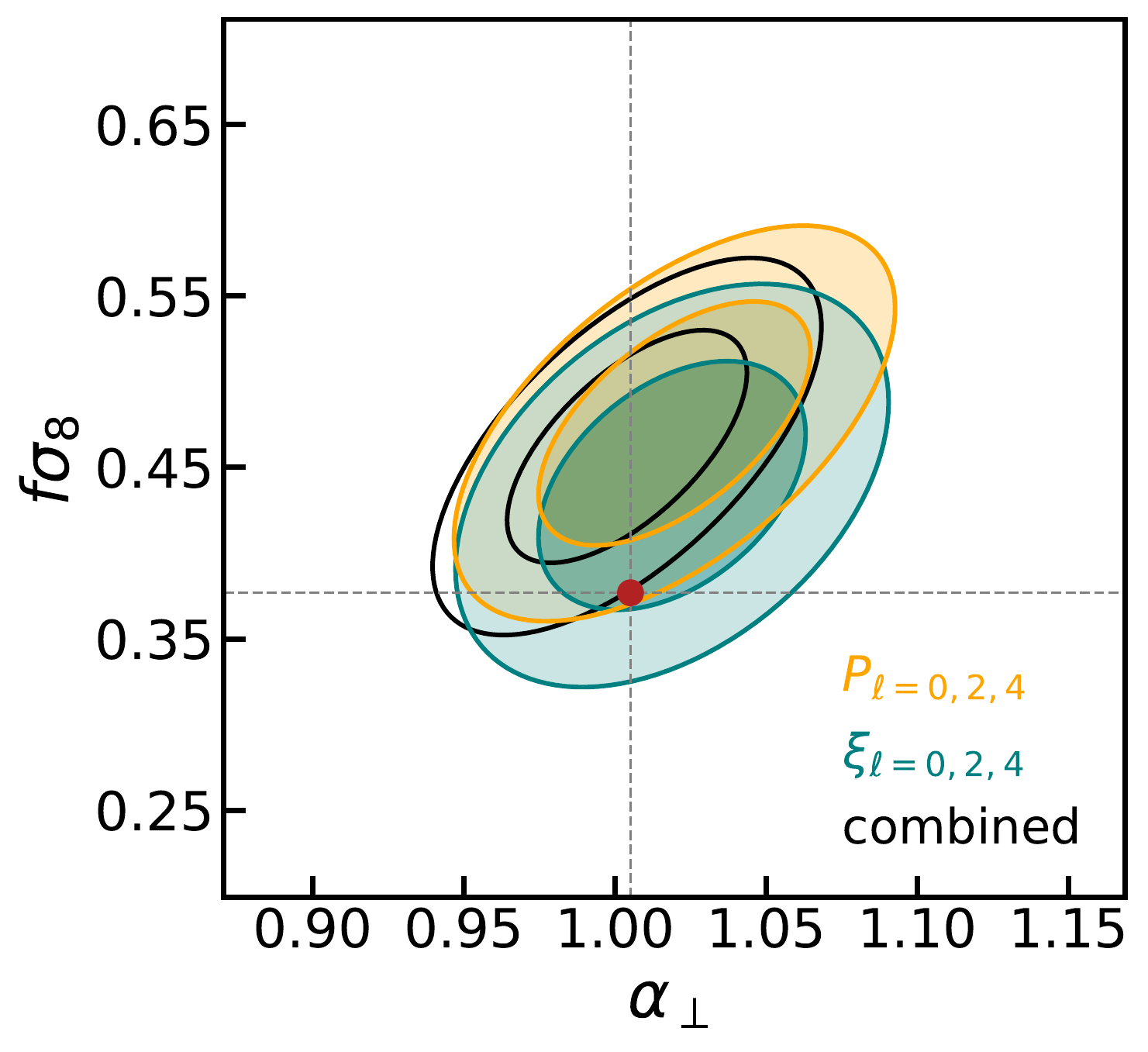}
    \includegraphics[width=0.6\columnwidth]{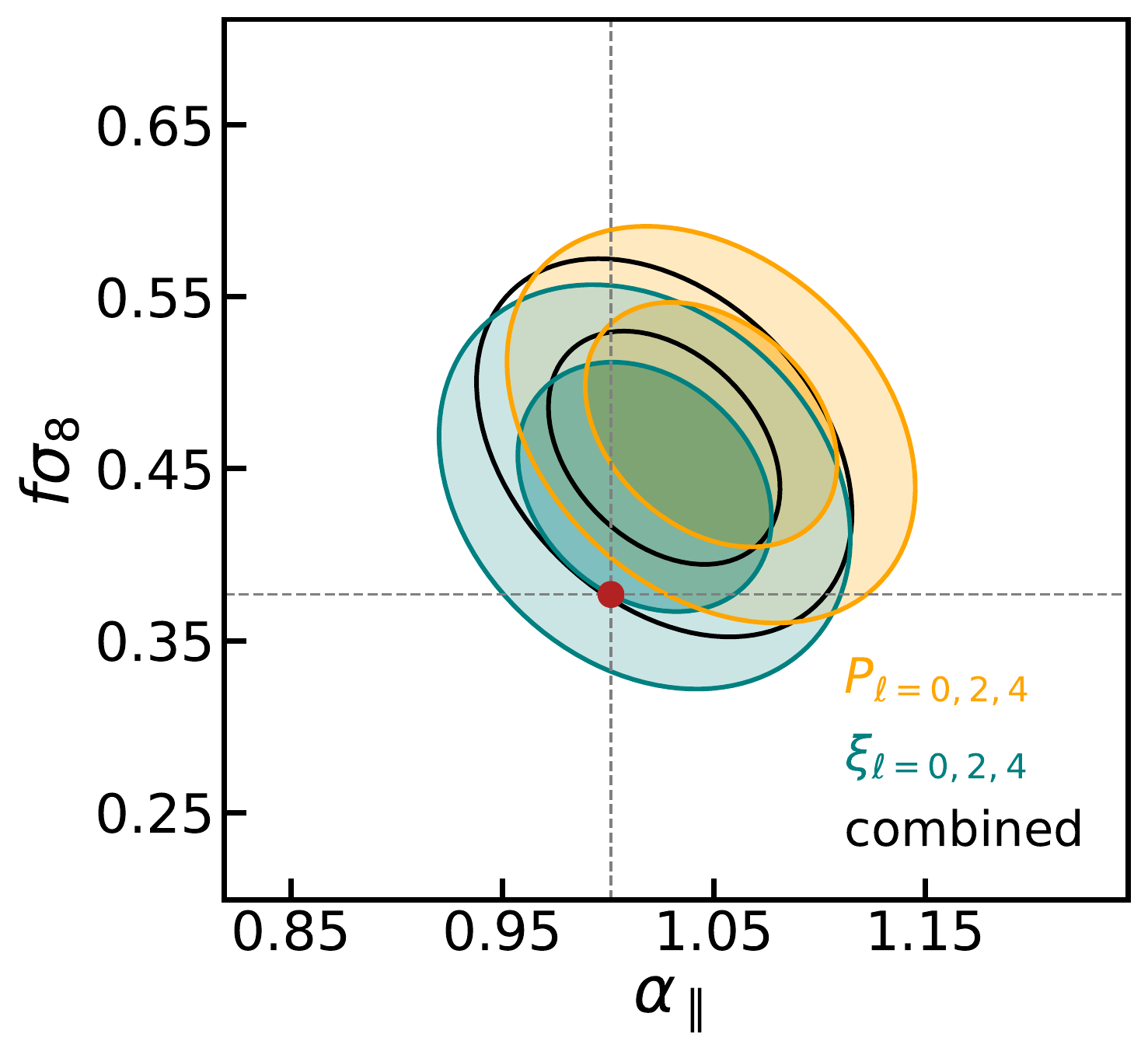}
    \caption{Posterior for $\alpha_{\perp}$, $\alpha_{\parallel}$ and $f\sigma_8$ configuration space, Fourier space and the combined results using the method described in ~\citep{Sanchez2016a}. The filled contours are derived from the MCMC chains in configuration space (green), and Fourier space(orange). The black solid ellipses are the combined constraints at the 68, 95 confidence limits. The red points denote the values that are inferred from the Planck constraints~\citep{collaboration2018planck}.}
    \label{fig:gaussian_posterior_combined_xipk}
\end{figure*}

Fig.~\ref{fig:gaussian_posterior_combined_xipk} shows the posterior for $\alpha_{\perp}$, $\alpha_{\parallel}$, and $f\sigma_8$ in configuration space (green), Fourier space (orange), and the combined results using the method described in~\citet{Sanchez2016a}. The black solid ellipses represent the combined constraints at the 68 and 95 confidence limits. As summarized in Table~\ref{tab:combined_xipk_constraints}, by combining the configuration and Fourier space results, we find an improvement in the statistical uncertainty of $\sim 7\% ~\sigma$ in $\alpha_{\perp}$, $\sim 3\% ~\sigma$ in $\alpha_{\parallel}$, $\sim 5\% ~\sigma$ in $f\sigma_{8}$.

\begin{table}
\caption{Table summarizes the values on the final DR16 data for the $\alpha_{\perp}$, $\alpha_{\parallel}$ and $f\sigma_8$ in configuration space, Fourier space and the combined results. }
\begin{tabular}{@{}cccc@{}}
\toprule
Full-shape & $\alpha_{\perp}$ & $\alpha_{\parallel}$ & $f\sigma_8$ \\ \midrule
$\xi_{\ell}$ & 1.019 $\pm$ 0.028 & 1.017 $\pm$ 0.038 & 0.439 $\pm$ 0.046 \\
$P_{\ell}$ & 1.020 $\pm$ 0.029 & 1.049 $\pm$ 0.038 & 0.476 $\pm$ 0.045 \\
combined & 1.004 $\pm$ 0.026 & 1.027 $\pm$ 0.035 & 0.462 $\pm$ 0.043 \\ \hline \hline
BAO-only & $\alpha_{\perp}$ & $\alpha_{\parallel}$ & $\alpha_{\rm iso}$ \\ \midrule
$\xi_{\ell}$ & 1.024 $\pm$ 0.026 & 1.026 $\pm$ 0.042 & 1.026 $\pm$ 0.016 \\ \bottomrule
\end{tabular}
\label{tab:combined_xipk_constraints}
\end{table}

To quantify the combined systematic error, we use the non-blind mocks that include the effects of redshift smearing and catastrophic redshifts and the blind mocks with various implementations of HODs (see Section~\ref{subsec:blind_mock_challenge}). We combine the configuration and power spectrum multipoles for each of the boxes, and calculate the correlation coefficients using the 100 realizations for each box. The systematic error is derived from the rms of the difference with respect to the true cosmology. The combined statistics on the \textsc{OuterRim} mocks is summarized in Table~\ref{tab:combined_or_syst_dmh}. 
The observational systematics inferred from the \textsc{EZmocks} are directly added to the diagonal terms of the data covariance matrix.
Finally, we arrive at the combined result in terms of comoving angular diameter distance, Hubble parameter, and $f\sigma_8$,
\begin{align}
&D^{\bf{c}}_{\rm M}(z_{\rm eff}=1.48)/r_{\rm drag} = 30.21 \pm 0.79, \\
&D^{\bf{c}}_{\rm H}(z_{\rm eff}=1.48)/r_{\rm drag} = 13.23 \pm 0.47, \\
&f\sigma_8^{\bf{c}}(z_{\rm eff}=1.48) = 0.462 \pm 0.045,
\end{align} 
where the errors include both the statistical and systematic uncertainties. The final covariance matrix for the combined data reads 

\begin{equation}
\mathbf{C}^{\bf{c}}_{\lambda} = \begin{blockarray}{ccc}
D_{\rm M}/r_{\rm drag} & D_{\rm H}/r_{\rm drag} & f\sigma_{8} \vspace{1mm} \\
\begin{block}{(ccc)}
  6.227\times 10^{-1} & 1.424\times 10^{-2} & 2.257\times 10^{-2} \\
  -                   & 2.195\times 10^{-1} & -7.315 \times 10^{-3} \\
  -                   & -                   & 2.020\times 10^{-3}\\
\end{block}
\end{blockarray}
\end{equation}{}

\begin{table}
\caption{Table summarizes the combined systematics for configuration and the Fourier space results, calculated from the non-blind and blind \textsc{OuterRim} mocks. We use the 20 sets of non-blind \textsc{OuterRim} mocks, and the 24 sets of blind mocks, in 8 different cosmologies.}
\centering
\begin{tabular}{@{}ccc@{}}
\toprule
combined & non-blind & blind \\ \midrule 
$\delta D_{\rm M}/r_{\rm drag}$ & 0.079 & 0.129 \\ 
$\delta D_{\rm H}/r_{\rm drag}$ & 0.053 & 0.094\\ 
$\delta f\sigma_{8}$ & 0.009 & 0.008\\ 
\bottomrule
\end{tabular}
\label{tab:combined_or_syst_dmh}
\end{table}

\section{Robustness tests on the data analysis}
\label{sec:robustness_tests}
In this section we describe the various systematic tests we perform on the data to check the robustness of our inferred cosmological constraints. We consider alternative definitions of the systematic weights, model for the two-point correlation function, definition on the effective redshift, and the impact of the fibre collision correction. The final results are summarised in Table~\ref{tab:syst_test_dr16}, which shows how the final measurements of $\alpha_{\perp}$, $\alpha_{\parallel}$ and $f\sigma_8$ shift with different choices for the systematic corrections.

\subsection{List of tests performed on the data}
\subsubsection{Redshift efficiency weights}
\label{subsec:spec_weight}
The redshift detection efficiency depends, for example, on the efficiency of the spectrograph, observational conditions, position of the objects with respect to the focal plane, and the intrinsic properties of the objects. To account for the inhomogeneity in the redshift detection efficiency, we identify the trends in $n_{\rm good}/n_{\rm total}$ as a function of the fibre number ID and the spectral SNR, where $n_{\rm good}$ stands for the number of good objects and $n_{\rm total}$ is for the total objects. An inverse weighting is assigned to each object to correct for the trend. As discussed in Section~\ref{sec:overview_eboss}, the efficiency in detecting the redshift of the objects is not uniform across different fibres (see figure 4 in the companion paper \citet{Ross2020}). The detection efficiency is lower near the edge of the CCDs, as well as near the locations of the CCD amplifiers.  While the trend as a function of the spectral SNR is weak for the quasar sample, we include the correction to remove any dependency. Both effects are accounted for in the final redshift failure weighting. In~\citet{Ata2017} the correction was performed by up-weighting objects by the success rate of the sectors. In Table~\ref{tab:syst_test_dr16} we show the impact of weighting based on the success rate of each sector (denoted as ``$w_{\rm noz, ssr}$"). In addition, we also show a weighting scheme that only corrects for the trend in fibre ID number, without considering the spectral SNR (denoted as ``$w_{\rm noz, id}$").

\subsubsection{Photometric weights}
\label{subsec:photometric weights}
In DR14 QSO analysis~\citep{Gil-Marin2018,Hou2018,Zarrouk2018}, the trend in $n_{\rm data}/n_{\rm random}$ was calibrated against the extinction corrected $g$-band depth and the extinction coefficients $E(B-V)$. In fact, the QSO data also shows trends in the sky background and seeing, in the i-band (see figure 9 of~\citet{Ross2020}). In the final data catalogue we correct for all of these trends. In Table~\ref{tab:syst_test_dr16}, we show the impact of using photometric weights that omit the trends in the i-band, which we denote as ``$w_{\rm photo, no{\text -}{\it{i}}}$".

\subsubsection{Close pair correction}
The finite radius of the fibre leads to objects in close pairs being missed. Our fibre collision correction, which models the impact on the two-point correlation function, is described in Section~\ref{subsec:obs_sys_ezmocks}. An alternative treatment of this effect can be found in~\citet{Bianchi2017} and \citet{Mohammad2018}, where correlation function measurements are corrected using pairwise inverse probability (PIP) weights. The idea is to up-weight the pair counts based on the probability that each pair can be observed. This probability is inferred by running the fibre assignment algorithm many times (on the corresponding eBOSS input target catalogue)
to find how often each pair can be observed. The detailed description of catalogue with PIP weights that we use can be found in ~\citet{Mohammad2020}.
Table ~\ref{tab:syst_test_dr16} shows the impact of using the PIP weighting, which is denoted as ``$w_{\rm cp, pip}$".

\subsubsection{Impact on the combination of NGC and SGC}
\label{appendix:caps}
\begin{figure}
       \centering
        \includegraphics[width=\columnwidth]{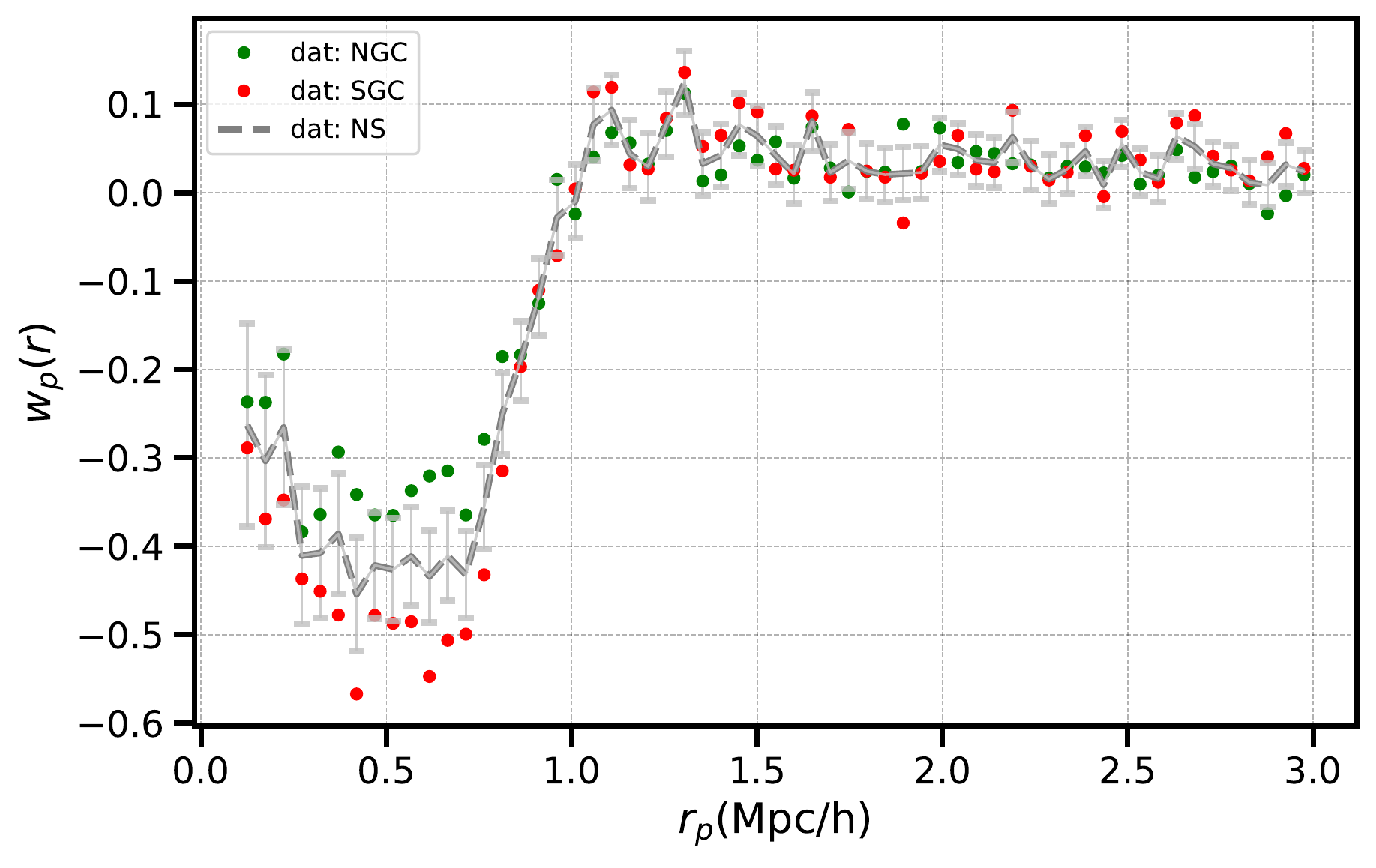}
        \caption{Projected correlation function on the data for north cap (green), south cap (red), and the combined pair counts (grey).  The error bars are derived from the EZmocks.}
        \label{fig:wp_fc_dr16_ns}
    \end{figure}

We compare two methods for combining the data from the NGC and SGC. In the first method, which is done in our final analysis, the pair counts from the north and south caps are combined. In the second method, the north and south caps are fitted separately, and the posterior distributions are combined. Given that the north and south caps are statistically independent, the second method would correspond to applying fits simultaneously to both caps, but leaving all the fitting parameters free at the same time (including the AP parameters, $f\sigma_8$, bias parameters, etc). To determine the correction factor of the fibre collision, $f_{\rm s}$ (see Section~\ref{subsec:obs_sys_ezmocks}), for the north and south caps, we measure the projected correlation function. To increase the signal-to-noise ratio, we integrate over the full depth of the QSO sample along the radial direction. Fig.~\ref{fig:wp_fc_dr16_ns} shows the projected correlation function for the NGC, the SGC and the combination. When fitting the NGC and SGC separately, we find $f_{\rm s}|_{\rm NGC} = 0.36$ and $f_{\rm s}|_{\rm SGC} = 0.45$, which is consistent with~\citet{Neveux2020}. In Table~\ref{tab:syst_test_dr16} we show the effect on our results of combining independent fits to the NGC and SGC. The shifts are small compared to the total systematic uncertainty. 
 
    {}    
    
\subsubsection{Alternative estimation of the correlation coefficients}
\label{subsec:test_gaussian_comb}
    As discussed in Section~\ref{subsec:comb_xi_pk}, to estimate the correlation between cosmological parameters measured using different methods, our only option is to use the 1000 \textsc{EZmocks} (the set that includes the systematic effects. To estimate the correlation between cosmological parameters within the same method, we have two options: use either the \textsc{EZmocks} or use a covariance matrix that is inferred directly from the data. The latter option is justified if, on average, the error inferred from a single realization matches that from the ensemble of the mocks. 

    Fitting the 1000 \textsc{EZmocks} in configuration space, we find a good agreement between the mean of the standard deviation and the scatter of the best-fit values for the 1000 realizations.
    For presenting the results, we select the first option of estimating the correlation coefficients using the mocks. Although the correlation coefficients are cosmology dependent, the estimation from an ensemble of mocks is expected to be more robust and less sensitive to statistical fluctuations. To further confirm the combining method, we performed test on the 1000 \textsc{EZmocks} for the first option, we arrive at the mean of the standard deviation of the 1000 realizations: $ \overline{\rm{std} (\alpha_{\perp})} = 0.038$, $\overline{\rm{std}(\alpha_{\parallel})} = 0.052$ and $\overline{\rm{std}( f\sigma_{8})} = 0.049$, which is in good agreement with the scatter in the best fitting parameters for the 1000 realizations (include the correction factor given by equation (22) in~\citet{Percival2014}): $\delta\alpha_{\perp} = 0.039$, $\delta\alpha_{\parallel} = 0.052$, $\delta f\sigma_{8} = 0.049$. The effect of choosing the second option of using the data to infer the correlation coefficients is shown in Table~\ref{tab:syst_test_dr16}.

\subsection{Summary of the robustness test}

Table~\ref{tab:syst_test_dr16} shows how the measurements of $\alpha_{\perp}$, $\alpha_{\parallel}$ and $f\sigma_8$ are shifted, for alternative choices of weighting schemes,
compared to the one used in the final data catalogue. In the spectroscopic weighting, the effect of correcting for the trend in the spectral SNR has a marginal impact on the parameter constraints. In addition, the difference when using the ``SSR" weights applied to the DR14 data is at the sub-percent level compared to the statistical error. Similarly, the correction in the photometric weights by including the sky background and seeing in the $i$-band also induces changes at a sub-percent level, and therefore this does not influence the conclusions drawn from the DR14 release. The close pair correction using the PIP algorithm has a larger impact on $\alpha_{\parallel}$ and $f\sigma_8$, where the latter one accounts for $44\%$ of the statistical error. Given the statistical properties of the two close-pair treatment schemes, this is difference is statistically not significant; nevertheless, it would be worth exploring for future denser samples.
The table also lists miscellaneous tests including the impact of setting $f_{\rm s}=0$ in our modelling of the fibre collision effect, a different definition of the effective redshift (see Section~\ref{appendix:zeff_definition}), constraints derived using the gRPT model, a different method to combine the NGC with SGC, and an alternative estimation of the correlation coefficients. These tests all show a much smaller variation compared to the statistical uncertainty, which demonstrates the robustness of our analysis.

\begin{table*}
\centering
\caption{Table compares the impact on potential systematics, that includes alternative weighting schemes for redshift efficiency weights ($w_{\rm noz, \{id, ssr \}}$), photometric ($w_{\rm photo, no{\text -}{\it{i}}}$), and fibre collision weight ($w_{\rm cp, pip}$). The table also includes the definition of the effective redshift ($5$-th row), the correction on the fibre collision effect ($6$-th row), the difference using the gRPT model ($7$-th row), the impact on the combination of the caps ($8$-th row) and alternative estimation of the correlation coefficients when combining the data ($9$-th row).}
\begin{tabular}{@{}cccc@{}} \toprule
  & $\Delta\alpha_{\perp}$ & $\Delta\alpha_{\parallel}$ & $\Delta f\sigma_8$ \\ \midrule
$w_{\rm noz, id}$ & 0.000 $\pm$ 0.026 & 0.001 $\pm$ 0.037& 0.000 $\pm$ 0.043 \\ \midrule
$w_{\rm noz, ssr}$ & 0.002 $\pm$ 0.026 & -0.002 $\pm$ 0.036& 0.004 $\pm$ 0.046 \\ \midrule
$w_{\rm photo, no{\text -}{\it{i}}}$ &
-0.002 $\pm$ 0.027 &
-0.001 $\pm$ 0.036 &
-0.002 $\pm$ 0.042 \\ \midrule
$w_{\rm cp, pip}$ & -0.007 $\pm$ 0.026 & 0.012 $\pm$ 0.034& -0.019 $\pm$ 0.043 \\ \midrule
fibre collision \\ $f_{\rm s} = 0$ & 0.003 $\pm$ 0.027 & -0.004 $\pm$ 0.036 & 0.007 $\pm$ 0.044 \\
\midrule
$z_{\rm eff}=1.52$ & 0.000 $\pm$ 0.028 & -0.001 $\pm$ 0.036 & 0.001 $\pm$ 0.044 \\
\midrule
model \\ gRPT & 0.002 $\pm$ 0.027 & -0.001 $\pm$ 0.037 & 0.002 $\pm$ 0.044 \\ \midrule
\begin{tabular}[c]{@{}l@{}}NGC+SGC\\ independent \end{tabular} & 0.002 $\pm$ 0.028 & 0.013 $\pm$ 0.037 & -0.005 $\pm$ 0.043 \\ \midrule
correlation coeff.  & 0.003 $\pm$ 0.026 & 0.007 $\pm$ 0.035 & -0.002 $\pm$ 0.043 \\ 
\bottomrule
\end{tabular}
\label{tab:syst_test_dr16}
\end{table*}

\section{Conclusions}
\label{sec:conclusions}
In this paper we presented the full-shape and BAO-only analysis of the eBOSS DR16 QSO clustering sample. We measured the two-point correlation function of the quasar sample, which we decomposed into Legendre multipoles, $\xi_{\ell}(s)$, with $\ell=0,2,4$. In our full-shape analysis, we incorporated a new recipe to describe the correlation function. 
The matte power spectrum is calculated using \textsc{RESPRESSO}~\citep{Nishimichi2017}, whose original python code was implemented in \textsc{Fortran}. The power spectra that involve the velocities were computed using the fitting formulae provided by~\citet{Bel2019}.

In the final data release, we doubled the number of objects and the survey area compared to the DR14 sample, leading to a 6$\sigma$ detection of the BAO signal in configuration space \citep[consistent with the Fourier space analysis of ][]{Neveux2020}. Compared to the DR14 analysis, the final sample represents a reduction of $\sim 45\%$ in the statistical 
uncertainties of our distance measurements, and $\sim 30\%$ for the growth rate measurement. 
We obtained the comoving angular diameter distance
$D_{\rm M}(z_{\rm eff})/r_{\rm drag} = 30.66\pm0.88$,
the Hubble distance 
$D_{\rm H}(z_{\rm eff})/r_{\rm drag} = 13.11\pm0.52$,
and the cosmic structure growth rate
$f\sigma_8(z_{\rm eff}) = 0.439\pm0.048$.
Our analysis in the configuration space combined with the analysis in the Fourier space~\citep{Neveux2020} allowed us to obtain a tighter constraints in the cosmological distance and growth rate parameters:
$D^{\bf{c}}_{\rm M}(z_{\rm eff})/r_{\rm drag} = 30.21 \pm 0.79$,
$D^{\bf{c}}_{\rm H}(z_{\rm eff})/r_{\rm drag} = 13.23 \pm 0.47$, and
$f\sigma_8^{\bf{c}}(z_{\rm eff}) = 0.462 \pm 0.045$.

The measuremnts of the AP parameters are found to be within $1\sigma$ to the best-fitting $\Lambda$CDM model to the combination of Planck and previous BAO 
measurements~\citep{collaboration2018planck}.
The growth rate measurement $f\sigma_8$ in configuration space is found to agree at the $1.4\, \sigma$ level with the same $\Lambda$CDM prediction. Meanwhile, when 
combined with the results in the Fourier space, the inferred growth rate $f\sigma_8$ is ${\sim}2\sigma$ higher than the $\Lambda$CDM model with the best fit from the Planck measurements. The tendency of higher $f\sigma_8$ was observed in DR14 analysis~\citep{Hou2018, Gil-Marin2018, Zarrouk2018}.

We performed extensive tests to quantify potential systematics and focused on testing observational effects as well as the modelling of the two-point correlation function.
We tested observational systematics using fast mocks including various angular effects (such as fibre collision, photometric, and redshift failure effects). We corrected for
the largest angular systematics (fibre collision) using a modified form following~\citet{Hahn2017}.
We also corrected for the radial integral constraints as described in~\citet{deMattia2019}. Based on these tests, the residual observational systematics on the inferred 
parameters are shown to be at sub-percent level.
Based on a set of HOD mocks built on N-body simulation~\citep{Smith2020}, we examined the modelling of the two-point correlation function. In these mocks we checked the 
impact of various HODs and also included different redshift uncertainty distribution, as well as catastrophic redshift failure objects (potentially important for future 
surveys). Our model can account for these effects, and we can recover 1 percent accuracy for the distance measurement and 3 percent for the growth rate measurement.
A larger systematics turned out to be the impact of the fiducial cosmology and is the dominant source of our systematic error budget that accounts for up to $30$ percent 
of the statistical error.

As a consistency check for our constraints on the data, we also performed a BAO-only analysis, which was proven to be more robust to the assumption of the fiducial cosmology~\citep{Carter2019}. We found good agreement between the full-shape and BAO-only analyses, which demonstrates the robustness of the methodology given the current statistical precision. In the line with our findings from DR14~\citep{Hou2018, Gil-Marin2018, Zarrouk2018}, we demonstrate that quasars are robust tracers of the underlying matter field. 

Our work has several points in common with those of our companion papers. The configuration-space BAO-only analyses of \citet{Tamone2020} and \citet{Bautista2020} are based on a similar method as the one used here. Regarding the modelling of the full-shape of two point statistics, the predictions of RESPRESSO used here were also tested in the analysis of the LRG sample \citep{Bautista2020}. All analyses use a consistent definition of the total systematic error budget. The differences in the statistical uncertainties of the results inferred from each sample are mainly due to their different volumes and number densities. 

The distance and growth of structure measurements inferred from all samples are summarised in Table 3 by~\citet{Mueller2020}. A comparison of these measurements with the predictions of the best-fitting $\Lambda$CDM model to Planck CMB data shows good agreement. The largest deviations in the distance measurements are given by the LRG BAO measurements, which are ${\sim}1.7\sigma$ lower. Regarding $f\sigma_8$, the consensus QSO measurement is ${\sim}1.9\sigma$ higher, while the ELG analysis is lower by ${\sim} 1.4\sigma$. However, these measurements cover a wide range of the redshift and showed no clear deviation from the $\Lambda$CDM paradigm. 

The different eBOSS samples overlap in redshifts and can be studied using the multi-tracer technique~\citep{Seljak2009}. \citet{Wang2020} and~\citet{ZhaoGB2020} show that the combination of the eBOSS tracers can further help to tighten the constraints on cosmological parameters.

The cosmological implications of our results and those of our companion papers will be explored in \citet{Mueller2020}.

\section*{Acknowledgements}

JH and AGS would like to thank Daniel Farrow, Martha Lippich, and Agne Semenaite for the helpful discussions. JH would like to thank Hao Ding for the support. This research was supported by the Excellence Cluster ORIGINS, which is funded by the Deutsche Forschungsgemeinschaft (DFG, German Research Foundation) under Germany's Excellence Strategy - EXC-2094 - 390783311.

G.R. acknowledges support from the National Research Foundation of Korea (NRF) through Grants No. 2017R1E1A1A01077508 and No. 2020R1A2C1005655 funded by the Korean Ministry of Education, Science and Technology (MoEST), and from the faculty research fund of Sejong University.

Funding for the Sloan Digital Sky Survey IV has been provided by the Alfred P. Sloan Foundation, the U.S. Department of Energy Office of Science, and the Participating Institutions. SDSS-IV acknowledges
support and resources from the Center for High-Performance Computing at
the University of Utah. The SDSS web site is www.sdss.org.

SDSS-IV is managed by the Astrophysical Research Consortium for the 
Participating Institutions of the SDSS Collaboration including the 
Brazilian Participation Group, the Carnegie Institution for Science, 
Carnegie Mellon University, the Chilean Participation Group, the French Participation Group, Harvard-Smithsonian Center for Astrophysics, 
Instituto de Astrof\'isica de Canarias, The Johns Hopkins University, Kavli Institute for the Physics and Mathematics of the Universe (IPMU) / 
University of Tokyo, the Korean Participation Group, Lawrence Berkeley National Laboratory, 
Leibniz Institut f\"ur Astrophysik Potsdam (AIP),  
Max-Planck-Institut f\"ur Astronomie (MPIA Heidelberg), 
Max-Planck-Institut f\"ur Astrophysik (MPA Garching), 
Max-Planck-Institut f\"ur Extraterrestrische Physik (MPE), 
National Astronomical Observatories of China, New Mexico State University, 
New York University, University of Notre Dame, 
Observat\'ario Nacional / MCTI, The Ohio State University, 
Pennsylvania State University, Shanghai Astronomical Observatory, 
United Kingdom Participation Group,
Universidad Nacional Aut\'onoma de M\'exico, University of Arizona, 
University of Colorado Boulder, University of Oxford, University of Portsmouth, 
University of Utah, University of Virginia, University of Washington, University of Wisconsin, 
Vanderbilt University, and Yale University.

In addition, this research relied on resources provided to the eBOSS
Collaboration by the National Energy Research Scientific Computing
Center (NERSC).  NERSC is a U.S. Department of Energy Office of Science
User Facility operated under Contract No. DE-AC02-05CH11231.

\section*{Data availability}
The correlation functions, covariance matrices, and resulting likelihoods for cosmological parameters are (will be made) available (after acceptance) via the SDSS Science Archive Server (https://sas.sdss.org/), with the exact address tbd.




\bibliographystyle{mnras}
\bibliography{ref} 




\appendix

\section{Consistency check on the potential systematics}
\label{appendix:miscellaneous_syst}

\subsection{Impact of the effective redshift definition}
\label{appendix:zeff_definition}

We define the effective redshift, $z_{\rm zeff}$, of the quasars using Eq.~\eqref{eqn:zeff_eboss}, which matches the definition used for the other eBOSS tracers. The main motivation for this definition is that, in practice, the correlation function is measured in terms of weighted pairs. Taylor expanding the correlation function about $z_{\rm zeff}$ gives
 
\begin{equation}
\xi(z)=\xi(z_{\rm zeff})+\left.\frac{d \xi}{d z}\right|_{z=z_{\rm zeff}}(z_{\rm zeff}-z)+O\left[(z_{\rm zeff}-z)^{2}\right],
\label{eqn:corr_zeff_taylor_expand}
\end{equation}
and we effectively measure the correlation function
\begin{equation}
\hat{\xi}=\frac{\sum_{i,j} \xi\left(z\right)|_{z=(z_i+z_j)/2} w_{i}w_{j}}{\sum_{i,j} w_{i}w_{j}}, \quad \textrm{for} \quad \Delta {\bf s}_{i,j} \in [s_{\rm min}, s_{\rm max}].
\end{equation}
The first order term in Eq.~\eqref{eqn:corr_zeff_taylor_expand} vanishes if we define the effective redshift as in Eq.~\eqref{eqn:zeff_eboss}. However, there is some ambiguity in the definition of the effective redshift. First, objects at higher redshifts are more likely to receive a larger weight, and the effect of this should in principle also being taken into account. Secondly, there is also potential ambiguity in the range of pair separations, $\Delta {\mathbf s}_{i,j}$, that are summed over.  
In Table~\ref{tab:zeff_def} we list the value of $z_{\rm zeff}$ obtained using different definitions. We find that the range of pair separations has only a marginal impact, while the pair-defined $z_{\rm zeff}$ differ by ${\sim}3\%$ compared to the definition we used for the DR14 analysis. Nevertheless, we compare the inferred cosmological parameters obtained using either of the effective redshift definitions, and the difference is small compared to the statistical error (see Table~\ref{tab:syst_test_dr16}).

\begin{table}
\caption{Effective redshift, $z_{\rm zeff}$, of the NGC, SGC, and combined NGC$+$SGC, for different definitions of $z_{\rm zeff}$. The first row uses the definition of $z_{\rm zeff}$ used in the DR14 analysis. The second and third rows show the definition used in our DR16 analysis, with different ranges of pair separations.}
\begin{tabular}{@{}clll@{}}
\toprule
\multicolumn{1}{c}{ $z_{\rm eff}$} & 
NGC & SGC & NS\\ \midrule
$\frac{\sum_i(w_i*z_i)}{\sum_i w_i }$ & 1.512 & 1.520 & 1.515 \\ \cmidrule(r){1-1}
\begin{tabular}[c]{@{}c@{}}
$\sum_{i,j} (w_i w_j*(z_i+z_j)/2)\sum_{i,j} w_i w_j$
 \\ s $\in$ {[}25, 120{]} $\mpch$\end{tabular} & 1.474 & 1.491 & 1.480 \\ \cmidrule(r){1-1}
\begin{tabular}[c]{@{}c@{}}
$\sum_{i,j} (w_i w_j*(z_i+z_j)/2)\sum_{i,j} w_i w_j$
 \\ s $\in$ {[}20, 160{]} $\mpch$\end{tabular} & 1.474 & 1.491 & 1.480 \\ \bottomrule
\end{tabular}
\label{tab:zeff_def}
\end{table}

\section{Analytical covariance matrix estimation with Gaussian approximation}
\label{appendix:analytical_gaussian_covmat}

Following the prescription in \citet{Grieb2016}, the covariance of the Legendre multipoles in configuration space can be expressed as
\begin{equation}
C_{\ell_{1} \ell_{2}}^{\xi}\left(s_{i}, s_{j}\right)=\frac{\mathrm{i}^{\ell_{1}+\ell_{2}}}{2 \uppi^{2}} \int_{0}^{\infty} k^{2} \sigma_{\ell_{1} \ell_{2}}^{2}(k) \bar{j}_{\ell_{1}}\left(k s_{i}\right) \bar{j}_{\ell_{2}}\left(k s_{j}\right) \mathrm{d} k,
\label{eqn:gaussian_cov_mul}
\end{equation}
where the $\bar{j}_{\ell}$ is the bin-averaged spherical Bessel function over a volume $V_{s_i}=4 \uppi\left(s_{i, \max }^{3}-s_{i, \min }^{3}\right) / 3$ around a bin $s_i$
\begin{equation}
\bar{j}_{\ell}\left(k s_{i}\right) \equiv \frac{4 \uppi}{V_{s_{i}}} \int_{s_{i}-\Delta s / 2}^{s_{i}+\Delta s / 2} s^{2} j_{\ell}(k s) \mathrm{d} s.
\end{equation}
The per-mode covariance in Eq.~\eqref{eqn:gaussian_cov_mul} is given by

\begin{equation}
\sigma_{\ell_{1} \ell_{2}}^{2}(k) \equiv \frac{\left(2 \ell_{1}+1\right)\left(2 \ell_{2}+1\right)}{V_{\mathrm{eff}}} \int_{-1}^{1}\left[P(k, \mu)+\frac{1}{\overline{n}}\right]^{2} \mathcal{L}_{\ell_{1}}(\mu) \mathcal{L}_{\ell_{2}}(\mu) \mathrm{d} \mu.
\end{equation}
In the case of cubic simulation box the volume is estimated by $V_{\rm eff} = L^3$. For a survey with selection function and varying radial number density, the volume $V_{\rm eff}$ can be defined as the integral of the amplitude squared of the survey window function, $Q(\mathbf{k})$, in Fourier space
\begin{equation}
V_{\mathrm{eff}}^{-1} \equiv \int \frac{d^{3} \mathbf{k}}{(2 \uppi)^{3}}|Q(\mathbf{k})|^{2}= \frac{\int d^{3} \mathbf{x} \overline{n}^{4}(\mathbf{x}) w^{4}(\mathbf{x})}{\left[\int d^{3} \mathbf{x} \overline{n}^{2}(\mathbf{x}) w^{2}(\mathbf{x})\right]^{2}}.
\label{eqn:veff}
\end{equation}
The window function is designed to maximize the signal-to-noise ratio by weighting the density fluctuation field. For the non-blind mock analysis, the weight $w(\bf{x})$ is simply taken to be the  FKP weight defined by Eq.~\eqref{eqn:wfkp},
with $P_{0}=6000 \, h^{-3} \mathrm{Mpc}^{3}$. 
Here we have neglected the effects from super-survey modes, such as the beat-coupling and the local average effect. The first effect is induced by the survey window that mixes the small- and large-scale modes. The second effect is caused by the zero-mode-modulated average density that is estimated from a limited survey. These two effects cancel each other and leaves only up to $\sim 10\%$ excess in the original variance~\citep{Putter2012}. In the fitting scales we consider in this paper, the analytical Gaussian covariance should remain a quite good approximation. This was also observed in ~\citet{Lippich2019} from comparing to a set of fast mocks. Currently, the downside of the method is that it does not include the survey geometry or the window function, therefore we only use it when the analysing the \textsc{OuterRim} N-body mocks.

\section{Post-processing the EZmocks}
\label{appendix:postprocess_ezmocks}
We start with mocks that include the angular selection function. In the first step, contamination from the data is added to the mocks, which includes stars, objects with redshift failures, wrong objects classes, objects that have no chances to receive a good redshifts (e.g. due to unplugged fibres, etc.) and the not tiled objects. For the photometric correction, a fit is applied to the minimize the trend in the ratio $n_{\rm QSO}/n_{\rm random}$ in the extinction corrected $g$-band depth and the stellar density. The sample of quasars used in the data analysis includes a special Legacy group (see last paragraph in Section~\ref{sec:overview_eboss}). These objects do not receive any spectroscopic or collision correction and are separately assigned with the tag ${\rm IMATCH} = 2$ to distinguish from the CORE sample. The fibre collision effect is added to the \textsc{EZmocks} using the \textsc{FiberCollision} module from \textsc{Nbodykit}~\citep{Hand2018}. After identifying the close pairs, the objects that fall in the same collision groups are up-weighted by the ratio $n_{\rm hasfibre}/n_{\rm total}$. The redshift failure corrections are applied using the same definition as for the real data (see discussion in Section~\ref{sec:robustness_tests}), and the mock objects inherit the property from the data objects through closest angular-matching.

\section{Results expressed in alternative basis}
\label{appendix:alter_basis}
In this appendix, we consider the results given in different alternative basis. First, we list the results expressed in the basis of comoving angular diameter $D_{\rm M}/r_{\rm drag}$\footnote{The comoving angular diameter is defined as $D_{\mathrm{M}}=\int_{0}^{z} \frac{c d z^{\prime}}{H\left(z^{\prime}\right)}$.}, Hubble parameter, $H r_{\rm drag}$ and $f \sigma_8$. Second, we list the results expressed in the basis of comoving angular diameter $D_{\rm M}/r_{\rm drag}$, Hubble distance, $D_{\rm H}/r_{\rm drag}$ and $f \sigma_{12}$.

\subsection{$D_{\rm M}/r_{\rm drag}$-$H r_{\rm drag}$-$f \sigma_8$ basis}
\subsubsection{BAO-only fits}
We convert the BAO $\alpha_{\perp}$ and $\alpha_{\parallel}$ results to constraints on the comoving angular diameter and Hubble parameter with:
\begin{align}
&D_{\rm M}(z_{\rm eff}=1.48)/r_{\rm drag} = 30.82 \pm 0.80 \pm 0.21, \\
&H(z_{\rm eff}=1.48) \, r_{\rm drag} = (2.267 \pm 0.093 \pm 0.025) \times 10^{4}\, \rm km/s.
\end{align} 
The first error denotes the statistical uncertainty, which is rescaled by $\sqrt{M}=1.010$, the second error denotes the uncertainty inferred from the \textsc{OuterRim} mock challenge (without the observational uncertainties).

\subsubsection{Full-shape analysis in configuration space}
\begin{align}
&D_{\rm M}(z_{\rm eff}=1.48)/r_{\rm drag} = 30.66 \pm 0.84 \pm 0.25, \\
&H(z_{\rm eff}=1.48) \, r_{\rm drag} = (2.289 \pm 0.085 \pm 0.029) \times 10^{4}\, \rm km/s, \\
&f\sigma_8(z_{\rm eff}=1.48) = 0.439 \pm 0.046 \pm 0.014,
\end{align} 
where the first error denotes the statistical uncertainty, which is rescaled by $\sqrt{M}=1.036$, the second error denotes the systematics uncertainty inferred from the \textsc{OuterRim} mock challenge as well as the observational systematics inferred from the \textsc{EZmocks} by adding them in quadrature.
The covariance matrix is given by
\begin{equation}
\mathbf{C}_{\lambda} = \begin{blockarray}{ccc}
D_{\rm M}/r_{\rm drag} & H \, r_{\rm drag} & f\sigma_{8} \vspace{1mm} \\
\begin{block}{(ccc)}
  7.709\times 10^{-1} & 9.780\times 10^{1} & 1.750\times 10^{-2} \\ 
  -                   & 8.007\times 10^{5} & 1.085 \times 10^{1} \\
  -                   & -                  & 2.308\times 10^{-3}\\
\end{block}
\end{blockarray}
\end{equation}

\subsubsection{Combined full-shape analysis}
\begin{align}
&D^{\bf{c}}_{\rm M}(z_{\rm eff}=1.48)/r_{\rm drag} = 30.21 \pm 0.79, \\
&H^{\bf{c}}(z_{\rm eff}=1.48) \, r_{\rm drag} = (2.255 \pm 0.079) \times 10^{4}\, \rm km/s, \\
&f\sigma_8^{\bf{c}}(z_{\rm eff}=1.48) = 0.459 \pm 0.045,
\end{align} 
where the errors include the statistical and systematic uncertainties (inferred from \textsc{OuterRim} mock challenge and \textsc{EZmocks}). The covariance matrix for the combined data is given by  
\begin{equation}
\mathbf{C}^{\bf{c}}_{\lambda} = \begin{blockarray}{ccc}
D_{\rm M}/r_{\rm drag} & H \, r_{\rm drag} & f\sigma_{8} \vspace{1mm} \\
\begin{block}{(ccc)}
  6.222\times 10^{-1} & -4.374\times 10^{1}& 2.223\times 10^{-2} \\
  -                   & 6.216\times 10^{5} & 1.191 \times 10^{1} \\
  -                   & -                  & 2.010\times 10^{-3}\\
\end{block}
\end{blockarray}
\end{equation}{}

\subsection{$D_{\rm M}/r_{\rm drag}$-$D_{\rm H}/r_{\rm drag}$-$f \sigma_{12}$ basis}

\citet{Sanchez2020} showed that the constraints on $f\sigma_8(z)$ depend on the particular value of $h$ assumed in the full 
shape analysis, and that a more correct constraints on this combination should be marginalized over the uncertainties on $h$.
Then, using the growth rate measurements expressed in terms of $f\sigma_8(z)$ to constrain cosmological parameters leads to 
both a potential systematic bias and an underestimation of the uncertainties. This problems is solved if the growth rate 
measurements are expressed instead in terms of the combination $f\sigma_{12}$, where $\sigma_{12}$ represents the rms of the 
mass contained in a sphere of radius $12\,{\rm Mpc}$. As the BAO-only measurements are not affected by this choice, 
we list here only the results for the full-shape analysis in the configuration-space and combined cases.

\subsubsection{Full-shape analysis in configuration space}

\begin{align}
&D_{\rm M}(z_{\rm eff}=1.48)/r_{\rm drag} = 30.66 \pm 0.84 \pm 0.25, \\
&D_{\rm H}(z_{\rm eff}=1.48)/r_{\rm drag} = 13.11 \pm 0.49 \pm 0.17 , \\
&f\sigma_{12}(z_{\rm eff}=1.48) = 0.435 \pm 0.046 \pm 0.012.
\end{align} 
Using the same convention as above: the first error denotes the statistical uncertainty, which is rescaled by $\sqrt{M}=1.036$, 
the second error denotes the systematic uncertainty by adding the error inferred from \textsc{OuterRim} mock challenge as 
well as the one from the \textsc{EZmocks} in quadrature. The blind mock challenge test shows that the relative error on the 
growth rate measurement improves from $2.8\%$ to $2.5\%$ when presenting the results in the $f\sigma_{12}$ basis.
Finally, the covariance matrix is given by
\begin{equation}
\mathbf{C}_{\lambda} = \begin{blockarray}{ccc}
D_{\rm M}/r_{\rm drag} & D_{\rm H}/r_{\rm drag} & f\sigma_{12} \vspace{1mm} \\
\begin{block}{( ccc )}
  7.709\times 10^{-1} & -5.656\times 10^{-2} & 1.733\times 10^{-2} \\ 
  - & 2.640\times 10^{-1} & -6.145\times 10^{-3} \\
  - & -      & 2.227\times 10^{-3}\\
\end{block}
\end{blockarray}
\end{equation}

\subsubsection{Combined full-shape analysis }
\begin{align}
&D^{\bf{c}}_{\rm M}(z_{\rm eff}=1.48)/r_{\rm drag} = 30.21 \pm 0.79, \\
&D^{\bf{c}}_{\rm H}(z_{\rm eff}=1.48)/r_{\rm drag} = 13.23 \pm 0.47 \\
&f\sigma_{12}^{\bf{c}}(z_{\rm eff}=1.48) = 0.458 \pm 0.044,
\end{align} 
where the errors include both the statistical and systematic uncertainties. For the combined results, there is also a slight 
improvement in the relative error from $2\%$ to $1.8\%$ on the growth rate parameter based on the blind mock challenge results.
The final covariance matrix for the combined data reads as the following 
\begin{equation}
\mathbf{C}^{\bf{c}}_{\lambda} = \begin{blockarray}{ccc}
D_{\rm M}/r_{\rm drag} & D_{\rm H}/r_{\rm drag} & f\sigma_{12} \vspace{1mm} \\
\begin{block}{(ccc)}
  6.227\times 10^{-1} & 1.424\times 10^{-2} & 2.235\times 10^{-2} \\
  -                   & 2.195\times 10^{-1} & -7.246 \times 10^{-3} \\
  -                   & -                   & 1.958\times 10^{-3}\\
\end{block}
\end{blockarray}
\end{equation}{}


\bsp	
\label{lastpage}
\end{document}